\documentclass[11pt]{article}

\usepackage{arxiv}
\usepackage[utf8]{inputenc} 
\usepackage[T1]{fontenc}    
\usepackage{hyperref}       
\usepackage{url}            
\usepackage{booktabs}       
\usepackage{tabularx}
\usepackage{amsfonts}       
\usepackage{nicefrac}       
\usepackage{microtype}      
\usepackage{amssymb}
\usepackage{amsmath}
\usepackage{pifont}
\usepackage{tikz}
\usetikzlibrary{positioning}
\usepackage{subcaption}
\usepackage[linesnumbered,ruled]{algorithm2e}
\usepackage[normalem]{ulem}
\hypersetup{colorlinks,urlcolor=blue}
\usepackage[bbgreekl]{mathbbol}
\usepackage{etoolbox}
\usepackage{mathtools}
\usepackage{color}
\usepackage{xcolor}
\usepackage[stable]{footmisc}
\definecolor{PineGreen}{rgb}{0.0, 0.65, 0.35}  
\usepackage{placeins}
\usepackage{xspace}
\usepackage[style=numeric-comp,backend=biber, giveninits=true, sorting=none,maxbibnames=999, maxcitenames=999]{biblatex}
\usepackage{ntheorem}
\theoremstyle{remark}
\usepackage{cancel}
\newtheorem{remark}{Remark}[section]
\DeclareSymbolFontAlphabet{\mathbb}{AMSb}
\DeclareSymbolFontAlphabet{\mathbbl}{bbold}
\DeclarePairedDelimiterX{\infdivx}[2]{}{}{%
  {#1}\;\big\|\;{#2}%
}

\newcommand{\eg}{e.\,g.,\xspace}
\newcommand{\ie}{i.\,e.,\xspace}
\newcommand{\Eg}{E.\,g.,\xspace}

\newcommand{\wrt}{w.\,r.\,t.\xspace}
\newcommand{\iid}{i.\,i.\,d.\xspace}

\newcommand{\dd}{\,\mathrm{d}} 

\newcommand{\bs}{\boldsymbol}
\newcommand{\bx}{{\bs{x}}}
\newcommand{\bxele}{{\bs{x}_{\mathrm{ele}}}}
\newcommand{\by}{{\bs{y}}}

\newcommand{\txgt}{\tilde{x}_{\mathrm{gt}}}

\newcommand{\yobs}{{y_{\text{obs}}}}

\newcommand{\Yobs}{{Y_{\text{obs}}}}

\newcommand{\ylf}{{y_{\text{LF}}}}

\newcommand{\yhf}{{y_{\text{HF}}}}

\newcommand{\Ylf}{{Y_{\text{LF}}}}
\newcommand{\Yhf}{{Y_{\text{HF}}}}

\newcommand{\byhf}{\boldsymbol{y}_{\text{HF}}}

\newcommand{\byhfi}{\boldsymbol{y}_{\text{HF},i}}

\newcommand{\byobsi}{\boldsymbol{y}_{\text{obs},i}}

\newcommand{\Mlf}{\mathfrak{M}_{\text{LF}}}
\newcommand{\Mhf}{\mathfrak{M}_{\text{HF}}}

\newcommand{\zlf}{{z_{\text{LF}}}}

\newcommand{\Zlf}{{Z_{\text{LF}}}}

\newcommand{\Kmat}{K}

\newcommand{\ntrain}{\mathrm{n}_{\text{train}}}
\newcommand{\nrefine}{\mathrm{n}_{\text{refine}}}

\newcommand{\numobs}{\mathrm{n}_{\text{obs}}}

\newcommand{\Ex}[2]{{\mathbb{E}}_{#1}\left[#2\right]}

\newcommand{\be}{\begin{equation}}
\newcommand{\ee}{\end{equation}}
\newcommand{\bd}{\begin{description}}
\newcommand{\ed}{\end{description}}
\newcommand{\bi}{\begin{itemize}}
\newcommand{\ei}{\end{itemize}}

\newcommand{\bc}{\boldsymbol{c}}
\newcommand{\bci}{\boldsymbol{c}_{i}}

\newcommand{\phf}{p_{\text{HF}}}
\newcommand{\plf}{p_{\text{LF}}}
\newcommand{\pmf}{p_{\text{MF}}}
\newcommand{\qmf}{q_{\text{MF}}}

\newcommand{\lhf}{\mathcal{L}_{\text{HF}}}
\newcommand{\llf}{\mathcal{L}_{\text{LF}}}
\newcommand{\lmf}{\mathcal{L}_{\text{MF}}}

\newcommand{\bmu}{\boldsymbol{\mu}}

\newcommand{\diag}{\mathrm{diag}}
\newcommand{\Kmf}{K_{\text{MF}}}

\newcommand{\bu}{\boldsymbol{u}}
\newcommand{\bdisp}{\boldsymbol{d}}
\newcommand{\bn}{\boldsymbol{n}}
\newcommand{\bN}{\boldsymbol{N}}
\newcommand{\bnil}{\boldsymbol{0}}
\newcommand{\bF}{\boldsymbol{F}}
\newcommand{\bS}{\boldsymbol{S}}
\newcommand{\bE}{\boldsymbol{E}}
\newcommand{\bC}{\boldsymbol{C}}

\newcommand{\bz}{\boldsymbol{z}}

\newcommand{\bxgt}{\boldsymbol{x}_{\mathrm{gt}}}
\newcommand{\bphi}{\boldsymbol{\phi}}

\newcommand{\br}{\boldsymbol{r}}

\newcommand{\nsamples}{n_{\mathrm{samples}}}
\newcommand{\ntau}{n_{\tau}}

\newcommand{\Alf}{\mathfrak{A}_{\mathrm{LF}}}
\newcommand{\lfm}{$\mathrm{LF}_{\mathrm{m}}\ $}
\newcommand{\lfb}{$\mathrm{LF}_{\mathrm{b}}\ $}
\newcommand{\Yhfgt}{Y_{\mathrm{HF,gt}}}
\newcommand{\bw}{\boldsymbol{w}}
\newcommand{\commentout}[1]{}
\newsavebox{\blueline}
\savebox{\blueline}{\tikz[baseline=-0.5ex]  \draw[blue, thick] (0,0) -- (0.3,0);}

\definecolor{mydarkgreen}{RGB}{0,100,0}
\newsavebox{\darkgreenline}
\savebox{\darkgreenline}{\tikz[baseline=-0.5ex]  \draw[mydarkgreen, thick] (0,0) -- (0.3,0);}

\newsavebox{\bluedashed}
\savebox{\bluedashed}{\tikz[baseline=-0.5ex]  \draw[blue, thick, dashed] (0,0) -- (0.3,0);}

\newsavebox{\greydot}
\savebox{\greydot}{\tikz[baseline=-0.5ex] \fill[gray] (0,0) circle (0.2ex);} 

\definecolor{customLightGrey}{rgb}{0.6, 0.6, 0.6} 

\newsavebox{\lightgreybox}
\savebox{\lightgreybox}{\tikz\fill[customLightGrey] (0,0) rectangle (0.4,0.2);} 

\newsavebox{\reddot}
\savebox{\reddot}{\tikz[baseline=-0.5ex] \fill[red] (0,0) circle (0.3ex);}

\newcommand{\refeqp}[1]{Equation \eqref{#1}}

\usepackage{soul}
\newcommand{\add}[1]{#1}

\newlength{\commentindent}

\allowdisplaybreaks 
\title{Efficient Bayesian multi-fidelity inverse analysis for expensive and non-differentiable physics-based simulations in high stochastic dimensions}

\author{
  Jonas Nitzler\\
  Institute for Computational Mechanics\\
  Technical University of Munich\\
  D-85748 Garching b. München\\
  \texttt{jonas.nitzler@tum.de} \\
  \And
  Bu\u{g}rahan Z. Temür\\
  Institute for Computational Mechanics\\
  Technical University of Munich\\
  D-85748 Garching b. München\\
  \texttt{bugrahan.temuer@tum.de} \\
    \And
 Phaedon-S. Koutsourelakis \\
 Professorship of Data-driven Materials Modeling\\
   Technical University of Munich\\
  D-85748 Garching b. München\\
  \texttt{p.s.koutsourelakis@tum.de} \\
  \And
  Wolfgang A. Wall\\
Institute for Computational Mechanics\\
  Technical University of Munich\\
  D-85748 Garching b. München\\
  \texttt{wolfgang.a.wall@tum.de} \\
}

\begin{document}

\maketitle

\begin{abstract}
High-dimensional Bayesian inverse analysis ($\dim \gg 100$) is mostly unfeasible for computationally demanding, nonlinear physics-based high-fidelity (HF) models. Usually, the use of more efficient gradient-based inference schemes is impeded if the multi-physics models are provided by complex legacy codes. Adjoint-based derivatives are either exceedingly cumbersome to derive or nonexistent for practically relevant large-scale nonlinear and coupled multi-physics problems. Similarly, holistic automated differentiation \wrt primary variables of multi-physics codes is usually not yet an option and requires extensive code restructuring if not considered from the outset in the software design. This absence of differentiability further exacerbates the already present computational challenges.
To overcome the existing limitations, we propose a novel inference approach called \emph{Bayesian multi-fidelity inverse analysis (BMFIA)}, which leverages simpler and computationally cheaper lower-fidelity (LF) models that are designed to provide model derivatives. 
BMFIA learns a simple, probabilistic dependence of the LF and HF models, which is then employed in an altered likelihood formulation to statistically correct the inaccurate LF response. From a Bayesian viewpoint, this dependence represents a multi-fidelity \add{(MF)} conditional density (discriminative model). We demonstrate how this \add{MF} conditional density can be learned robustly in the \emph{small data regime} from only a few HF and LF simulations (50 to 300), which would not be sufficient for naive surrogate approaches. 
The formulation is fully differentiable and allows the flexible design of a wide range of LF models. 
We demonstrate that BMFIA solves Bayesian inverse problems for scenarios that used to be prohibitive, such as finely-resolved \add{and hence high-dimensional} spatial reconstruction problems \add{in two-dimensional Euclidean domains with static posteriors,}  \add{given} nonlinear and transient coupled poro-elastic media physics. We show that the resulting \add{static MF} posteriors are in excellent agreement with the (usually inaccessible) \add{HF} posteriors or ground-truth data \add{and note that extending the framework to arbitrary three-dimensional domains is a natural and important direction for future work}.
\end{abstract}


\section{Introduction}
\label{sec:introduction}
We are interested in solving Bayesian inverse problems (BIPs) \cite{biegler_large-scale_2010} for computationally expensive, large-scale, coupled, nonlinear, multi-physics problems in high stochastic dimensions. Specifically, we solve \add{static} Bayesian spatial reconstruction problems for forward models often available as mature, albeit black-box, legacy codes. \add{By \emph{static} BIPs, we mean that the posterior distribution is not time-dependent; however, the forward model might be transient}. High-dimensional \add{BIPs} necessitate gradient-based inference methods to converge efficiently. Unfortunately,  HF legacy codes usually do not supply model gradients \wrt parameterizations (spatial fields) or primary variables. Adjoint formulations, which would provide such gradients, are typically cumbersome to derive or simply unavailable for multi-physics models. The computational demands of complex nonlinear multi-physics models and the extensive simulations required by state-of-the-art Bayesian inference schemes hinder accurate inference. We provide a short review of state-of-the-art approaches \add{focusing on MF methods,} to solve high-dimensional  \add{BIPs}, which is organized in the following  \add{three} categories. \add{Since our focus is on very high-dimensional scenarios with computationally expensive models, we omit a discussion of plain surrogate-based approaches, which are not practical in this setting and known to deteriorate with increasing dimension (curse of dimensionality); more involved variants (e.g., physics-informed surrogates or surrogates embedded in parts of an inference algorithm) are, however, considered in the following.}

\add{The} idea of exploiting computationally cheaper model versions directly in the Bayesian inference procedure, especially in Markov Chain Monte Carlo (MCMC) and sequential Monte Carlo (SMC) samplers, has been proposed \add{for a long time}. \add{MF} MCMC approaches in the context of delayed-rejection mechanisms \cite{tierney1999some,christen2005markov}, or \add{MF} approaches to motivate inexpensive proposal densities in MCMC \cite{fox1997sampling,peherstorfer2019transport} are two examples. Furthermore, multi-level Monte Carlo (MLMC) methods exploit the (linear) correlation structure between model hierarchies \cite{scheichl2017quasi,giles2008multilevel, beskos2018multilevel,latz2018multilevel, heinrich2001multilevel, peherstorfer2019multifidelity, peherstorfer2018multifidelity}. Similar ideas were also employed by \add{MF} importance sampling \cite{peherstorfer2016multifidelity} and \add{MF} sequential Monte Carlo methods \cite{koutsourelakis2009multi, richter2024bayesian}. While the strategies above recover the HF posterior, their savings depend on the error between \add{LF and HF} models, which restricts the flexibility in choosing appropriate LF models. Furthermore, one usually has very little control over the necessary number of HF model calls and bias correction. Mostly, such \add{MF} sampling approaches are of a frequentist nature, \ie they offer only asymptotic metrics of their error estimator, which is prohibitive for high-dimensional inverse problems in combination with expensive legacy codes. Beyond that, these approaches mostly do not support gradient-boosting and require a prohibitive number of solver calls for high-dimensional settings.

A \add{second} strategy involves methods that do not require the HF model per se but directly \emph{learn} from first principles and residuals of the underlying governing equations \cite{zhu2019physics, raissi2019physics}. In the context of \add{MF} surrogate modeling, leveraging physical domain knowledge has proven highly effective in augmenting surrogate models, leading to improved accuracy, reduced data requirements, and enhanced extrapolation capabilities \cite{zhu2019physics, rixner2021probabilistic, patel2022solution, xia2022bayesian, hou2019solving, mucke2023markov}.
However, (physics-informed) approaches can face challenges in high-dimensional stochastic settings and are currently more commonly applied to problems involving simpler physical models. In particular, extending physics-informed learning methods and surrogates to complex multi-physics scenarios, especially in contexts where mature legacy codes cannot easily be replaced, remains an open and active area of research. While residuals can be a helpful information source, the number of such evaluations can become comparable to multiple expensive forward model solves. 

\add{Third}, dimensionality-reduction techniques \cite{cui2015data, marzouk2009dimensionality} attempt to mitigate computational demands by reducing the dimensionality of input parameter spaces. These approaches typically learn the dimensionality reduction mappings offline and do not adapt dynamically to the posterior distribution, often capturing only a single posterior mode. Furthermore, dimensionality reduction alone is usually insufficient to achieve substantial computational savings and thus must be combined with additional surrogate models. 

We propose a novel, efficient method for high-dimensional \add{BIPs} called \emph{multi-fidelity Bayesian inverse analysis (BMFIA)} to address some of the existing challenges. 
In particular, \add{BMFIA} handles three significant shortcomings of existing methods to solve high-dimensional nonlinear \add{BIPs}. First is the computational burden of HF models, and second is the absence of model gradients for coupled physics-based codes. BMFIA builds upon ideas of our Bayesian multi-fidelity Monte Carlo (BMFMC) method \cite{nitzler_generalized_bmfmc, koutsourelakis2009accurate, biehler2015towards} for efficient uncertainty quantification, initially developed in \cite{koutsourelakis2009accurate}. BMFIA facilitates using gradient-based inference algorithms by requiring only gradients from simpler LF models. This capability significantly enhances the solvability of challenging coupled nonlinear physics problems. A third strength of BMFIA lies in its flexibility to incorporate \add{a wide range of} LF models, \add{with great flexibility towards their actual quality, due to a probabilistic treatment}.
\add{BMFIA relies on four key assumptions: (i) A cheaper low‑fidelity (LF) model must be available (or is easy to implement). The latter is assumed to be differentiable with respect to the parameters of interest, typically through an adjoint formulation or, alternatively, as a physics‑informed surrogate. (ii) The LF model outputs must exhibit at least a weak statistical relationship with the HF outputs. While absolute deviations, noise, or (moderate) nonlinear deviations are acceptable, the LF model should capture the overall rough statistical trend of the HF to enable informative posterior approximations. (iii) The generation of a relatively small set of paired HF-LF training data (typically 200–300 samples for good results) is feasible (embarrassingly parallel task), with the primary constraint being the cost of HF evaluations and available computational resources. (iv) Sufficient computational resources, such as a GPU, are available to train the probabilistic regression model for the multi‑fidelity conditional. These costs are usually negligible compared to the cost of HF simulations for the problems considered (at least in 2D).}
\add{While \emph{BMFIA} is conceptually inspired by our \emph{BMFMC} method for forward uncertainty quantification \cite{nitzler_generalized_bmfmc}, it represents a substantial generalization to a far more challenging setting. In forward UQ, using an MF expansion is mathematically equivalent to the original formulation, but in \emph{BMFIA}, introducing the MF conditional into the likelihood represents a novel and genuine approximation of the original HF likelihood. Moreover, the high-dimensional BIPs we investigate require end-to-end differentiability to allow gradient-based inference, which we achieve by combining LF model adjoints with stabilized probabilistic surrogates for entire random fields. Meanwhile, our former work on BMFMC only operated on single scalar outputs and their probability densities. Furthermore, the novel presentation presented in this work automatically infers optimal informative features from the simulation input to enhance the learning process of the MF conditional, a task that was still cumbersome and partially manual in our previous work on BMFMC.}

The remainder of the paper is structured as follows: In Section \ref{sec:methodology}, we first introduce the fundamental concepts of BMFIA. Specifically, Section \ref{sec:small_data_approx} demonstrates the efficient approximation of the \add{MF} conditional in a \emph{small data regime}. Subsequently, we outline an efficient sparse variational inference (VI) strategy in combination with Variational Bayes Expectation Maximization \add{(VB-EM)} in Subsection \ref{sec: svi} and derive the gradients of the \add{MF} log-likelihood function and its LF model. Section \ref{sec:demonstration} demonstrates the capabilities of our approach through two numerical examples. The first example in Section \ref{sec: darcy_flow} is a more controllable porous media flow example for which we calculate the reference HF posterior. Afterwards, the second example in Section \ref{sec: poro_elastic} is an actual coupled poro-elastic problem for which an HF posterior is not readily available. We summarize our findings and give a short outlook for BMFIA in Section \ref{sec:conclusion_outlook}.

\section{Methodology}
\label{sec:methodology}
We introduce our notation using the standard \add{BIP}, given in Equation \eqref{eqn: bayes}, as an example. Lowercase regular letters represent scalars, while lowercase boldface letters denote vector-valued quantities. Capital regular letters are used for matrices or tensors. We do not explicitly differentiate between random variables, their realizations, and deterministic variables, but we provide clarification when the context does not make it clear.
Let $\bx$ be a potentially high-dimensional input vector of an \add{HF}, deterministic, computational, physics-based model $\Mhf$ (\eg given as a legacy code)\add{, usually governed by a set of (coupled) partial differential equations (PDEs)}. We are particularly interested in, but the approach is not limited to, cases where $\bx$ represents a parameterization of a spatially varying field, such as they arise in FE formulations.
We further denote with $\byhfi=\Mhf(\bx,\bci)$ the vector-valued outputs of the HF model at coordinate $\bci$\footnote{\add{In particular, we mean a \emph{spatial coordinate} in the context of spatial reconstruction problems}}, which are arranged row-wise in a matrix $\Yhf=\Mhf(\bx)$.
Similarly, we consider a \add{deterministic} lower-fidelity model (LF) with the input $\bx$ and denote with $\Ylf=\Mlf(\bx)$ the corresponding outputs at the same coordinates $C=\{\bci\}$.
Finally, we denote with $\Yobs$ the matrix of observations, each row $i$ of which is associated with coordinates $\bci$, and $\byhfi$. \add{While the LF and HF models represent deterministic mappings, the observations are contaminated by random noise. 
In a Bayesian context, this uncertainty is modeled in the likelihood function while input uncertainties, before data is observed, are modeled by a prior density $p(\bx)$.}

In a Bayesian calibration problem, we seek the posterior distribution $\phf(\bx|\Yobs)$ of the model inputs given the observations $\Yobs$:
\begin{equation}
    \label{eqn: bayes}
    \underbrace{\phf(\bx|\Yobs)}_{\text{posterior}} = \frac{\int \overbrace{p(\Yobs|\Yhf)}^{\text{Noise model}} \cdot \overbrace{p(\Yhf|\bx)}^{\text{HF mapping}}\dd \byhf \cdot \overbrace{p(\bx)}^{\text{prior}}}{\int\int p(\Yobs|\Yhf)\cdot p(\Yhf|\bx)\dd\byhf \cdot p(\bx) \dd\bx}=\frac{\overbrace{\phf(\Yobs|\bx)}^{\text{likelihood}}\cdot \overbrace{p(\bx)}^{\text{prior}}}{\underbrace{\phf(\Yobs)}_{\text{evidence}}}
\end{equation}
We use subscripts such as HF, LF, or MF if the dependence of a distribution on a specific model is not directly apparent,   
as is the case with the likelihood $\phf(\Yobs|\bx)=\int p(\Yobs|\Yhf)~p(\Yhf|\bx)\dd\byhf$. In particular, we assume a deterministic simulation mapping $\Yhf=\Mhf(\bx)$, such that the HF conditional collapses to a Dirac distribution: $p(\Yhf|\bx)=\delta(\Yhf-\Mhf(\bx))$.
Furthermore, we define the logarithm of the likelihood as  $\lhf(\bx)=\log \phf(\Yobs|\bx)$, and similarly $\llf(\bx)=\log \plf(\Yobs|\bx)$
for the LF case, and $\lmf(\bx)=\log \pmf(\Yobs|\bx)$ 
for the BMFIA or \add{MF} log-likelihood. 

\add{
\paragraph{Problem setup}
In this work, we focus on high-dimensional spatial Bayesian inverse problems for complex, nonlinear, and potentially coupled multi-physics systems, where the unknowns (the model input for the BIP) are discretized constitutive fields, material parameters, or boundary conditions. 
Let $\bx \in \mathbb{R}^d$ denote the unknown input vector, which may represent such a discretized field. In general, capturing fine-scale details necessitates high-dimensional discretizations, \ie we are interested in cases that $d\gg1$. A prior distribution $p(\bx)$ encodes uncertainty in these inputs before data is observed.
We now consider two deterministic forward models:
\bi
\item The high-fidelity (HF) model $\Mhf(\bx)$ is accurate but computationally expensive. Very often in practice, this is in the form of a black-box, legacy code which is generally computationally costly but, more importantly, does not provide derivatives of the output with respect to $\bx$, which are, however, essential for high-dimensional BIPs. 
\item The low-fidelity (LF) model, $\Mlf(\bx)$, is cheaper but approximate. While we provide concrete examples later, we note that LF models can, in general, be easily constructed depending on the application, and examples include simplified physics-based models (\eg neglect of a physical coupling), reduced-order models, statistical surrogates, analytical approximations, and lower-resolution FE simulators with lower spatio-temporal resolutions. For single-physics LF models, adjoint or gradient formulations are, on the other hand, often available or at least simple to derive and implement, which is particularly important in high-dimensional settings. 
\ei
We aim to retain statistical similarity to the HF outputs while enabling efficient gradient computation. We will see in the numerical demonstration in Section~\ref{sec:demonstration} that BMFIA only requires a weak statistical similarity between LF and HF models, allowing substantial modeling flexibility while enabling accurate MF posterior approximations.
}

\subsection{The Bayesian multi-fidelity inverse problem (BMFIA)}
\label{sec:bmfia_conti}
\add{
To overcome the aforementioned difficulties of high-dimensional BIPs in combination with non-differentiable and expensive HF models, we introduce the LF feature vector
\[
\Zlf(\bx) = [\Ylf(\bx), X(\bx)],
\]
which in general consists  of the LF model outputs $\Ylf$ as well as a set of features of the input $\bx$ , which we denote with $\mathbf{X}(\mathbf{x})$. Here, $X$ represents the values of the (spatial) input fields at the same spatial coordinates $\{\bc_i\}|_{i=1}^{n}$ at which the output fields $\Ylf(\bx)$ are evaluated, and $\bx$ is a high-dimensional parameterization of these (input) fields. The discrete input field values $X$ and the field's parameterization $\bx$ (our model input) are related by a FE representation (see Remark \ref{rem: fe_approx_fields} for more details). We posit the density $p(\Yhf | \Zlf(\bx))$, \ie a stochastic map that, given the lower-fidelity features for a certain input value, attempts to predict the corresponding HF outputs. We call this stochastic map the multi-fidelity conditional (MF conditional). Since $\Zlf$ filters the full input $\bx$, resulting in information loss, the predictions will be fraught with uncertainties, which this density quantifies.  With the help of this density, we define the \textbf{multi-fidelity likelihood}:
\begin{equation}
\pmf\big(\Yobs | \Zlf(\bx)\big) = \int p(\Yobs | \Yhf) \cdot p(\Yhf | \Zlf(\bx)) \dd\byhf
\end{equation}
}

\begin{remark}[Finite element representations of random fields]
\label{rem: fe_approx_fields}
FE representations offer a flexible, geometry-adaptable approach for approximating random fields. The localized FE basis leads to sparse matrices, simplifying adjoint computations, as gradients are only supported locally. Moreover, the posterior field's covariance, when approximated with the local FE bases, typically exhibits a diagonal-dominant structure, making Gaussian \add{VI} (see Section~\ref{sec: svi}) both efficient and scalable. In contrast to global expansions (e.g., Fourier or Karhunen–Lo\`eve), FE bases can represent arbitrary, high-frequency, or even discontinuous fields.

A \add{FE} representation of a random (potentially vector-valued) field implies the following parameterization: 
\begin{subequations}
\begin{align}
    \label{eqn: rf_fe}
    \underbrace{\tilde{\bx}(\bx,\bc)}_{\substack{\text{Continuous}\\ \text{(scalar) field}}} &= \underbrace{S_\mathrm{ele}(\bc)}_{\substack{\text{Local}\\ \text{interpolation}\\ \text{matrix}}}\cdot\underbrace{\bxele}_{\substack{\text{Local}\\ \text{DoF}\\ \text{vector}}}\ ,\quad \mathrm{for}\ \bc\in\Omega_{\mathrm{ele}}\\ 
    \underbrace{\bxele}_{\substack{\text{Local}\\ \text{DoF}\\ \text{vector}}} &= \underbrace{\mathfrak{R}}_{\text{Restrictor}} [\underbrace{\bx}_{\substack{\text{Global}\\ \text{DoF}\\ \text{vector}}}] \\
    X(\bx)&:=\tilde{\bx}(\bx,C)=
    \begin{bmatrix}
    \tilde{\bx}(\bx,\bc_1)\\
    \vdots\\
    \tilde{\bx}(\bx,\bc_n)\\
    \end{bmatrix}\label{eqn: input_field} 
\end{align}
\end{subequations}
In Equation \eqref{eqn: rf_fe}, the local (element-wise) DoF vector $\bxele$ represents the discrete \add{DoFs} for a specific \add{FE} with local spatial support on the coordinate $\bc$. The local DoF vector $\bxele$ is extracted from the global DoF vector $\bx$ by a restriction operation (local to global DoF mapping), represented by the restriction operator $\mathfrak{R}$. Hence, $\bx$ holds, in the general case, the nodal values of the FE representation. In the \add{BIP}, we infer the posterior distribution of the global DoF vector $\bx$. The input field matrix $X(\bx)$ (see Equation \eqref{eqn: input_field}) represents the row-wise collection of (interpolated) input field (vector) values $\tilde{\bx}(\bc)$ evaluated at the same spatial coordinate $\{\bc_i\}|_{i=1}^n$ as the spatial point in the output fields collected in $\Ylf$ or $\Yhf$.
\end{remark}

For each input field parameterization $\bx$ we predict the corresponding $\Yhf$ with the help of $\Zlf(\bx)$, \ie instead of $p(\Yhf|\bx)=\delta_{\Yhf}\left(\Yhf-\Mhf(\bx)\right)$ we employ $p(\Yhf|\Zlf(\bx))$. The latter accounts for the fact that $\Zlf$ does not necessarily contain all the information $\bx$ contains, in which case uncertainty with regard to $\Yhf$ is introduced \cite{bilionis2013solution}. This uncertainty gives rise to the \add{MF} posterior $\pmf(\bx|\Yobs)$:
\be
\pmf(\bx|\Yobs) \propto \underbrace{\int \overbrace{p(\Yobs|\Yhf)}^{\substack{\text{conditional}\\ \text{noise model}}} \cdot\overbrace{p(\Yhf|\Zlf(\bx))}^{\substack{\text{\add{MF}}\\ \text{conditional}}} \dd \byhf}_{\pmf\left(\Yobs|\Zlf(\bx)\right)}\cdot p(\bx)
\label{eqn:bmfia_conti}
\ee
We denote with $\lmf(\bx)=\log\left(\pmf\left(\Yobs|\Zlf(\bx)\right)\right)$ the logarithm of the \add{MF} likelihood  $\pmf\left(\Yobs|\Zlf(\bx)\right)=\int p(\Yobs|\Yhf) p(\Yhf|\Zlf(\bx)) \dd \byhf$
which replaces the expensive HF log-likelihood $\lhf(\bx)$. While its introduction introduces additional uncertainty, we note that each evaluation involves the much less expensive $\Zlf(\bx)$, making Bayesian inference much more manageable.
\add{
\paragraph{Interpretation}
The MF likelihood can be viewed as a convolution of the HF likelihood with the conditional density $p(\Yhf | \Zlf(\bx))$ (instead of the Dirac-delta $p(\Yhf|\bx)=\delta(\Yhf-\Mhf(\bx))$ in \refeqp{eqn:bmfia_conti}). This smoothing effect produces a slightly blurred posterior relative to the HF posterior. As we demonstrate in the subsequent sections, the computational savings typically outweigh the (mild) loss of sharpness.
The apparent advantage of the MF formulation is the increased computational scalability due to the reduced computational cost of evaluating the LF features and potentially the availability of derivatives of the latter with respect to $\bx$, which can significantly expedite inference.
On the other hand, the disadvantage is that the MF posterior is (slightly) blurred compared to the reference HF posterior, leading to reduced sharpness in uncertainty estimates. One would expect that this phenomenon is amplified as the LF features selected are more myopic to the input's $\bx$ details. Nevertheless, if the associated accurately represents $p(\Yhf|\Zlf(\bx))$, the MF posterior will not exhibit undesirable biases. This uncertainty can guide model refinements and data acquisition, as we explain in the sequel. 
In order to fully exploit the potential of the multi-fidelity formulation, two key challenges must be addressed: 
(i) efficiently learning the conditional density 
$p(\Yhf|\Zlf(\bx))$, and (ii) ensuring that this density faithfully captures the associated uncertainty. We outline our approach to these challenges in the following section.
}

\begin{remark}[MF conditional acts as a smoothing kernel]
\label{rem:smoothing_filter}
\add{A smoothing operation for an arbitrary function $f(\alpha)$, given a (non-negative and normalized) smoothing kernel $g(\alpha)$, is mathematically described by a convolution: $(f*g)(\alpha)=\int f(\alpha-t)\cdot g(t) \dd t$. In our case of the MF likelihood, we can interpret the function $f(\alpha)$ as the noise model $p(\Yobs|\Yhf)$ (we show in Section \ref{sec:small_data_approx} that we employ an additive Gaussian noise model $p(\Yobs|\Yhf)=\mathcal{N}\big(\Yobs|\Yhf,\tau^{-1}\cdot I\big)=\mathcal{N}\big(\Yobs-\Yhf|\boldsymbol{0},\tau^{-1}\cdot I\big)$ with precision $\tau$). Furthermore, the positive and normalized smoothing kernel $g(\alpha)$ is given by the MF conditional $p(\Yhf|\Zlf(\bx))$. This gives rise to the following convolution: $[p(\Yobs|\Yhf)*p(\Yhf|\Zlf(\bx))](\Yobs)=\int \mathcal{N}\big(\Yobs-\Yhf|\boldsymbol{0},\tau^{-1}\cdot I\big)\cdot p(\Yhf|\Zlf(\bx))\dd\byhf$, with $\Yhf$ being the shifting variable (before: $t$) and $\Yobs$ the variable over which the smoothing is defined (before: $\alpha$). This results in a smoothing or blurring of the MF likelihood, compared to the HF likelihood. The MF posterior inherits this smoothing from the MF likelihood (being proportional to the MF likelihood function): Sharp peaks in the original HF posterior become now wider and less pronounced. BMFIA keeps the MF posterior \emph{honest} by spreading probability mass according to the modeled HF-LF mismatch, which acts out as a blurring of the MF posterior compared to the HF posterior.} 
\end{remark}

\add{The idea is that most of the complex structure in the mapping $\bx\mapsto\Yhf$ is, at least in a statistical sense, captured by the LF model output $\Ylf$. Hence, the MF conditional $p(\Yhf|\Zlf)$ becomes a simple stochastic function that can be learned reliably even in a small data regime, \ie for little training data $\mathcal{D}=\{\Yhf_i, \Zlf_i\}|_{i=1}^{\ntrain}$. Evaluating the MF conditional involves two steps: i) Realizing an LF simulation to yield $\Yhf_i=\Mlf(\bx_i)$ for a sample $\bx_i$.  ii) Plugging $\Ylf$ into the (much simpler) MF conditional to get a distribution (instead of a deterministic prediction) $p(\Yhf|\Zlf)$. We argue that the total error in the resulting posterior approximation, for a limited computational budget, is smaller than any direct attempt of sampling the HF model or building a high-dimensional surrogate for $\bx\mapsto\Yhf$. Please see also Remark \ref{rem:when_conditional}.} \add{We furthermore emphasize that we condition on the entire LF output vector $\Zlf$ in the MF conditional $p(\Yhf|\Zlf)$, not only on an output at one spatial point. Conditioning on a single spatial output point would miss crucial correlation patterns and coherent features that are essential to describe the complexity of the HF model output field $\Yhf$.}

BMFIA employs a probabilistic regression approach to model this conditional distribution efficiently.
\add{To} reduce the information loss introduced by filtering through $\Ylf$, we augment the \add{MF} input by incorporating the interpolated representation $X$ of the input field. These additional features\footnote{\emph{Features} are informative quantities derived from the input, either as deterministic functions or as latent variables, that capture aspects of the data relevant for modeling the target distribution. In probabilistic models, features need not be uniquely defined or deterministic; instead, they may summarize uncertainty or structure that is useful for inferring $p(\Yhf|\Ylf)$.} are not used to reconstruct the deterministic mapping but rather to \emph{guide the probabilistic embedding}-providing complementary information such as low-frequency components of $X$ that help reduce the uncertainty in $p(\Yhf | \Ylf)$. The resulting enriched representation $\Zlf = [\Ylf, X]$ leads to a refined conditional distribution $p(\Yhf | \Zlf)$
with lower conditional variance and improved predictive accuracy. 

\begin{remark}[When and why should we learn an MF conditional and use BMFIA?]
\label{rem:when_conditional}
\add{Classical surrogates ($\bx\mapsto\Yhf$) and MF surrogates ($(\bx, \Ylf) \mapsto \Yhf$)  (\eg see \cite{perdikaris2017}), attempt to learn the full input space, retaining the high dimensionality of $\bx$ and leading to rapidly increasing errors in high‑dimensional, small‑data regimes. In contrast, BMFIA reformulates the problem as learning the conditional $p(\Yhf|\Zlf)$, projecting it onto the lower‑complexity $\Ylf \times \Yhf$ representation (augmented with $X$). Here, $X$ serves only as rough, low‑frequency guidance to reduce conditional variance. The conditional is of simpler functional complexity due to the statistical similarity between $\Ylf$ and $\Yhf$. Therefore, a sufficient approximation can be learned from limited data.
While we sacrifice full determinism, we at least explicitly model and incorporate the induced uncertainty in terms of the conditional density. This treatment avoids a bias in the posterior while accepting a loss of detail (blurring) in the posterior approximation. 
This idea in BMFIA is motivated by an error trade‑off: For low‑dimensional $\bx$ and simple mappings $\bx\mapsto\Yhf$, classical surrogates are preferable and have probably a smaller overall approximation error. However, as $\dim(\bx)$ and functional complexity increase, the conditional view achieves lower overall error and higher reliability in a small data regime.}
\end{remark}

In Section \ref{sec:small_data_approx}, we discuss how good approximations of $p(\Yhf|\Zlf(\bx))$ can be obtained with a \emph{small amount} of HF-model evaluations.  While we \add{investigate} potential error sources and extreme cases more deeply in Section \ref{sec:errors_and_extremes}, we note at this stage that the utility of $p(\Yhf|\Zlf(\bx))$ does not depend on offsets between HF and LF models nor on the correlation (\ie linear dependence) between HF and LF model outputs as, \eg in Multi-Level Monte Carlo and other pertinent techniques \cite{tierney1999some,christen2005markov,scheichl2017quasi,giles2008multilevel, beskos2018multilevel,latz2018multilevel}.
\FloatBarrier

\begin{figure}[htbp]
    \centering
    \begin{tikzpicture}
        \node[anchor=south west,inner sep=0] (background) at (0,0) {\includegraphics[scale=0.6]{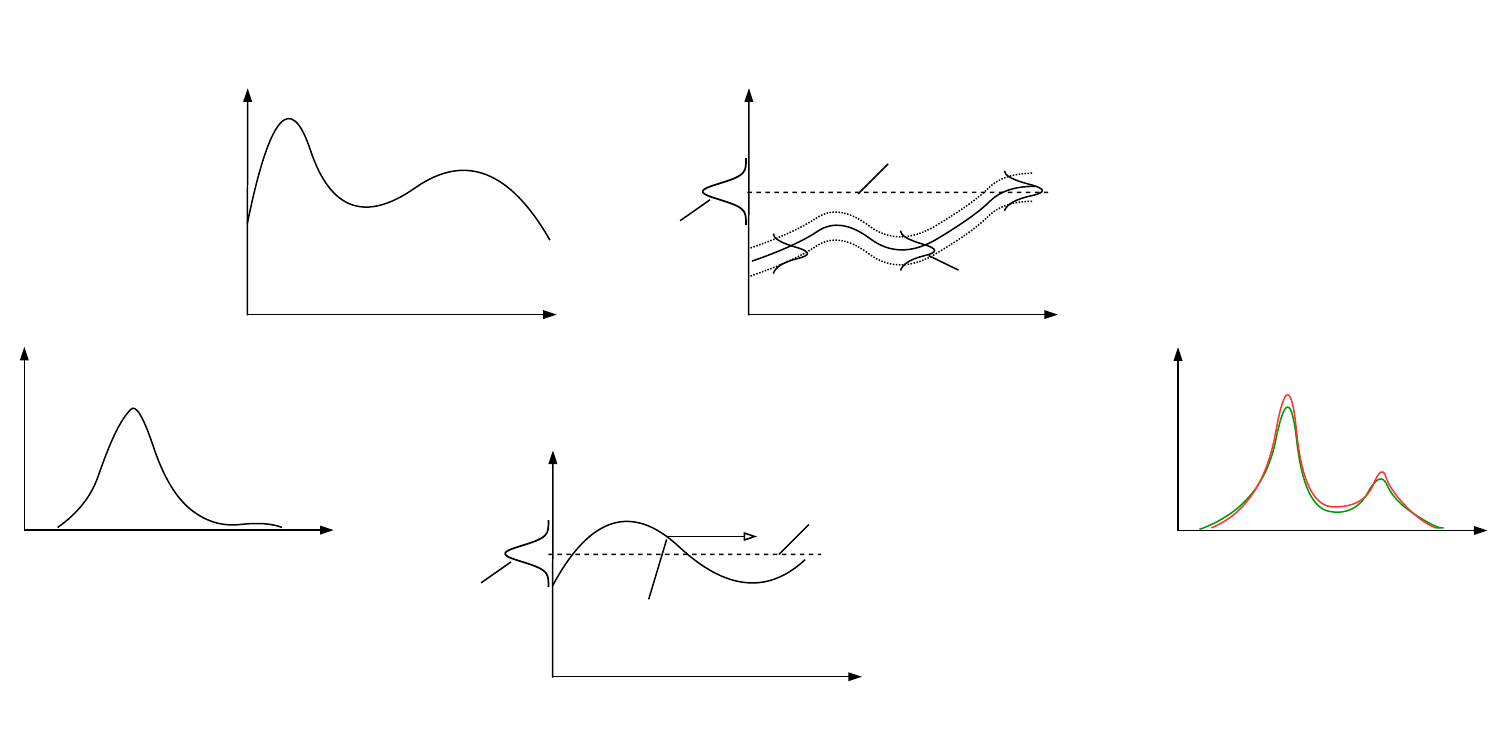}}; 
        \begin{scope}[x={(background.south east)},y={(background.north west)}]
            \node at (-0.02, 0.5) {$p(x)$}; 
            \node at (0.23, 0.27) {$x$}; 
            
            \node at (0.72, 0.38) {$\color{red}\phf(x|\yobs)$}; 
            \node at (0.72, 0.45) {$\color{PineGreen}\pmf(x|\yobs)$}; 
            \node at (1.0, 0.27) {$x$}; 
            
            \node at (0.59, 0.07) {$x$}; 
            \node at (0.34, 0.38) {$\yhf$}; 
            \node at (0.30, 0.2) {$p(\yobs|\yhf)$}; 
            \node at (0.55, 0.32) {$\yobs$}; 
            \node at (0.55, 0.18) {$p(\yhf|x)=\delta_{\yhf}\left(\yhf-\Mhf(x)\right)$}; 
            
            \node at (0.71, 0.55) {$\zlf$}; 
            \node at (0.44, 0.67) {$p(\yobs|\yhf)$};
            \node at (0.475, 0.85) {$\yhf$};
            \node at (0.71, 0.64) {$p(\yhf|\zlf)$};
            
            \node at (0.61, 0.8) {$\yobs$}; 
                        
            \node at (0.38, 0.55) {$x$}; 
            \node at (0.14, 0.85) {$\zlf$}; 

            \node at (0.08, 0.53) {\boxed{Prior}}; 
            \node at (0.87, 0.53) {\boxed{Posterior}}; 
            \node at (0.61, 0.9) {\boxed{MF Likelihood}}; 
            \node at (0.27, 0.9) {\boxed{LF Mapping}}; 
            \node at (0.48, 0.42) {\boxed{HF Likelihood}}; 

              \draw[->, PineGreen, ultra thick] (0.04,0.65) -- (0.12,0.75) 
                node[midway, above, sloped] {\small{Eval. LF}} 
                node[midway, below, sloped] {\small{BMFIA}}; 
                
              \draw[->, PineGreen, ultra thick] (0.75,0.72) -- (0.85,0.62) 
                node[midway, above, sloped] {\small{MF Posterior}}; 
                
              \draw[->, PineGreen, ultra thick] (0.35,0.9) -- (0.45,0.9) 
                node[midway, above, sloped] {\small{Eval. MF}} 
                node[midway, below, sloped] {\small{Cond.}}; 
                
              \draw[->, red, ultra thick] (0.04,0.22) -- (0.15,0.18) 
                node[midway, above, sloped] {\small{Eval. HF Lik.}} 
                node[midway, below, sloped] {\small{BIA}}; 
                
              \draw[->, red, ultra thick] (0.75,0.18) -- (0.85,0.22) 
                node[midway, above, sloped] {\small{HF Posterior}}; 
        \end{scope}
    \end{tikzpicture}
    \caption{Schematic representation of \textbf{BMFIA at the top}, following the green arrows, and Bayesian inverse analysis \textbf{(BIA) at the bottom}, following the red arrows. \textbf{Bottom:} Iteratively calculating the HF posterior involves the multiplication of the prior values $p(x)$ by the HF likelihood value $p(x|\yobs)=\int p(\yobs|\yhf)\cdot p(\yhf|x)\dd\yhf$. The HF likelihood is dependent on the expensive HF model $\Mhf(x)$. \textbf{Top:} To overcome the computational burden, we propose following the top path using BMFIA along the green arrows. Here, we replace the expensive dependence on the HF model with the \add{MF} conditional: $p(\yhf|x)\approx p(\yhf|\zlf(x))$.}
    \label{fig:bmfia_graph}
\end{figure}

\textbf{Bayesian inverse analysis versus BMFIA} We summarize the idea behind BMFIA in the schematic Figure \ref{fig:bmfia_graph}, which shows the involved steps for a one-dimensional example. The algorithm is actually iterative, but for simplicity, we demonstrate the steps as one blend from prior to posterior: The usual way of solving \add{BIPs} would follow the red arrows at the bottom of the figure. The inference algorithm of choice proposes an input sample $x$ (based on the multiplication of former likelihood and prior values), for which it evaluates the likelihood function $\phf(x)=\int p(\yobs|\yhf)\cdot p(\yhf|x)\dd \yhf$, triggering an expensive forward call to the HF model $\yhf=\Mhf(x)$ in the HF Dirac conditional $p(\yhf|x)=\delta\left(\yhf-\Mhf(x)\right)$. The resulting HF likelihood value triggers, in combination with the prior $p(x)$, a following sample candidate $x$ until convergence of the inference process. Note that for high-dimensional inference, the proposal of a next candidate usually involves gradients of the log-likelihood function (and therefore the computational model) \wrt the inputs $x$.

Instead of pursuing this expensive and non-differentiable (we assume that the HF legacy codes are not differentiable) procedure, BMFIA follows the computationally cheaper and differentiable green path on top of Figure \ref{fig:bmfia_graph}: Starting again from a sample $x$, we now evaluate a \add{LF} model $\ylf=\Mlf(x)$ and concatenate the response along with the corresponding input in $\zlf$. The \add{LF} output $\zlf$ then serves as an input to the \add{MF} conditional $p(\yhf|\zlf)$. An initial approximation for this distribution has to be learned beforehand from a \emph{small amount of HF and LF data} (50 to 300 simulation runs) in the \emph{initial training phase} of the algorithm. Later on, the first approximation can be updated continuously in the inference phase of the algorithm by drawing further samples from the current \add{MF} posterior approximation to enrich the data-driven conditional.
In the new \add{MF} likelihood $\pmf(x|\yobs)=\int p(\yobs|\yhf)\cdot p(\yhf|\zlf(x))\dd\yhf$ (see Equation \eqref{eqn:bmfia_conti}), the dependence on $x$ moved to the cheap LF model, which we assume is fully differentiable. Following this idea, BMFIA accurately approximates the true HF posterior at a fraction of the HF costs. Especially for gradient-based inference schemes, BMFIA has the significant advantage that it only requires model gradients of the LF model. The latter can be designed flexibly, as we only require a weak stochastic link between the LF and the HF model, which gives a lot of design freedom for the LF in practice (\eg see our demonstrations \ref{sec:demonstration}). \add{To get a better intuitive feeling for BMFIA and the idea of the MF conditional, we compiled a physical analogy in Appendix \ref{sec:physical_analogy} for the interested reader.}

\FloatBarrier

\subsection{Modeling and learning the MF conditional and the MF likelihood in the small data regime}
\label{sec:small_data_approx}
The \add{MF} likelihood, which lies at the core of the proposed formulation in Equation \eqref{eqn:bmfia_conti}, consists of two components.
First, the noise model in the \add{MF} likelihood $p(\Yobs|\Yhf)$ (the likelihood of the observables given the HF model output) is the same noise model that appears in the standard HF likelihood formulation (see \refeqp{eqn: bayes}). In particular, we assume that the data are contaminated by additive independent and identically distributed (\iid) Gaussian noise with an isotropic variance of $\tau^{-1}$, which is a common modeling choice:
\be
p(\Yobs|\Yhf)=\mathcal{N}\left(\Yobs~|~\Yhf, \tau^{-1}\cdot I \right)
\label{eqn:like1}
\ee
With regards to the second component, \ie the \add{MF} conditional $p(\Yhf|\Zlf(\bx))$ that links HF and LF model outputs, we postulate a further Gaussian model of the form\footnote{In the actual implementation, we flatten the involved matrices to vectors to avoid using higher order tensors.}: 
\be
p(\Yhf|\Zlf(\bx))\approx\mathcal{N}\left(\Yhf|M\left(\Zlf(\bx)\right), \Kmat\left(\Zlf(\bx)\right)\right),
\label{eq:like2}
\ee
\add{where $M(\Zlf(\bx))$ and $\Kmat(\Zlf(\bx))$ denote a learnable mean and covariance function, whose parameterizations are detailed in the next section.}
The combination of two Gaussians enables us to marginalize $\Yhf$ in the integral of \refeqp{eqn:bmfia_conti} to obtain a \add{MF} (log-)likelihood $\pmf(\Yobs|\bx,\tau)$\add{, that allows for efficient gradient-based inference}\footnote{We explicitly include here the hyper-parameter $\tau$ relating to the precision of the observation noise.}:
\begin{subequations}
\begin{align}
    \label{eqn:MFGaussianLik}
    \log \pmf(\Yobs|\bx,\tau)&:=\lmf(\bx,\tau)=\log\bigg(\int \underbrace{\mathcal{N}\left(\Yobs|\Yhf, \tau^{-1}\cdot I \right)}_{p(\Yobs|\Yhf)}\cdot ~\underbrace{
\mathcal{N}\left(\Yhf|M\left(\Zlf(\bx)\right), \Kmat\left(\Zlf(\bx)\right)\right)}_{\approx p(\Yhf|\Zlf(\bx))}~\dd\Yhf
    \bigg)\\
    \begin{split}
    \label{eqn:margMFGauss}
   &= \log\bigg(\mathcal{N}\big(\Yobs\Big|M\left(\Zlf(\bx)\right),\underbrace{\tau^{-1}\cdot I+\Kmat\big(\Zlf(\bx)\big)}_{\Kmf(\Zlf(\bx),\tau)}\big)\bigg) \\
   &= -\frac{\numobs}{2}\log(2\pi)-\frac{1}{2}\log\big(\det\left(\Kmf(\Zlf,\tau)\right)\big)\\
   &\quad-\frac{1}{2}\left(M\left(\Zlf(\bx)\right)-\Yobs\right)^T \cdot \left(\Kmf\left(\Zlf(\bx),\tau\right)\right)^{-1} \cdot (M\left(\Zlf(\bx)\right)-\Yobs)
   \end{split}
\end{align}
\end{subequations}
Equation \eqref{eqn:margMFGauss} now depends on the \add{MF} mean function $M\left(\Zlf(\bx)\right)$, the \add{MF} covariance function $\Kmat\left(\Zlf(\bx)\right)$, and the precision hyper-parameter $\tau$. We defer discussions on the hyper-parameter $\tau$ to Section \ref{sec: svi} and focus first on modeling the Gaussian \add{MF} conditional with a probabilistic regression model.
\begin{remark}[Non-Gaussian MF conditionals]
\label{remark:non_gauss_cond}
The choice of a Gaussian \add{MF} conditional in Equation \eqref{eqn:margMFGauss} introduces an approximation (a modeling error) but leads to a closed-form expression for the \add{MF} log-likelihood, which simplifies and accelerates the subsequent inference procedure. In practice, this modeling error is negligible in our experience. Nevertheless, we highlight that a non-Gaussian regression model for the \add{MF} conditional could be employed. However, this would render the integral in Equation \eqref{eqn:MFGaussianLik} intractable and require inference of the latent variables $\Yhf$, which can be of (very) high dimension. While Monte Carlo integration could approximate the integral, if $p(\Yhf|\Zlf(\bx))$ can be sampled, the outer logarithm complicates the gradient calculation \wrt $\bx$ drastically. To mitigate this problem, one could use Jensen's inequality to find a lower bound to the expression, which would again introduce an approximation error. This procedure could be promising, but should be studied more deeply. \add{From an information-theoretic perspective, the Gaussian is the maximum-entropy distribution consistent with the first two moments, making it a principled choice in data-scarce settings.}
\end{remark}

\subsubsection{Training data generation and iterative refinement}
\label{sec: training_data}

\add{
Learning the MF conditional density $p(\Yhf|\Zlf(\bx))$ lies at the core of the proposed formulation. We propose an adaptive procedure that is based on a small initial training dataset
\[
\mathcal{D} = \{(\Zlf_i, \Yhf_i)\}_{i=1}^{\ntrain},
\]
obtained by sampling inputs $\bx_i $ from the prior $p(\bx)$\footnote{For hierarchical priors where direct sampling is infeasible, we sample from the joint prior before marginalization.} and evaluating the corresponding LF and HF models:
\[
\Yhf_i = \Mhf(\bx_i),\quad \Ylf_i=\Mlf(\bx_i) \quad \Zlf_i = [\Ylf_i, X(\bx_i)].
\]
Empirical evidence suggests that  50-300 HF samples suffice for an initial approximation, although the exact number depends on LF model quality, input dimensionality, and dependence structure between the LF and HF models. 
This data is used to learn $p(\Yhf|\Zlf(\bx))$ as explained in the sequel. We can refine the MF conditional and expand the dataset $\mathcal{D}$ by:
\begin{enumerate}
    \item Computing the MF posterior using the current conditional.
    \item Sampling additional inputs $\bx$ from the posterior estimate. We choose a batch size of 10 as a heuristic compromise between refinement effectiveness and computational efficiency: while this is likely at the lower end of beneficial batch sizes, it limits early HF evaluation costs. Adaptively determined batches could improve refinement, but more elaborate strategies are outside the scope of this paper. The interested reader is also referred to the convergence study in Appendix \ref{sec:convergence_study} for the effect of different batch and training sizes.    
    \item Evaluating LF and HF models for these new points and augmenting the training set.
\end{enumerate}
This procedure, illustrated in Algorithms \ref{alg:offline_bmfia} to \ref{alg:online_bmfia}, balances computational efficiency with improved model accuracy.
}

\subsubsection{Approximating the MF conditional with a probabilistic convolutional autoencoder}
\label{sec:p_cae}
Accurate modeling of the conditional \add{MF density} $p(\Yhf|\Zlf)$ represents one of the core challenges in \add{BMFIA}. \add{In this work, we focus on static spatial posteriors, meaning the posteriors of the reconstructed spatial fields are not time-dependent. The underlying forward models might, however, be transient. In the case of time-dependent observables $\Yobs$, it can be beneficial to consider a time-dependency in the multi-fidelity conditional, even for static posteriors. This step adds further complexity and will be studied in future extensions of BMFIA.} 
To ensure stable inference with the proposed probabilistic regression model, we introduce the following requirements, which we discuss in detail below:
\begin{itemize}
\item Stable and smooth gradients of the probabilistic regression model \wrt $\bx$, \add{for gradient-based inference} 
\item Robust and well-regularized in the \emph{small data regime} (simple but flexible enough)
\item Preservation of spatial features
\item \add{Probabilistic output: Explicit variance prediction allows robust uncertainty estimates}
\end{itemize}

\add{In principal any} probabilistic machine learning models \add{could} approximate \add{the MF conditional $p(\Yhf|\Zlf)$}, notably Fourier Neural Operators (FNOs) \cite{li2020fourier}, convolutional neural operators \cite{raonic2023convolutional}, or more general operator-learning frameworks for PDE-driven problems include \cite{kovachki2023neural} (please see also the Introduction \ref{sec:introduction} for a more complete review). Additionally, methods using fast Fourier convolutions \cite{fft_convolution} and image-to-image learning techniques such as physics-informed convolutional architectures \cite{zhu2019physics, winovich2019convpde} have shown promise. While variational convolutional autoencoders \cite{Kingma_Welling_2022}, conditional variational autoencoders \cite{mirza2014conditional,zhu2019physics}, and graph-based variational convolutional autoencoders \cite{niepert2016learning,pfaff2020learning} offer flexible and advanced modeling capabilities, we empirically encountered complications regarding gradient instabilities \wrt $\bx$, computational complexity, or underestimating conditional variance, especially in small-data scenarios. \add{Future work should systematically explore such architectures within the BMFIA framework, particularly for handling more flexible geometrical domains or fully (geometrically) three-dimensional settings. In this regard, graph-based methods and operator-learning approaches appear especially promising, as they can naturally incorporate irregular meshes and topologies. These investigations, however, are beyond the scope of the present introductory study.}

\begin{figure}[htbp]
    \centering
    \includegraphics[scale=1.3]{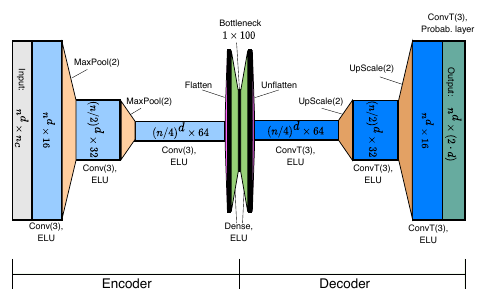}
    \caption{Architecture of the employed convolutional autoencoder with a probabilistic output layer, approximating the \add{MF} conditional $p(\Yhf|\Zlf)$.}
    \label{fig:cae}
\end{figure}

We found that a probabilistic convolutional autoencoder with a Gaussian output layer, illustrated in Figure~\ref{fig:cae}, offers a good practical compromise between accuracy and complexity while also handling the laid-out smoothness \add{requirements, gradient stability, and preservation of spatial features, at least for the rectangular, two-dimensional geometric domains we investigate in this manuscript}. 
\add{By preservation of spatial features, we refer to retaining correlation structures and coherent patterns in the spatial output fields when mapping from LF to HF, including shifts, deformations, or amplitude changes. Providing the full spatial field to a model that can exploit such correlations enables learning a more accurate MF conditional model than training separate models per spatial location, which discards cross-location dependencies. In this work, we treat each time point independently and do not model spatio-temporal correlations, although extending the approach to capture them, \eg via recurrent, hybrid convolutional, or attention-based models, is a promising avenue for future research.}
The proposed model prioritizes gradient robustness and allows for several simple and easy-to-control regularization measures, which will be listed subsequently.
The model performs probabilistic regression between the composed \add{LF} output $\Zlf = [\Ylf, X]$ and the HF model output $\Yhf$. The probabilistic autoencoder predicts the mean $M(\Zlf)$ and variance $V(\Zlf)$ of a Gaussian distribution that approximates the \add{MF} conditional\footnote{\add{For compactness, we slightly abuse notation by using matrix symbols $\Ylf, \Yhf, \Zlf, M, V$ to represent both their matrix form (with rows as spatial points and columns as components), and their reshaped tensor form used in the convolutional autoencoder}}:
\begin{equation}
\label{eqn:autoencoder_prob_layer}
p(\Yhf | \Zlf) \approx \mathcal{N}\left[\Yhf | M(\Zlf), \diag\left(V(\Zlf)\right)\right]
\end{equation}
For simplicity, we assume a square spatial domain with a regular grid structure, allowing the use of standard convolutions. However, this is not a limitation: The approach can be extended to irregular domains using graph-based autoencoder variants.

\FloatBarrier
\add{
In summary, the main features of the adopted architecture are enumerated below, and additional details are provided in Appendix \ref{sec:details_autoencoder}:
\begin{itemize}
    \item Encoder: three convolutional layers (feature dimensions 16, 32, 64) with ELU activations and max-pooling, followed by flattening and a fully connected bottleneck (dimension 200).
    \item Decoder: mirrored convolutional architecture reconstructing mean and variance for each spatial location.
    \item Output: Gaussian layer with diagonal covariance per spatial point; variance reparameterized for stability, small nugget added ($10^{-5}$).
    \item Training: negative log-likelihood loss, SGD or Adam optimizer, standardized data, dropout (30\%), batch normalization.
\end{itemize}
}

\subsection{Probabilistic model and variational inference strategy for the Bayesian inverse problem}
\label{sec: svi}

After introducing BMFIA and the \add{MF} log-likelihood in the previous sections, we briefly define the prior model and \add{inference approach} used in this work. We are interested in inferring posterior distributions of spatially varying fields, \ie constitutive or material fields in (bio)mechanical problems. Specifically, we focus on problems discretized and solved with the \emph{\add{FE} method}. Therefore, we also represent the spatial fields of interest with a \emph{continuous \add{FE} representation using linear shape functions} (see Remark \ref{rem: fe_approx_fields} for more details).

We use a hierarchical \emph{Gaussian Markov random field (GMRF) prior} \cite{markov_priors, rue2005gaussian} for the discretized fields  \cite{dempster1977maximum,Koutsourelakis_2012,Franck_Koutsourelakis_2016}, which is in the log-space given by:
\begin{subequations}
\begin{align}
\label{eqn: marginalized_prior}
    \begin{split}
    \log p(\bx) &=\log \int \mathcal{N}\left(\bx|\bmu_0, \delta\cdot P\right)\cdot p(\delta) \dd \delta\\
    &=\log \Ex{p(\delta)}{\mathcal{N}\left(\bx|\bmu_0, \delta\cdot P\right)}
    \end{split}\\
    &=\Ex{p(\delta|\bx)}{\log \left(\frac{\mathcal{N}\left(\bx|\bmu_0, \delta\cdot P\right)\cdot p(\delta)}{p(\delta|\bx)}\right)}\\
   \text{with } p(\delta)&=\Gamma(\delta|a_0,b_0) \label{eqn: marginalized_prior_last}
\end{align}
\end{subequations}
\add{The GMRF prior can be motivated from the continuous variational problem of minimizing field roughness, whose Euler–Lagrange operator is the Laplacian $\Delta\cdot$. Its FE discretization leads to the FE Laplace matrix, which acts here as the precision matrix $P$ in the Gaussian. The latter naturally enforces smoothness in high-dimensional reconstructions. The convergence of this discretization follows standard FE theory, and since the random field is resolved on the same mesh as the governing PDEs, the prior is already sufficiently accurate relative to the physics.} The sparsity property allows for the treatment of very large and fine-resolved domains and parameterizes the precision matrix (inverse of the covariance matrix) directly, which avoids the solution of an additional equation system in the log-likelihood. Please see \cite{markov_priors} for more details. The Gaussian Markov prior prefers structured fields with structured patterns at different length scales, which offers a good compromise between smoothness, correlation, and irregular structures.
We model the hyper-prior on $\delta$ by a conjugate Gamma distribution with $a_0=b_0=10^{-9}$ and $\bmu_0$ as a fixed value that reflects the average belief of the random field. This Gamma distribution prefers smaller values for $\delta$, the scaling factor for the precision matrix $P$ (effectively preferring a higher variance in the resulting random field) \cite{Franck_Koutsourelakis_2016}\footnote{Usually, the constitutive fields we seek to infer are strictly positive. We conduct an exponential transform as part of the forward solve to operate on the unconstrained variable $\bx$ in the \add{BIP}.}.
In analogy to the likelihood model in Equation \eqref{eqn: effective_mf_likelihood_em}, we use the \emph{Variational Bayes-Expectation Maximization} (VB-EM) algorithm \cite{dempster1977maximum, Beal_Ghahramani_2003} to iteratively calculate the log prior and its gradient \wrt $\bx$. Figure \ref{fig: markov_prior_samples} shows three exemplary samples of the Markov prior in a two-dimensional unit square domain discretized with $100\times100$ elements:
\begin{figure}[htbp]
    \centering
    \includegraphics[scale=0.4]{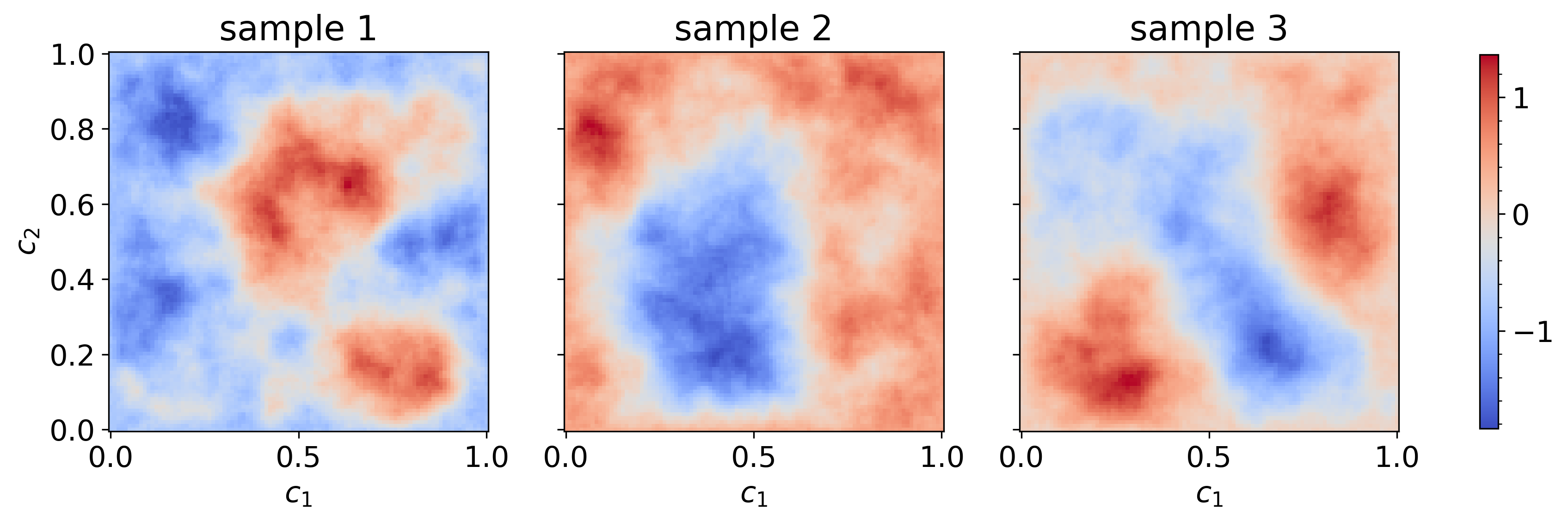}
    \caption{\add{Three different, exemplary realizations (sample 1, sample 2, sample 3)} \add{sampled from} a Gaussian Markov \add{scalar field} on a two-dimensional unit square \add{over the coordinates $c_1\times c_2$} with periodic boundary conditions (\add{BCs}) and $\bmu_0=\boldsymbol{0}$ and $\delta=10^{-3}$.}
    \label{fig: markov_prior_samples}
\end{figure}

After the prior model (Equation \eqref{eqn: marginalized_prior}) and the \add{MF} likelihood (Equation \eqref{eqn: effective_mf_likelihood}) are defined, the log-unnormalized posterior is given by:
\begin{equation}
    \label{eqn: unnormalized_posterior}
    \log \pmf(\bx,\Yobs)=\lmf(\bx) + \log p(\bx)
\end{equation}
The prior and its gradient \wrt $\bx$ can be employed by any state-of-the-art iterator to solve \add{BIPs}. In this work, we use the gradient-based \emph{SVI} procedure \cite{hoffman2013stochastic, blei2017variational} due to its efficiency in high stochastic dimensions, especially for calculating posterior approximations of spatial random fields \cite{koutsourelakis2016variational, Franck_Koutsourelakis_2016, rixner2021probabilistic}. 
More specifically, we use a sparse multivariate Gaussian, whose covariance matrix $K$ has a fixed bandwidth, as the variational distribution $\qmf(\bx|\bphi)$ to approximate the \add{MF} posterior $\pmf(\bx|\Yobs)$:
\begin{align}
\label{eqn: var_approx}
\begin{split}
\pmf(\bx|\Yobs)&\approx \qmf(\bx|\bphi)=\mathcal{N}\left(\bx|\bphi_1, L(\bphi_2)\cdot L^T(\bphi_2)\right)\\
\text{with } \bphi&=[\bphi_1, \bphi_2]^T
\end{split}
\end{align}
In Equation \eqref{eqn: var_approx}, $\bphi_1$ corresponds to the parameterization of the mean vector, and $\bphi_2$ to the parameterization of the lower Cholesky factor $L$ of the Gaussian's covariance matrix. The latter is modeled as a sparse matrix with controllable off-diagonal width (controllable bandwidth), such that we only account for the correlation structure of neighboring DoFs. 
SVI and other gradient-based inference schemes require the gradient $\nabla_{\bx} \pmf(\bx,\Yobs)$ of the unnormalized posterior in Equation \eqref{eqn: unnormalized_posterior}, and hence of the \add{MF} log-likelihood $\nabla_{\bx}\lmf(\bx)$. The latter is implicitly dependent on a physics-based simulation model $\Mlf(\bx)$ (in our case, a FEM model), such that we additionally have to implement the LF adjoint model $\Alf\left(\bx,\nabla_{\Ylf}\lmf(\Ylf)\right)$, which is usually easy to derive. Our open-source software framework QUEENS \cite{queens} automatically manages the involved models and gradient evaluations. Specifically for BMFIA, we furthermore clip the partial gradient of the \add{MF} log-likelihood $\frac{\partial \lmf(\bx,\Zlf)}{\partial \Zlf}$ that serves as a right-hand side for the adjoint solve to avoid instabilities in the solution process.

Once we have learned the \add{MF} conditional from $\mathcal{D}$, we evaluate the approximation within Equation \eqref{eqn:margMFGauss}. For the additional hyper-parameter $\tau$, we prescribe an uninformative Gamma prior $p(\tau)=\Gamma(\tau|a_0,b_0)$ with $a_0=b_0=10^{-9}$, which prefers smaller values for $\tau$, respectively, larger noise variances in the likelihood formulation. Afterwards, we marginalize the parameter $\tau$ as a latent variable model \cite{Franck_Koutsourelakis_2016} to yield the following \add{MF} log-likelihood, which we iteratively calculate with the (variational Bayes) Expectation-Maximization (VB-EM) algorithm \cite{dempster1977maximum, Beal_Ghahramani_2003} in the form of \eqref{eqn: effective_mf_likelihood_em}:
\begin{subequations}
\begin{align}
    \label{eqn: effective_mf_likelihood}
    \begin{split}
    \log \pmf(\Yobs|\bx) &= \log \int \pmf(\Yobs|\bx, \tau)\cdot p(\tau) \dd\tau\\
                         &= \log \Ex{p(\tau)}{\pmf(\Yobs|\bx,\tau)} 
    \end{split}\\
     \label{eqn: effective_mf_likelihood_em}   
                         &= \Ex{p(\tau|\Yobs,\bx)}{\log \left(\frac{\pmf(\Yobs|\bx,\tau) \cdot p(\tau) }{ p(\tau|\Yobs,\bx)}\right)}
\end{align}
\end{subequations}
Please see Appendix \ref{sec:vb_em_tau} for more details on how the EM algorithm is combined with the SVI procedure. 

\subsection{Algorithmic summary}
\label{sec: algo_summary}

To conclude the method section, we provide an overview of BMFIA in the form of pseudo-algorithms for its initial training phase in Algorithm \ref{alg:offline_bmfia} and its inference phase (with continuous updates for the \add{MF} conditional) in Algorithm \ref{alg:online_bmfia}. The VB-EM Algorithm \ref{alg:vb_em_tau} for the MF likelihood precision $\tau$ (which is part of the inference phase \ref{alg:online_bmfia}) is presented in Appendix \ref{sec:vb_em_tau}. 
In the first \emph{initial training phase} of BMFIA (Algorithm \ref{alg:offline_bmfia}), we generate training inputs $\bx_i$ sampled from the (joint) prior model $\bx_i\sim p(\bx)$ (see Section \ref{sec: training_data}). Afterwards, we realize the associated LF and HF simulations $\Ylf_i = \Mlf(\bx_i)$, and $\Yhf_i = \Mhf(\bx_i)$ to gather the training data $\mathcal{D}=[\Zlf_i,\Yhf_i]$, with $\Zlf_i=\{\Ylf_i,X_i\}$. We then train the probabilistic regression model, approximating the \add{MF} conditional $p(\Yhf|\Zlf)$ on $\mathcal{D}$. 
Once the probabilistic regression model is trained, we transition to the inference phase of BMFIA, which is given in Algorithm \ref{alg:online_bmfia}. Here, the probabilistic regression model approximates the \add{MF} conditional and gives rise to the \add{MF} log-likelihood. In this phase, we conduct sparse \add{VI} (see Section \ref{sec: svi} along with VB-EM (see Sections \ref{sec:small_data_approx} and \ref{sec: svi}) to iteratively calculate updates for the variational parameters $\bphi$ in the variational distribution $q_{\mathrm{MF}}(\bx|\bphi)$, which approximates the \add{MF} posterior $\pmf(\bx|\Yobs)$. In the inference phase, we only need to evaluate the LF model $\Mlf(\bx)$ and the adjoint LF model $\Alf\left(\bx,\nabla_{\Ylf}\lmf(\Ylf)\right)$ for the involved gradients, which is a crucial advantage of BMFIA. Optionally, we can update the \add{MF} conditional after some iterations of the SVI algorithm by adding additional training data sampled from the current variational posterior approximation using Algorithm \ref{alg:offline_bmfia} (with the current posterior approximation instead of the prior) and continuously retraining the \add{MF} approximation. 

\begin{algorithm}[htbp]
\SetAlgoLined

Generate initial samples from the joint-prior (or current variational distribution for refinements): 
\[
\{\bx_i\}|_{i=1}^{N_{\mathrm{init}}}\sim p(\bx,\delta), \qmf(\bx|\bphi)
\]

Evaluate LF and HF model (at coordinates $C=[\bc_i^T]$):
\[
\Ylf_i \gets \Mlf(\bx_i), \quad \Yhf_i \gets \Mhf(\bx_i), \quad i=1,\dots,N_{\mathrm{init}}
\]

Interpolate and evaluate the input field at the same spatial coordinates $C=[\bc_i^T]$ (see Equations \eqref{eqn: rf_fe} -  \eqref{eqn: input_field}):
\[
X_i(\bx_i) \gets [\tilde{\bx}_i(\bc_j)] = \left[S_{\mathrm{ele}}(\bc_j)\cdot \bx_{\mathrm{ele}}]=[S_{\mathrm{ele}}(\bc_j)\cdot\mathfrak{R}[\bx]\right]
\]

Compose \add{MF} training dataset:
\[
\mathcal{D} = \{(\Zlf_i, \Yhf_i)\},\quad\text{with}\quad\Zlf_i=[\Ylf_i,X_i]
\]

Train probabilistic regression model to approximate \add{MF} conditional density:
\[
p(\Yhf | \Zlf) \approx \mathcal{N}\left[\Yhf | M(\Zlf), \diag\left(V(\Zlf)\right)\right] \gets \mathcal{D},\ \text{untrained ProbabilisticRegression}
\]

\Return \add{MF} conditional approximation $p(\Yhf | \Zlf) \approx \mathcal{N}\left[\Yhf | M(\Zlf), \diag\left(V(\Zlf)\right)\right]$
\caption{\add{MF} Conditional Training (initial training phase)}
\label{alg:offline_bmfia}
\end{algorithm}

\begin{algorithm}[htbp]
\SetAlgoLined
Initialize variational parameters: $\bphi, \bphi_{\tau}\gets\bphi_0,\bphi_{0,\tau}$

\While{not converged}{
Sample $\nsamples$ samples from variational distribution $\qmf(\bx|\bphi)$ using reparameterization trick:
\[
\{\bx_i\}|_{i=1}^{\nsamples} =\bmu(\bphi) + L(\bphi)\cdot \{\br_i\}|_{i=1}^{\nsamples} \gets \{\br_i\}|_{i=1}^{\nsamples}\sim \mathcal{N}\left(\br|\boldsymbol{0},I\right)
\]

Evaluate LF model and input field for sample batch $\{\bx_i\}|_{i=1}^{\nsamples}$:
\[
\{\Zlf_i\}|_{i=1}^{\nsamples}\gets\{\Ylf_{i},X_i\}|_{i=1}^{\nsamples}\gets \{\Mlf(\bx_{i}),X(\bx_i)\}|_{i=1}^{\nsamples}
\]

VB-EM for likelihood precision $\tau$ (not given in closed-form), see Algorithm \ref{alg:vb_em_tau}:
\[
\{\nabla_{\Ylf}\lmf(\Zlf_i)\}|_{i=1}^{\nsamples}\gets \mathrm{Algorithm}\ \ref{alg:vb_em_tau}\ \left(\bphi_{0,\tau},q(\tau|\bphi_\tau), \lmf, \{\Zlf_i\}|_{i=1}^{\nsamples},p(\tau),\Yobs \right)
\]

Evaluate LF adjoint model for sample batch for MF log-likelihood partial gradients (right-hand side):
\[
\{\nabla_{\bx}\lmf(\bx_{i})\}|_{i=1}^{\nsamples}\gets \{\Alf\left(\bx_i, \nabla_{\Ylf}\lmf(\Zlf_i)\right)\}|_{i=1}^{\nsamples}
\]

Conduct VB-EM for prior $p(\bx)$, given in closed-form and get prior gradient $\nabla_\bx p(\bx)$:
\[
\{\nabla_{\bx} p(\bx_i)\}|_{i=1}^{\nsamples}\gets \mathrm{VB-EM}\left(p(\bx|\delta), p(\delta), \{\bx_i\}|_{i=1}^{\nsamples}\right)
\]

Apply chain rule from reparameterization to yield gradients \wrt $\bphi$:
\[
\{\nabla_{\bphi}\lmf(\bx_i(\bphi)), \nabla_{\bphi}p\left(\bx_i(\bphi)\right)\}|_{i=1}^{\nsamples}\gets \bx=\bx(\bphi)
\]

Compute ELBO gradient:
\[
\nabla_{\bphi}\mathrm{ELBO}\gets\{\nabla_{\bphi}\lmf(\bx_i(\bphi)), \nabla_{\bphi}p\left(\bx_i(\bphi)\right)\}|_{i=1}^{\nsamples}\]

Update variational parameters $\bphi$ (\eg SGD, ADAM):
\[
\bphi\gets \text{Stochastic optimizer step}\left(\bphi,\nabla_{\bphi}\mathrm{ELBO}\right)
\]

\If{refinement in iteration active}{
Draw additional samples from $\qmf(\bx|\bphi)$ and update the ProbabilisticRegression using Algorithm \ref{alg:offline_bmfia}\;
}
}
\Return optimized posterior approximation $\qmf(\bx|\bphi)$
\caption{BMFIA with Sparse Variational Inference and Expectation-Maximization (inference phase)}
\label{alg:online_bmfia}
\end{algorithm}

\FloatBarrier

\subsection{Error sources, information loss, and extreme cases of BMFIA}
\label{sec:errors_and_extremes}
We introduce a \textbf{model error} and an \textbf{information loss} by employing BMFIA. While neither can be avoided in practice, BMFIA leads to accurate HF posterior approximations. The two error sources are generally much smaller than the error that would arise when solving the inverse problem directly on the HF model with limited HF evaluations. 

The \textbf{model error} arises from the machine learning approach selected to approximate the \add{MF} conditional. In Equation \eqref{eqn:MFGaussianLik}, we postulated a Gaussian assumption for the \add{MF} conditional, which does, in general, not reflect the actual conditional distribution as it only estimates its first two moments. However, we note that a more flexible conditional model does not necessarily perform better in a \emph{small data regime} at the additional disadvantage of requiring more regularization, \eg through appropriate hyper-prior models. The Gaussian assumption allows the analytic calculation of the involved integral, which otherwise requires a Monte Carlo approximation. Investigating more flexible, non-Gaussian conditional models is an exciting direction for further research.

\begin{figure}[htbp]
    \centering
    \includegraphics[scale=0.4]{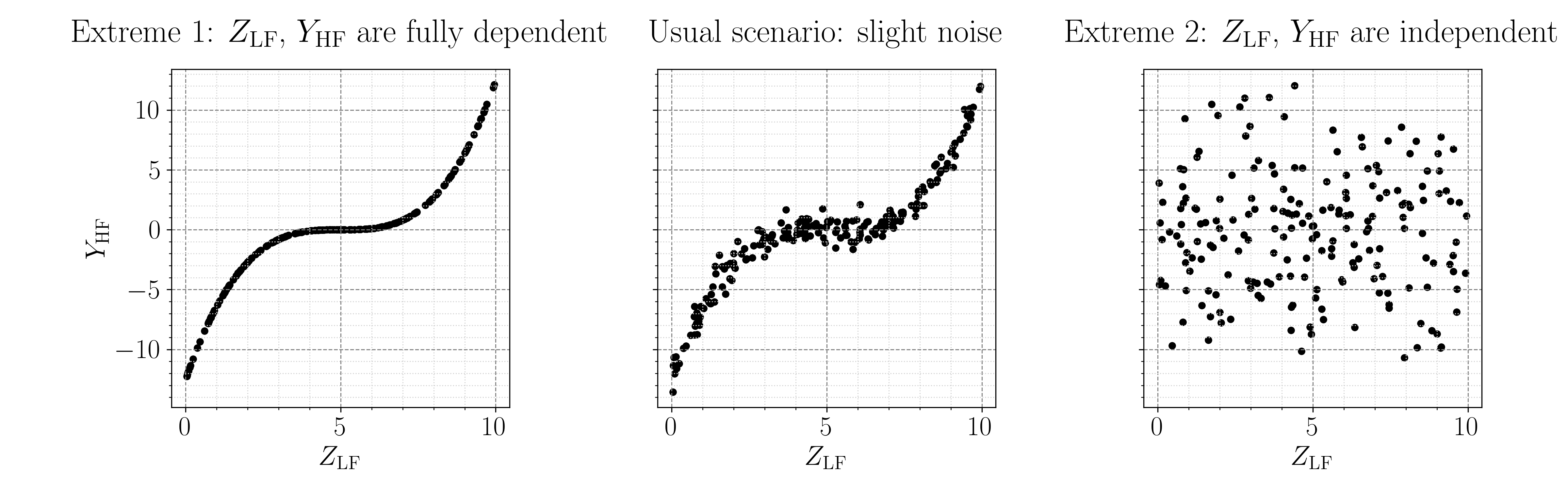}
    \caption{Extreme cases 1 and 2 and usual scenario for data distributions in the \add{MF} conditional of BMFIA. \textbf{Left}: LF and HF models are fully dependent, with no information loss and model error. \textbf{Middle}: Usual scenario in practice with a slightly noisy relationship between $\Zlf$ and $\Ylf$. \textbf{Right}: The LF model (and the augmenting input $X$) share no relationship with the HF model output $\Yhf$ in the small data regime, leading to 100\% information loss.}
    \label{fig: bmfia_extremes}
\end{figure}

Furthermore, BMFIA introduces an \textbf{information loss in the \add{MF} approximation} in Equation \eqref{eqn:bmfia_conti} (note here also the approximation ($\approx$)), as we lose the direct dependence on the input $\bx$ and only filter it through the LF model. Depending on the variance in the \add{MF} conditional, this information loss is more or less pronounced and leads to a blurring of the \add{MF} posterior and a loss of detail. The extreme cases 1 and 2 outline the limits of the information loss and are schematically illustrated in Figure \ref{fig: bmfia_extremes}: 
\textbf{Extreme 1}: The HF and LF models are \emph{entirely dependent}, \ie their relationship can be expressed as a (deterministic) function of the form $\Yhf=\boldsymbol{f}(\Zlf)$. In this case, the \add{MF} conditional $p(\Yhf|\Zlf)$ can be expressed by a deterministic mapping, and the \add{multi-fidelity} posterior $\pmf\left(\bx|\Yobs\right)$ coincides with the HF posterior $\phf(\bx|\Yobs)$. The \emph{model error} also vanishes in this case.
\textbf{Extreme 2}: The HF and the LF model are \emph{statistically independent}, \ie the \add{MF} conditional $p(\Yhf|\Zlf)$ disintegrates to $p(\Yhf|\Zlf)=p(\Yhf)$, meaning knowledge about the output of the LF model does not provide any insightful information about the HF model's output (and input features in $X$ do usually not provide sufficient information in the small data regime). The \add{MF} conditional would \emph{lean} toward pure noise in this scenario.
In practice, most scenarios lie between the two extremes. Additionally, using informative input features can somewhat mitigate the information loss as we learn (at least) low-frequency features of the input-to-output map.

An additional general challenge (which is present for all probabilistic machine learning approaches) is modeling the \textbf{epistemic uncertainty of the small data regime}, as it is generally unknown how the actual model behaves and how this uncertainty should be modeled. However, this error is not restricted to the BMFIA approach, and epistemic uncertainty arises in any problem with limited data.

\section{Numerical demonstration}
\label{sec:demonstration}
We now demonstrate BMFIA for two high-dimensional Bayesian inverse reconstruction problems from the field of porous (see Section \ref{sec: darcy_flow}) and coupled poro-elastic media flow \cite{vuong2015general, lewis1999finite} (see Section \ref{sec: poro_elastic}). Both describe common physical phenomena in biomechanics, particularly tissue modeling \cite{rauch2018coupled}, tumor modeling \cite{kremheller2018monolithic, sciume2013multiphase}, or modeling of the human lung \cite{wall2010towards}, but are also encountered in civil engineering applications such as oil reservoir modeling or groundwater flow applications \cite{zienkiewicz1999computational, rasmussen2021open, chavent1986mathematical, banaei2021numerical, schrefler1993fully}. A frequent inverse problem statement, which we also investigate with BMFIA, is the inference of the media's (isotropic) permeability field based on flow measurements. We generate a reference \add{HF} posterior for the first introductory porous media flow example (Section \ref{sec: darcy_flow}). The second coupled multi-physics example in Section \ref{sec: poro_elastic} demonstrates the Bayesian calibration of a more complex setting that does not exhibit model gradients. Here, we exploit the former porous-media flow problem from the first example as an \add{LF} model with simplified physics.

\add{Before we go into the details of the numerical demonstrations, we give an overview of the involved software frameworks that are freely available online and complement this manuscript:
\begin{itemize}
    \item The deal.II \cite{arndt2022deal} finite element library \url{https://github.com/dealii/dealii}, and our implementation of a 2D steady-state porous media flow along with its adjoint formulation \url{https://github.com/jnitzler/porous_media_flow_bmfia} in deal.II.
    \item 4C: A Comprehensive Multiphysics Simulation Framework \cite{4C}: \url{https://github.com/4C-multiphysics/4C}, which contains the implementation of the transient, nonlinear, and coupled poro-elastic media flow problem of the second numerical demonstration.
    \item QUEENS (Quantification of Uncertain Effects in Engineering Systems) \cite{queens}, our Python framework for solver-independent multi-query analyses of large-scale computational models: \url{https://github.com/queens-py/queens}. The latter is the framework in which we implemented the necessary modules and routines of BMFIA. We furthermore provide a separate tutorial repository \url{https://github.com/jnitzler/bmfia_queens_demo}, which relies on QUEENS (and uses the physics-based models above), to set up the BMFIA framework step by step.
\end{itemize}
}

\subsection{Introductory example: Bayesian calibration of a porous media flow's permeability field}
\label{sec: darcy_flow}

\begin{figure}[htbp]
\centering
\begin{tikzpicture}
    \node[anchor=south west,inner sep=0] (background) at (0,0) {\includegraphics[scale=1.0]{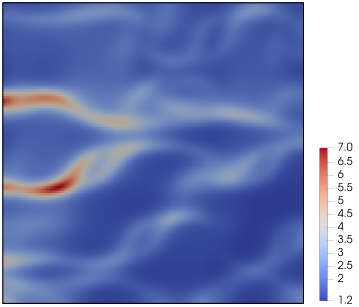}};

    \begin{scope}[x={(background.south east)},y={(background.north west)}]
        \draw[-latex, thick, black] (0.15 - 0.45, 0.03 + 0.7) -- (0.3 - 0.5, 0.03 + 0.7);        
        \draw[-latex, thick, black] (0.15 - 0.45, 0.03 + 0.7) -- (0.15 - 0.45, 0.21 + 0.65);        
        \node at (-0.16, 0.73) {$c_1$}; 
        \node at (-0.3, 0.89) {$c_2$}; 
        
        \draw[latex-latex, thin, black] (0.15-0.18, 0.03) -- (0.15-0.18, 0.98);               
        \draw (0.115-0.165, 0.012) -- (0.175-0.165, 0.012); 
        \draw (0.115-0.165, 0.99) -- (0.175-0.165, 0.99); 
        \node at (0.11-0.2, 0.5) {$1.0$};
       
        \draw[latex-latex, thin, black] (0.180-0.16, 0.035-0.058) -- (0.82+0.015, 0.035-0.058);               
        \draw (0.175-0.165, 0.07-0.058) -- (0.175-0.165, -0.012-0.045); 
        \draw (0.824+0.022, 0.07-0.058) -- (0.824+0.022, -0.012-0.045); 
        \node at (0.425, -0.015-0.058) {$1.0$};
        
        \node at (0.97, 0.6) {$k(\bxgt,\bc)$};
        
        \node at (0.12, 0.85) {$\Omega$};
        \node at (0.91, 0.84) {$\Gamma$};
        
        \draw (0.85, 0.8) -- (0.88, 0.82); 
        
    \end{scope}
\end{tikzpicture}
\caption{\add{Computational domain $\Omega$ of the porous media flow problem. Additionally, we show the ground-truth permeability field $k(\bxgt,\bc)$ by a color map.}}
\label{fig:poro_example}
\end{figure}

We first investigate the Bayesian reconstruction problem for a two-dimensional porous media flow in ${\Omega:=[0,1]\times[0,1]}$ with boundary $\Gamma$. The fluid flow is characterized by its velocity field $\bu(\bc)$ that emerges due to the pressure field $p(\bc)$, with $\bc$ denoting the spatial coordinates and the (isotropic) permeability tensor $K(\bx, \bc)=k(\bx, \bc)\cdot I$. Here, $\bx$ denotes a (high-dimensional) parameterization of the spatially varying permeability function $k(\bx, \bc)$. \add{An illustration of the computational domain $\Omega$ along with the ground-truth permeability field $k(\bxgt, \bc)$ is provided in Figure \ref{fig:poro_example}.} We prescribe a pressure BC $g(\bc)$ on the boundary $\Gamma$. \add{This first simple example was chosen to establish a controlled environment for which the HF posterior can be calculated (as the governing equations are simple enough to motivate adjoint-based model gradients) and several LF model variants of different quality can be generated. In particular, we will study the effect of deviating LF and HF physics by artificially altering the LF's boundary conditions \wrt the HF model. We emphasize that this artificial change of BCs was chosen to study the effect of deviating physics in isolation and to allow exact computation of reference posteriors. In realistic scenarios, LF models would typically keep boundary conditions close to the HF to remain informative while still being efficient. Please see also Remark \ref{rem:smoothing_filter} for further details on the mathematical effect of deviation model behavior on the MF posterior.}

\subsubsection{Governing equations and underlying physics}
The partial differential equations that describe the porous media flow are provided in mixed Laplacian formulation in the Equation system \eqref{eqn: strong_form_darcy}. The solution quantities are the vector-valued velocity field $\bu(\bc)$ and scalar pressure field $p(\bc)$:
\begin{align}
\label{eqn: strong_form_darcy}
    \begin{split}
    K^{-1}(\bx, \bc)\cdot \bu  + \nabla p &= \bnil, \qquad \text{ in } \Omega \ \text{(\add{Porous media} flow equation)}\\
    \text{div } \bu & = 0, \qquad \text{ in } \Omega\ \ \text{(incompressibility / continuity condition)}\\
    p(\bc) & = g(\bc), \quad \text{on } \Gamma\ \text{(pressure \add{BC})}
    \end{split}
\end{align}
For the HF model, we choose the following pressure \add{BC}:
\begin{equation}
    g_{\mathrm{HF}}(\bc) =  1 - c_1^2 + \left(c_2-\frac{1}{2}\right)^2.
    \label{eq:hfbc}
\end{equation}

\subsubsection{Unknowns of the Bayesian inverse problem}
Our goal is, given noisy velocity observations $\Yobs$, to identify the unknown  permeability field  
$k(\bx,\bc)$ which is expressed \wrt the  high-dimensional parameterization $\bx$ as explained in the sequel. 
To enable unconstrained Bayesian inference \wrt $\bx$, we employ an exponential reparameterization $k(\bx, \bc)=\exp\left(\txgt(\bx,\bc)\right)=\exp\left(S_{\mathrm{ele}}(\bc)\cdot\mathfrak{R}[\bx]\right)$ for the permeability field. Here $\mathfrak{R}$ is again the restrictor operator from Equations \eqref{eqn: rf_fe} to \eqref{eqn: input_field}, which selects the associated element DoFs from the global DoF vector $\bx$, and $S_{\mathrm{ele}}(\bc)$ represents the \add{FE} shape function matrix for the \add{FE} with support at coordinate $\bc$.

\subsubsection{Observables and ground-truth of the Bayesian inverse problem}
\begin{figure}[htbp]
    \centering
    \begin{tikzpicture}
    \node[anchor=south west,inner sep=0] (background) at (0,0) {\includegraphics[scale=0.6]{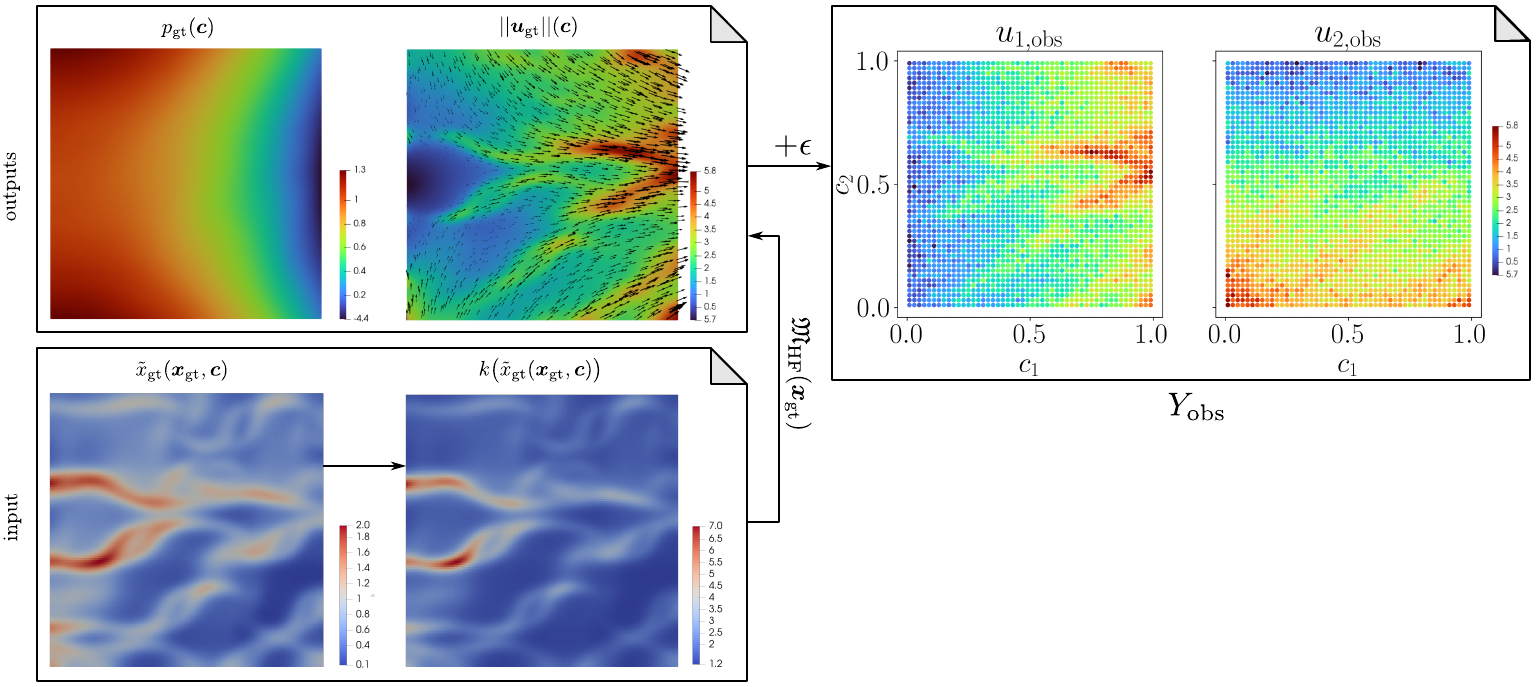}};
    \begin{scope}[x={(background.south east)},y={(background.north west)}]    
    \node[draw,circle,fill=white,inner sep=1.2pt,font=\normalsize] at (0.0,0.46) {1};    
    \node[draw,circle,fill=white,inner sep=1.2pt,font=\normalsize] at (0.0,0.95) {2};    
    \node[draw,circle,fill=white,inner sep=1.2pt,font=\normalsize] at (0.515,0.95) {3};    
    \end{scope}
    \end{tikzpicture}
    \caption{Generation \add{sequence} of \add{the} synthetic observation data $\Yobs$. \add{Starting at \ding{192}:} Computational domain $\Omega$ with ground-truth field $\txgt(\bxgt,\bc)$ (left) and exponentially transformed ground-truth permeability field $k(\bxgt,\bc)=\exp\left(\txgt(\bxgt,\bc)\right)$ \add{as the simulation input}. \add{Simulating the HF for input in \ding{192} leads to \ding{193}:} Associated HF ground-truth pressure field $p_{\mathrm{gt}}$ (left) and magnitude of the ground-truth velocity field $||\bu_{\mathrm{gt}}||$ (right). \add{Adding artifical noise $\epsilon$ and sub-sampling velocity field leads to \ding{194}:} Horizontal and vertical components of noisy, observed velocity data $\Yobs$ with an SNR of $50$, recorded on a $50 \times 50$ pixel grid.}
    \label{fig: darcy_obs_domain}
\end{figure}

We assume that noisy velocity data $\Yobs = \{\bu_{\mathrm{obs},i}\}$ is observed on a $50\times 50$ pixel grid ${C=\{(0.01:0.99) \times (0.01:0.99)\}}$ in the unit domain $\Omega$. For this study, we generate the observations $\Yobs$ synthetically from a ground-truth velocity field imposed with additive independent and identically distributed (\iid) Gaussian noise $\epsilon$. The ground-truth velocity field was simulated using the HF solver with a ground-truth permeability field $k\left(\txgt(\bx,\bc)\right)$. The example uses a \emph{medium} signal-to-noise ratio (SNR) of 50 for $\epsilon$ (see Equations \eqref{eqn: observation_darcy_1} to \eqref{eqn: observation_darcy_2}). This translates to a $2\%$ noise variance \wrt the signal variance, according to $\frac{1}{\dim(\by)\cdot\numobs}\sum_i^{\numobs}||\byobsi||^2/\sigma_{\mathrm{obs},2\%}^2=50$. 
\begin{subequations}
\begin{align}
    \label{eqn: observation_darcy_1}
    \Yobs &= \underbrace{[\Mhf(\bxgt, C)]_{\bu}}_{\Yhfgt} + \epsilon\\
    \epsilon &\sim \mathcal{N}(\hat{\epsilon}|0,\sigma_{\mathrm{obs},2\%}^2)
    \label{eqn: observation_darcy_2}
    \end{align}
\end{subequations}
The computational domain, the ground-truth input field $\txgt(\bxgt,\bc)$, and the corresponding \add{HF} simulation output $\Yhf$ are shown in Figure~\ref{fig: darcy_obs_domain}. The ground-truth permeability field $k(\bxgt,\bc)=\exp(\txgt(\bxgt,\bc))$ is available in numerical form and serves as the reference for evaluating the inferred posteriors. Although it lacks a closed-form analytical expression, it was constructed to exhibit complex spatial patterns by matching an isotropic field to the flow behavior of a separately simulated anisotropic case. This auxiliary step aims to generate a more intricate benchmark and does not influence the inference model presented in this work.
\FloatBarrier

\subsubsection{HF and LF solver and relative solver costs}
We solve the governing Equations \eqref{eqn: strong_form_darcy} in weak form using FEM. A very simple computationally faster LF model version, with a speed-up factor of roughly 6.9, is achieved by choosing a coarser numerical discretization. To emulate a slightly deviating physics of the LF model, we select different BCs for the latter (more on this in the following Section \ref{sec:lf_variants}).
An overview of the discretization and computational costs for the HF and LF forward model variants is given in Table \ref{tab: model_fem_details}. The computational costs of the respective LF adjoint models are roughly the same as those of the forward models.
\begin{table}[htbp]
\centering
\caption{Overview of employed FEM discretizations. Note that we evaluate both discretizations on the same observation grid (in this case, a $50\times50$ pixel grid) using the \add{FE} shape functions.}
\label{tab: model_fem_details}
\begin{tabular}{lcc}
\toprule
\textbf{Settings} & \textbf{HF} & \textbf{LF}\\
\midrule
Element type & \multicolumn{2}{c}{hexahedral}\\
Polynomial degree & \multicolumn{2}{c}{$\bu$: 2, $p$: 1}\\
Number of elements & 4096 & 1024\\
Number of DoFs & 37507& 9531\\
$\dim(\bx)$ & 4225 &1089\\
Approx. wall time [s] & 2.0 & 0.29\\
\midrule
\textit{HF/LF cost ratio} & \multicolumn{2}{c}{\textit{6.9}}\\
\bottomrule
\end{tabular}
\end{table}

\subsubsection{LF model variants}
\label{sec:lf_variants}
\begin{figure}[htbp]
    \centering
    \includegraphics[scale=0.32]{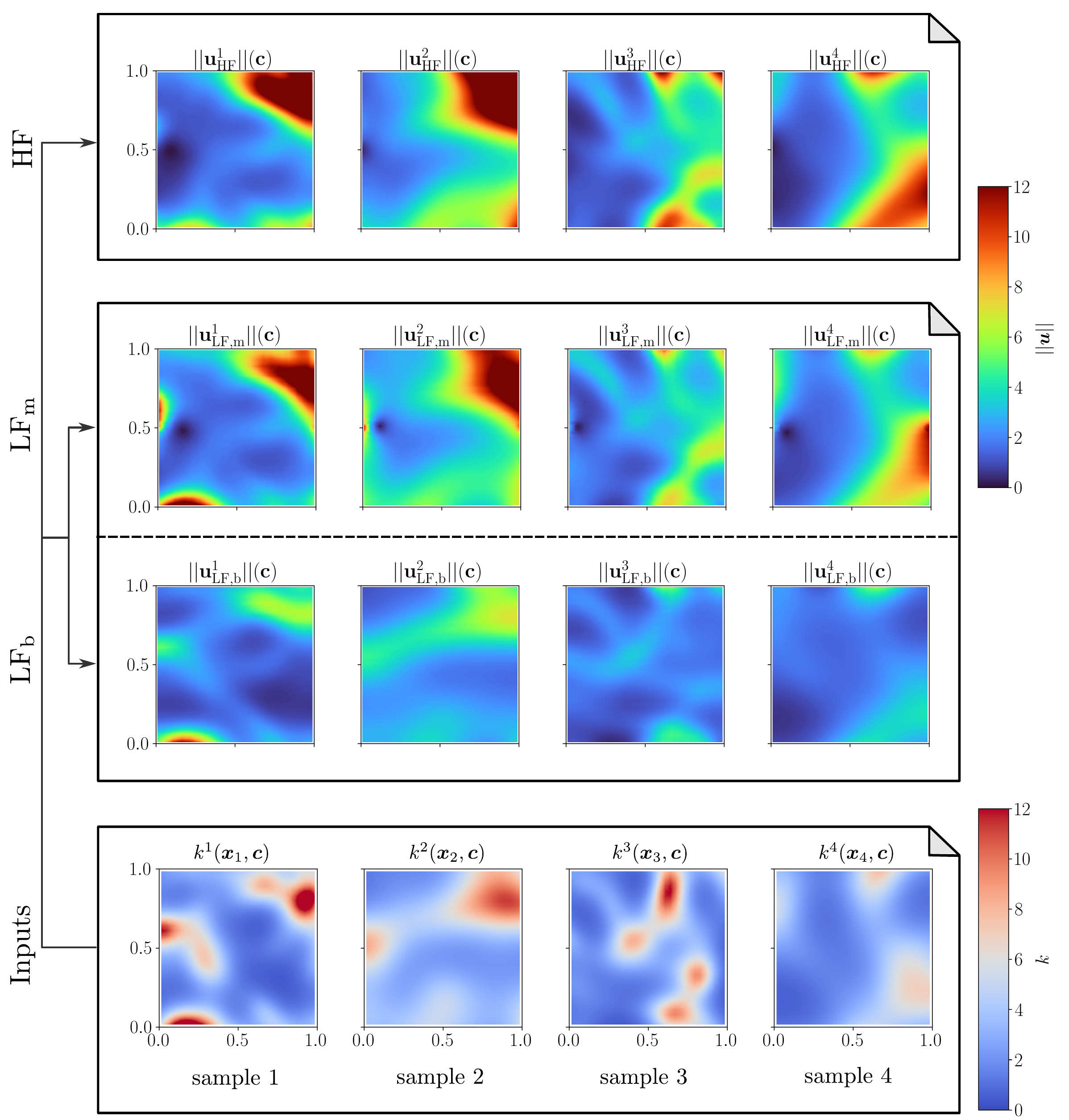}
    \caption{Comparison of the \emph{bad} \lfb, the \emph{moderate} \lfm, and the HF model responses (magnitude of the velocity field) for four different permeability input fields (samples 1 to 4). \textbf{Bottom block:} Exemplary permeability fields (inputs); \textbf{Middle block:} Model responses of the \emph{bad} \lfb model (bottom) and \emph{moderate} \lfm model (top); \textbf{Top block:} HF model response.}
    \label{fig: lf_hf_comparison}
\end{figure}

As the current porous media flow problem is simple, we can derive the adjoint formulation and infer the HF reference posterior (usually infeasible) without further physical simplifications. Moreover, to study the effect of model discrepancy between the LF and the HF model in BMFIA, we investigate two separate BMFIA approximations for 
a \emph{bad} \lfb and the \emph{moderate} \lfm by prescribing different pressure \add{BCs} than the HF model. In particular, compare with the HF BC in Equation \refeqp{eq:hfbc}:
\begin{align}
    \label{eqn: pressure_bcs}
     g_{\mathrm{LF}_{\mathrm{m / b}}}(\bc) & = \begin{cases}
    1 - c_1 + |c_2-\frac{1}{2}|, & \text{for the \emph{moderate} \lfm,}\\
    1 - \frac{2}{3}c_1, & \text{for the \emph{bad} \lfb,}\\
    \end{cases} \quad \text{on } \Gamma
\end{align}
The two LF variants, the \emph{bad} \lfb and the \emph{moderate} \lfm, are used in the following demonstration in this section to analyze the impact of a more or less noisy HF to LF relationship in the \add{MF} conditional.
For an actual, complex, nonlinear coupled problem, we would choose \add{BCs} that match the HF model version as closely as possible. The deviating physics of the LF model would arise here from the physical simplifications and numerical or geometrical coarsening. The moderate \lfm replaces the quadratic terms in the HF pressure \add{BC} by a linear expression. The bad \lfb additionally neglects the dependence on the $c_2$-coordinate and uses a different pressure gradient in the $c_1$-direction. Figure \ref{fig: lf_hf_comparison} shows the magnitudes of the  velocity fields predicted by the HF, the \emph{bad} \add{LF} model \lfb, and the moderate \add{LF} model \lfm, given four different permeability fields $k(\bx, \bc)$ (as input). One observes significant deviations, particularly with respect to \lfb, which differs strikingly from the reference HF model. 
In the subsequent examples in Section \ref{sec: porous_autoencoder} to \ref{sec: porous_costs}, we will continue to use the two LF model variants \lfb and \lfm as well as the HF porous media model to compare and highlight several aspects of BMFIA.

\FloatBarrier
\subsubsection{Probabilistic convolutional autoencoder and generation of training data}
\label{sec: porous_autoencoder}
For the \textbf{initial training phase of BMFIA}, we select suitable $\bx_i$ and obtain their corresponding outputs $\Yhf_i=\Mhf(\bx_i, C)$ and $\Ylf_i=\Mlf(\bx_i, C)$ by simulation \footnote{These simulations are embarrassingly parallel and  can be automatically performed by our freely available open-source software framework QUEENS \cite{queens}.}, \add{to generate training samples for the MF conditional}. To promote diversity in the training data, we draw samples $\bx_i$ from a \emph{modified prior} \add{$\bx_i\sim p(\bx | \delta)$}, where the original Gamma hyper-prior on $\delta$ is replaced by a uniform distribution $\tilde{p}(\delta) = \mathcal{U}(a_\delta, b_\delta)$ over the precision parameter. This encourages a more space-filling distribution of training inputs and improves the robustness of the learned regression model.
The bounds $a_\delta = 1 \cdot 10^{-3}$ and $b_\delta = 1 \cdot 10^{-2}$ are chosen heuristically based on the empirical distribution of prior samples. Since sampling from the prior is computationally inexpensive, we assess the resulting variability directly and select a range that yields a sufficiently rich and representative training dataset.
We furthermore use $\ntrain=100$ and two refinements of each $\nrefine=10$, after $100$ and $300$ SVI epochs, respectively.

\begin{figure}[htbp]
    \centering
    \includegraphics[scale=0.35]{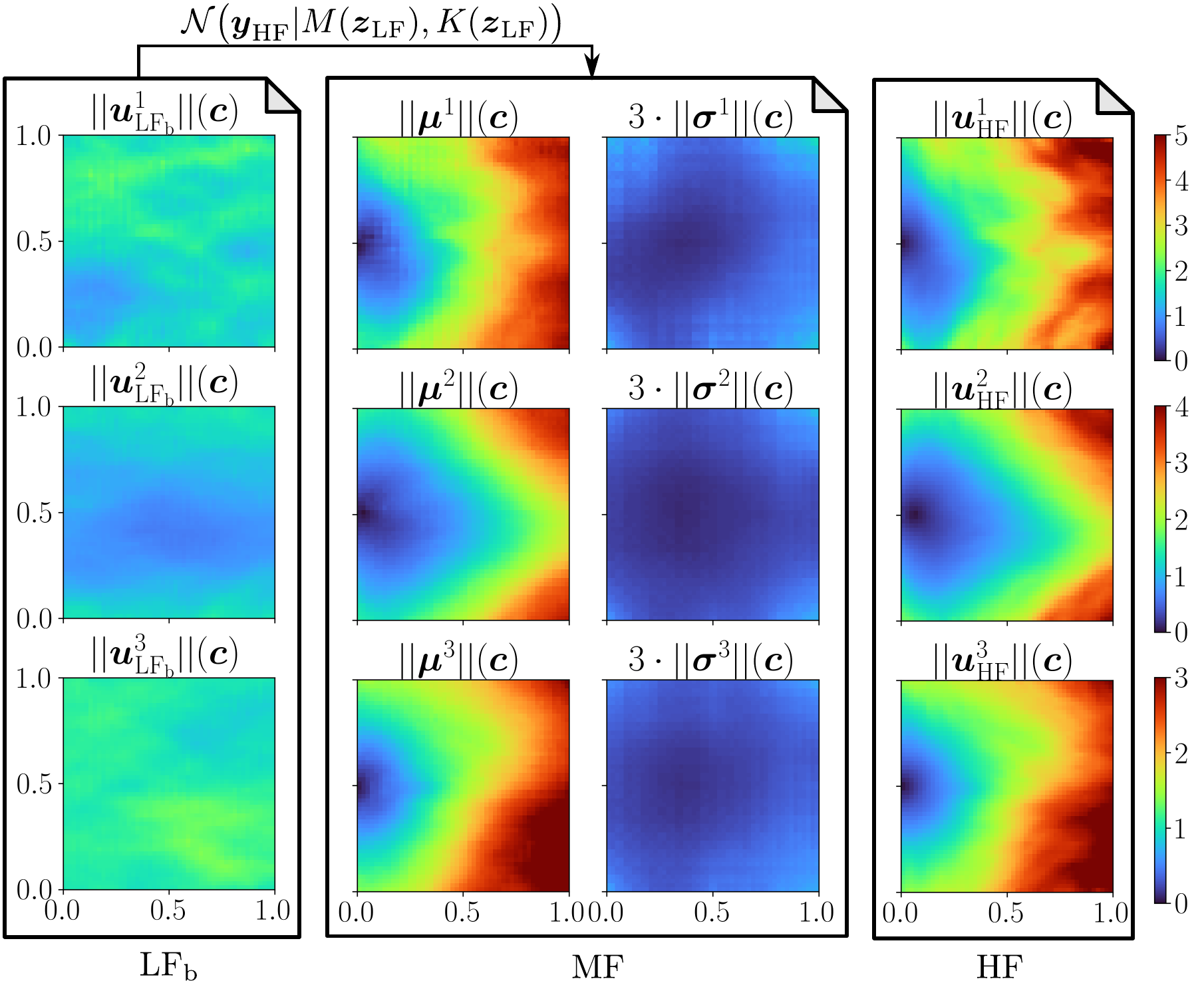}
    \caption{Demonstration of the conditional \add{MF} mapping ${p(\Yhf|\Zlf(\bx))\approx\mathcal{N}\left(\Yhf|M(\Zlf), \Kmat(\Zlf)\right)}$ for which we employ a probabilistic convolutional autoencoder, trained on (only) $\ntrain=100$ LF and HF simulations comprised in $\mathcal{D}:=\{\Zlf_i,\Yhf_i\}$. \textbf{Left box:} Three different \lfb  velocity outputs. \textbf{Middle box:} Predicted mean $M_i$ and three times standard deviation of the probabilistic convolutional autoencoder, approximating the HF output at the respective coordinate $\bc$. \textbf{Right box:} Corresponding HF model outputs for the same inputs $\bx_i$ are normally not calculated and only shown for comparison.}
    \label{fig: gaussian_nn_mapping}
\end{figure}

We subsequently employ the probabilistic convolutional autoencoder (see Section \ref{sec:p_cae}) within the \add{MF} likelihood function (see Equation \eqref{eqn:margMFGauss}) to \textbf{conduct inference in the inference phase of BMFIA}. \add{We use the training settings as specified in Table \ref{tab:bmfia_settings}. The training time for the convolutional autoencoder (here for a geometrical 2D domain) was in all investigated cases around two to five minutes on a single \emph{NVIDIA GeForce GTX 1080 Ti} GPU.} The demonstrations in this section use again the previously introduced porous media flow model with its HF and two \add{LF} model variants \lfb and \lfm (see Equation \eqref{eqn: pressure_bcs}).
Figure \ref{fig: gaussian_nn_mapping} illustrates the predictive capabilities of the trained convolutional autoencoder (the \add{MF} conditional) for three \lfb velocity fields that were not part of $\mathcal{D}$.
Despite the significant deviation between the \lfb and  the HF predictions, the \add{MF} conditional enables one to obtain very good approximations of the HF velocity fields. 
We predict the two-dimensional mean vector and the diagonal covariance matrix for the associated two-dimensional HF output. Learning a direct surrogate for the map $\bx\mapsto\Yhf$ without exploiting the LF model would not be expedient due to the curse of dimensionality.

\subsubsection{Employed SVI procedure}
We use sparse SVI with VB-EM steps according to Section \ref{sec: svi} to iteratively compute an approximation to the \add{MF} posterior with both LF models, \lfb and \lfm (see Equation \eqref{eqn: pressure_bcs}). Table \ref{tab:bmfia_settings} summarizes the settings. We provide the SVI procedure's convergence plots and iteration data in Appendix \ref{sec:convergence_plots} and Figure \ref{fig: svi_convergence_ex1}.
\begin{table}[htbp]
\centering
\caption{Settings for the BMFIA analysis}
\label{tab:bmfia_settings}
\begin{tabular}{lll} 
\toprule
\textbf{Settings} & \textbf{Sub-settings} & \textbf{Value} \\
\midrule
SVI settings & & \\
& Number of samples per batch & 6 \\
& Max number of solver calls & 4000 \\
& Variational distribution & Sparse Gaussian \\
& Off-diagonal bandwidth & 10 \\
& Stochastic optimizer & SGD \\
& Learning rate & $1\cdot 10^{-3}$\\
\addlinespace
\add{MF} likelihood & & \\
 & Type & Diagonal Gaussian \\
 & (Additive) precision parameter & Gamma hyper-prior\\
 & SVI epochs for refinements & [100, 300]\\
 & Number of refinement samples & 10 \\
 & Initial number of HF/LF simulations & 100 \\
 & Number of observation points & $50 \times 50 = 2500$ \\
\addlinespace
\add{MF} conditional & & \\
& Type & Gaussian convolutional autoencoder \\
& Inputs & $[U_{1, \mathrm{LF}}, U_{2,\mathrm{LF}}, X]$ \\
& Outputs & $[M_1, M_2, V_1, V_2]$\\
& Optimizer & SGD\\
& Hyper prior for $\mathcal{D}$ & $\delta\sim\mathcal{U}\left[10^{-3},10^{-2}\right]$\\
& Learning rate & $1\cdot10^{-3}$\\
& Number of epochs & 4000\\
& Batch size & $128$\\
\addlinespace
Prior & & \\
 & Type & Gaussian Markov field\\
& Prior mean & 1.0 \\
 & Hyper prior for precision $\delta$ & Gamma hyper-prior\\
 & Stochastic dimension LF & 1089\\
  & (Stochastic dimension HF) & (4225)\\
\bottomrule
\end{tabular}
\end{table}
In addition to the \add{MF} posteriors, we directly solve the (reference) \add{BIP} for the HF model. The HF posterior is prohibitive for more complex coupled physics-based problems and can only be provided here due to the simplicity of this demonstration example. Moreover, we infer the posteriors using the isolated LF models to compare their stand-alone capabilities for Bayesian inverse analysis. 

\FloatBarrier
\subsubsection{Discussion of posterior results}
\begin{figure}[htbp]
    \centering
    \begin{tikzpicture}
    \node[anchor=south west,inner sep=0] (background) at (0,0) {\includegraphics[scale=0.37]{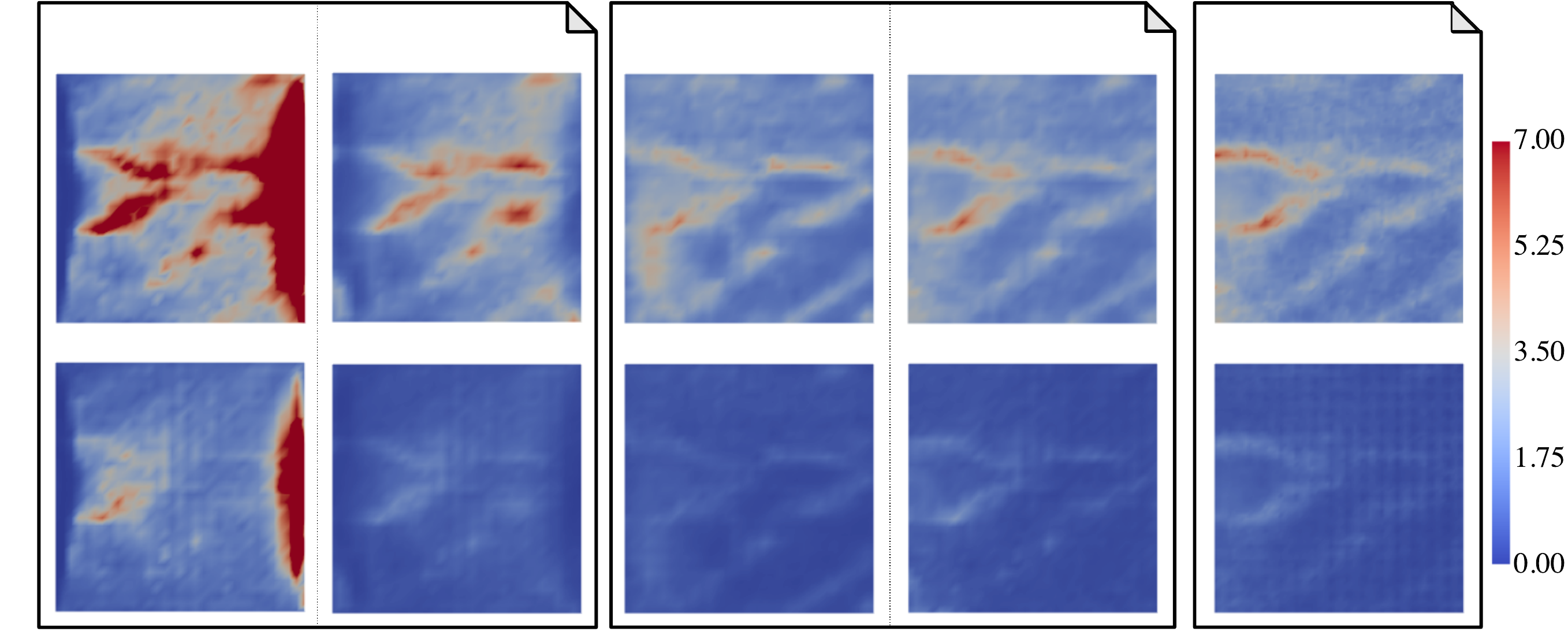}};
    \begin{scope}[x={(background.south east)},y={(background.north west)}]    
        \node at (0.11, 0.93) {\lfb};       
        \node at (0.3, 0.93) {\lfm};       
        \node at (0.48, 0.93) {BMFIA(\lfb)};       
        \node at (0.65, 0.93) {BMFIA(\lfm)};       
        \node at (0.85, 0.93) {HF};       
        \node at (0.96, 0.84) {$k$};       
        \node[rotate=90] at (0.01, 0.7) {mean};       
        \node[rotate=90] at (0.01, 0.23) {$2\cdot\mathbb{SDT}$};       
    \end{scope}
    \end{tikzpicture}
        \caption{Comparison of posterior distributions ($\mathrm{SNR}=50$) for the isotropic permeability field $k(\bx,\bc)$. The posteriors' mean functions (upper row) and $2\cdot \mathbb{STD}$ (bottom row) are depicted. \textbf{Left block:} Posterior distributions of LF models directly. \textbf{Middle block:} BMFIA posteriors for \lfb and \lfm using $\ntrain=100$ and $\nrefine=20$. \textbf{Right block:} HF posterior for comparison (normally unavailable). For a comparison of the computational costs, see Table \ref{tab: posterior_costs_ex1}.}
    \label{fig: bmfia_posterior_fields}
\end{figure}

\begin{figure}[htbp]
    \centering
    \includegraphics[scale=0.2]{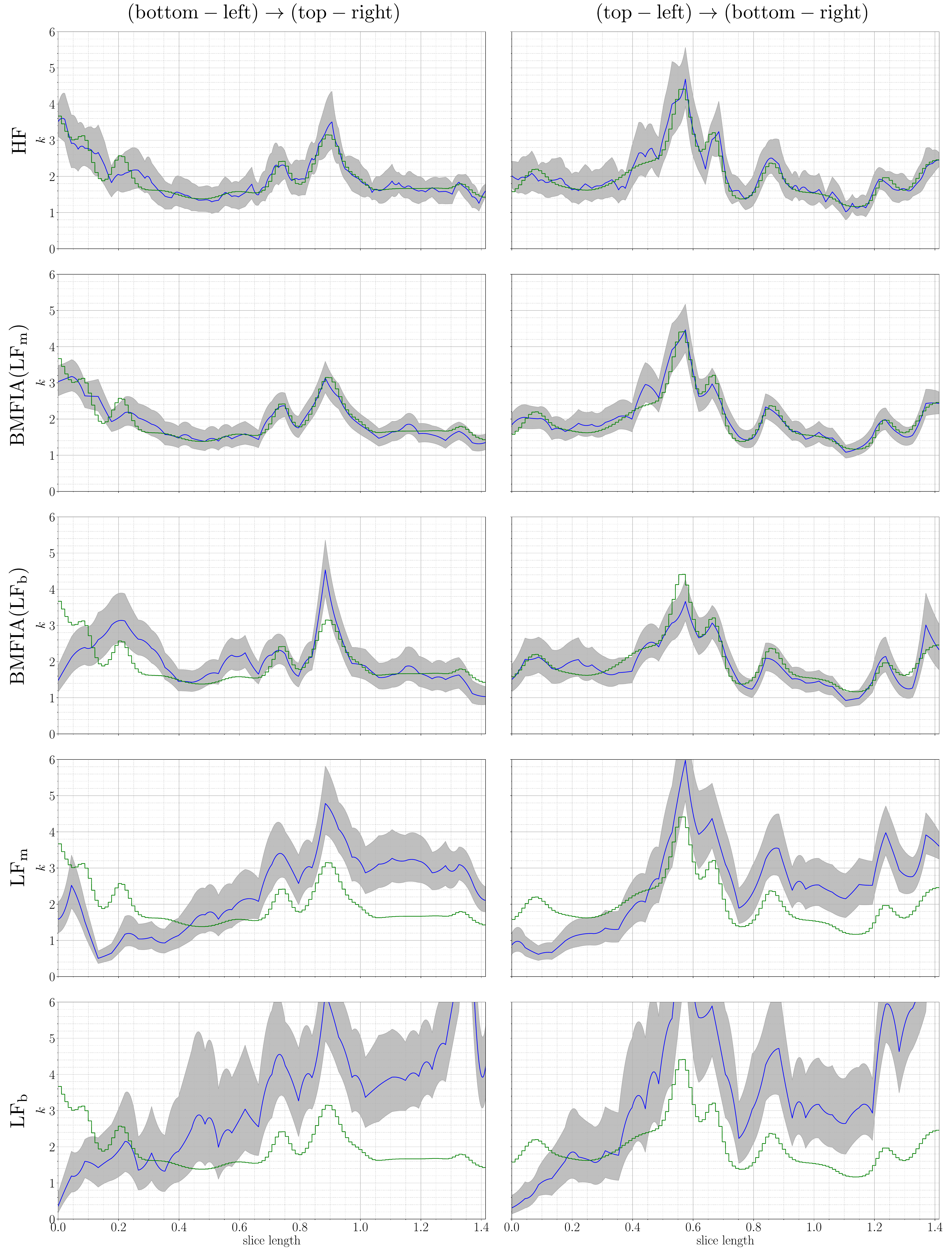}
    \caption{Diagonal slices through the posterior permeability field of Figure \ref{fig: bmfia_posterior_fields} with the ground-truth permeability (\usebox{\darkgreenline}), posteriors median (\usebox{\blueline}), and $90\%$ credible bounds (area between the 5\% and 95\% percentile) (\usebox{\lightgreybox}). \textbf{Top row:} HF posterior (normally unavailable!). \textbf{Second and third row to the top}: BMFIA posteriors with \lfm and \lfb model. \textbf{Two bottom rows:} Posterior slices when directly using the LF models. \textbf{Left column:} Slice from bottom left to top right. \textbf{Right column:} Slice from top left to bottom right.}
    \label{fig: bmfia_posterior_slices}
\end{figure}

Figure \ref{fig: bmfia_posterior_fields} compares the posteriors for the permeability fields $k(\bx)$ in the form of the posterior's mean and two times the standard deviation. Figure \ref{fig: bmfia_posterior_slices}  shows the (asymmetric) credible intervals along a diagonal slice of the problem domain.
The posteriors  corresponding to the  two  LF models (left columns  of Figure \ref{fig: bmfia_posterior_fields}) exhibit  substantial discrepancies with the HF reference posterior (right block). Their mean functions differ strongly from the HF posterior's mean function, while the uncertainty is vastly underestimated. Neither the \lfb posterior nor the \lfm posterior provides a good \emph{stand-alone} approximation of the HF posterior. Quantitatively, the posteriors are compared in Figure \ref{fig: bmfia_posterior_slices}, where we depict two diagonal slices (bottom left to top right and top left to bottom right) through the posterior field. The LF posteriors in the two bottom rows show a more than 350\% local error in their mean functions and significant bias. 

On the contrary, both BMFIA posterior approximations using the \lfb or the \lfm agree very well with the HF posterior and reflect its uncertainty (even for the small number of only 120 HF training data used in this example). In Figure \ref{fig: bmfia_posterior_slices}, the quantitative results resemble the HF posterior in very good agreement despite the low quality of the LF models and the 120 HF simulations used. This is especially impressive for the results obtained with the \emph{bad} \lfb. Figures \ref{fig: bmfia_posterior_fields} and \ref{fig: bmfia_posterior_slices} show only minor artifacts for the \lfb-BMFIA posterior, which are mostly resolved with more data, as shown in the convergence study in Appendix \ref{sec:convergence_study} and Figure \ref{fig: bmfia_posterior_training_points}. We believe that these remaining artifacts or differences are not rooted in the BMFIA approach itself but rather stem from the stochastic optimizer, and that gets partially trapped in local optima. As already pointed out in Section \ref{sec:methodology}, it would be very interesting to investigate more advanced (second-order) optimization routines and regularization approaches in the future to remedy this (local) phenomenon.
The BMFIA posterior using the moderate \lfm (second column from left in Figure \ref{fig: bmfia_posterior_fields} and second row in Figure \ref{fig: bmfia_posterior_slices}) follows the HF posterior almost perfectly. It has tighter uncertainty bounds than BMFIA(\lfb), as the information loss between the HF and the \lfm model is smaller. 

\FloatBarrier
\subsubsection{Comparison of computational costs}
\label{sec: porous_costs}
The computational costs for obtaining the posteriors in Figures~\ref{fig: bmfia_posterior_fields} and \ref{fig: bmfia_posterior_slices} are summarized in Table~\ref{tab: posterior_costs_ex1}, comparing their total wall time under different hardware scenarios with 6, 12, 30, and 60 available CPU cores. For this very simple demonstrative example, computational savings were not our main focus: Each simulation occupies a single core, with measured times of 0.29 seconds for an LF run and 2.0 seconds for an HF run (see Table~\ref{tab: model_fem_details})\footnote{For more demanding HF simulations, single-run parallelization would be employed. For this simplified test case, we omit this but do parallelize HF simulations in Section~\ref{sec: poro_elastic}.}. In the single-fidelity setting, the posterior inference requires 4000 forward model evaluations and 4000 adjoint solves, carried out in 1333 batch-sequential steps with a batch size of six. This sequential structure becomes particularly limiting when using expensive HF models. In contrast, BMFIA enables an embarrassingly parallel training phase with 100 LF and HF simulations. During inference, BMFIA requires only LF forward and adjoint evaluations (neglecting additional refinements of the \add{MF} conditional), making its inference cost comparable to the LF-only case. Table~\ref{tab: posterior_costs_ex1} reports the resulting wall times for each method across all core counts. BMFIA consistently achieves a six-to-seven-fold reduction in wall time compared to the HF baseline, with greater relative benefit as more cores become available. While this very simple test case is computationally light, the performance gains of parallelization become even more pronounced in more demanding scenarios (see Section~\ref{sec: poro_elastic}).

\begin{table}[htbp]
\centering
\caption{Computational costs for obtaining different posteriors in effective wall time [s]}
\label{tab: posterior_costs_ex1}
\begin{tabular}{ccccc} 
\toprule
\textbf{Num. avail. CPUs} & \textbf{LF posterior} [s] & \textbf{BMFIA posterior} [s] & \textbf{HF posterior} [s] & \textbf{HF/BMFIA cost ratio}\\
\midrule
6& 387 &  425 & 2666 & $6.3\times$\\
12& 387 & 406 & 2666 & $6.5\times$\\
30& 387 & 395 & 2666 & $6.7\times$\\
60& 387 & 391 & 2666 & $6.8\times$\\
\bottomrule
\end{tabular}
\end{table}


\subsection{Bayesian calibration of a dynamic and coupled nonlinear poro-elastic model without model gradient information}
\label{sec: poro_elastic}

In the second numerical demonstration, we conduct the Bayesian calibration of a dynamic and coupled nonlinear poro-elastic medium model (HF model) \cite{vuong2015general}, for which an adjoint formulation is unavailable. The Bayesian calibration aims to identify the poro-elastic medium's permeability field based on dynamic measurements of the two-dimensional flow. \add{The example considered here is motivated by actual continuum models used in patient-specific human lung simulations. During respiration, the lung undergoes large deformations, resulting in large strains that introduce both geometric and constitutive nonlinearities. Capturing these effects as well as the coupling between the fluid and elastic domain is essential, as linearized assumptions would lead to significant modeling errors \cite{vuong2015general}.}
We reuse the former steady-state porous media flow model as an LF model, \add{effectively neglecting the nonlinear coupling to the elastic domain}. This strategy has several advantages: This LF model has a straightforward adjoint formulation, as its governing equations are much simpler than those of the coupled nonlinear poro-elastic problem (HF). Hence, gradients of the LF are available and can be used in the context of BMFIA.
Furthermore, the computational cost of the LF is much lower due to the fewer unknowns and the complete omission of the nonlinear elasticity. The LF model is also steady-state compared to the transient HF model, reducing computational costs. Despite these drastic simplifications, a statistical relationship between the LF and HF models remains, as shown in the sequel. 
In the following, we demonstrate that BMFIA effectively leverages this structure to recover \add{HF} parameter posteriors with significant computational savings. \add{No HF adjoints or model gradients are available in this second numerical example, so we cannot compute a reference HF posterior in a feasible time. While one could in principle initialize a sampling-based method (e.g., MCMC) at the ground truth, this would be unrealistic in practice, still extremely costly, and would mix the effects of the sampling strategy with the variational approximation of BMFIA, particularly since we use a specifically designed VB–EM procedure in the SVI, which is not straightforward to replicate in MCMC; we therefore chose to present only the ground truth HF solution as a reference, mainly as we already discussed the HF-SVI approximation in the first numerical example.}

\subsubsection{Problem setup and governing equations of the transient coupled poro-elastic model}
In addition to the former porous media flow example (see Equation \eqref{eqn: strong_form_darcy}), the nonlinear poro-elastic problem introduces a two-way coupling between the fluid and the (nonlinearly modeled) solid domain in the porous medium. 
The fluid problem is defined on the spatial domain $\Omega_t$, which changes over time due to large elastic deformations \wrt its reference domain $\Omega_0$. The nonlinear elastic equations \eqref{eqn: strong_form_poroelasticity_c} are classically formulated \wrt the reference configuration $\Omega_0$.
The fluid pressure field $p$ exerts stresses on the elastic porous structure, leading to a nonlinear elastic deformation (described by the deformation gradient $\bF$). Assuming an incompressible skeleton material, any volumetric structural deformation changes the medium's porosity $\phi$. The latter induces a change in the flow field $\bu$ and couples back to the pressure field $p$. The governing equations of the transient poro-elastic problem read as follows: 
\begin{subequations}
\begin{align}
    \rho^\text{f} \frac{\partial \bu^\text{f}}{\partial t}
    - \rho^\text{f} \left( \bu^\text{s} \cdot \nabla \right) \bu^\text{f}
    + \mu \phi K^{-1}(\bx, \bc) \cdot \bu^\text{f}
    + \nabla p &= \bnil, \qquad
    \text{ in } \Omega_{t} 
\label{eqn: strong_form_poroelasticity_a} \\
    \frac{\partial \phi}{\partial t}
    + \phi \nabla \cdot \bu^\text{s}
    + \nabla \cdot \left(\phi \left(\bu^\text{f} - \bu^\text{s}\right)\right) & = 0, \qquad \text{ in } \Omega_{t}
\label{eqn: strong_form_poroelasticity_b} \\
    \rho^\text{s} \frac{\partial \bu^\text{s}}{\partial t}
    - \nabla_{0} \left(\bF \cdot \bS \right)
    - J \phi \bF^{-T} \nabla_{0} p
    - \mu J \phi^{2} K^{-1}(\bx, \bc) \cdot \left( \bu^\text{f} - \bu^\text{s} \right) & = \bnil, \qquad \text{ in } \Omega_{0}
\label{eqn: strong_form_poroelasticity_c}
\end{align}
\end{subequations}
Note that Equations \eqref{eqn: strong_form_poroelasticity_a} and \eqref{eqn: strong_form_poroelasticity_b} are the more general versions of the earlier steady-state, uncoupled porous media flow problem in Equation \eqref{eqn: strong_form_darcy}. Differences result from the transient formulation and the coupling terms between the deformable structure and the fluid. In the coupled formulation, we distinguish the velocities and densities of the fluid and solid phases, \ie $\bu^\text{f}, \bu^\text{s}$, and $\rho^\text{f}, \rho^\text{s}$, respectively, with their corresponding superscripts. Other physical quantities in the dynamic poro-elastic formulation are the fluid's dynamic viscosity $\mu$, which we assume is constant and homogeneous, and the porosity field $\phi = \phi(\bc)$. The additional Equation \eqref{eqn: strong_form_poroelasticity_c} describes the nonlinear momentum balance of the porous structure (including both the solid and the fluid phases). The elastic deformation is characterized by the deformation gradient $\bF$, its determinant $J$, the Green-Lagrange strain tensor $\bE$, and the second Piola-Kirchhoff stress tensor $\bS$. The constitutive relations for the second Piola-Kirchhoff stress tensor $\bS$ and pressure $p$ are given as follows:
\begin{align}
\label{eqn: constitutive_law_poroelasticity}
    \begin{split}
    \bS & = \cfrac{\partial \psi \left(\bE, J \phi\right)}{\partial \bE} - p J \bC^{-1} \\
    p & = \cfrac{\partial \psi \left(\bE, J \phi\right)}{\partial \left(J \phi\right)} \\
    \end{split}
\end{align}
The first equation in \eqref{eqn: constitutive_law_poroelasticity} is known as Terzaghi's principle. The choice of strain energy function $\psi$ incorporating the aforementioned incompressible skeleton material implies that the porosity is a function of the determinant $J$ of the deformation gradient $\bF$ and the initial porosity field $\phi_0$ only, and does not depend on the fluid pressure $p$, \ie
\begin{align}
    \phi = 1 - \frac{1 - \phi_0}{J}.
\end{align}
We employ a Neo-Hookean solid material law for this numerical example.
\begin{align}
    \cfrac{\partial \psi \left(\bE, J \phi\right)}{\partial \bE} = \frac{\partial}{\partial \bE} \left(C_1 \left( I_1 - 2 - \ln{J} \right) + C_2 \left(J - 1 \right)^2 \right)
\end{align}

\begin{figure}[htbp]
\centering
\begin{tikzpicture}
    \node[anchor=south west,inner sep=0] (background) at (0,0) {\includegraphics[scale=0.25]{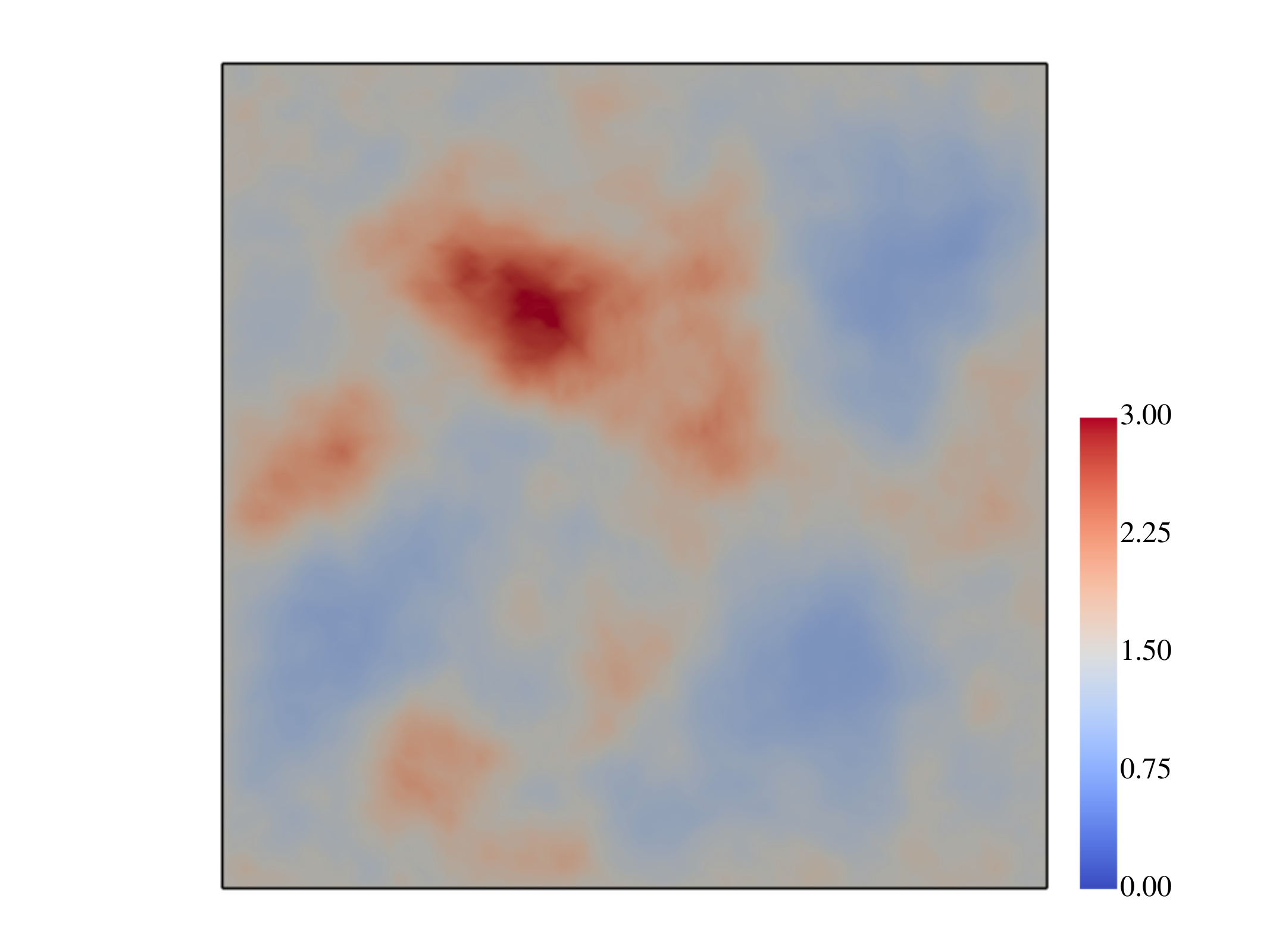}};

    \begin{scope}[x={(background.south east)},y={(background.north west)}]
        \draw[-latex, thick, black] (0.15 - 0.2, 0.03 + 0.7) -- (0.3 - 0.25, 0.03 + 0.7);        
        \draw[-latex, thick, black] (0.15 - 0.2, 0.03 + 0.7) -- (0.15 - 0.2, 0.21 + 0.65);        
        \node at (0.07, 0.73) {$c_1$}; 
        \node at (-0.05, 0.89) {$c_2$}; 
        
        \draw[latex-latex, thin, black] (0.15, 0.075) -- (0.15, 0.925);               
        \draw (0.115, 0.07) -- (0.175, 0.07); 
        \draw (0.115, 0.932) -- (0.175, 0.932); 
        \node at (0.11, 0.5) {$1.0$};
       
        \draw[latex-latex, thin, black] (0.180, 0.035) -- (0.82, 0.035);               
        \draw (0.175, 0.07) -- (0.175, -0.012); 
        \draw (0.824, 0.07) -- (0.824, -0.012); 
        \node at (0.5, -0.015) {$1.0$};
        
        \node at (0.89, 0.62) {$k(\bx,\bc)$};
        
        \node at (0.24, 0.85) {$\Omega$};
        \node at (0.9, 0.85) {$\Gamma_{\mathrm{right}}$};
        \node at (0.25, 0.7) {$\Gamma_{\mathrm{left}}$};
        \node at (0.3, 1.0) {$\Gamma_{\mathrm{top}}$};
        \node at (0.72, 0.11) {$\Gamma_{\mathrm{bottom}}$};
        
        \draw (0.825, 0.8) -- (0.86, 0.82); 
        \draw (0.18, 0.69) -- (0.21, 0.7); 
        \draw (0.23, 0.93) -- (0.26, 0.97); 
        \draw (0.46 + 0.17, 0.07) -- (0.48 + 0.17, 0.1); 
        
        \node[black] at (0.5, 0.93) {$\times$};
        \node[black, above] at (0.5, 0.94) {$P^{\text{s}}_0$};
        
    \end{scope}
\end{tikzpicture}
\caption{Computational domain $\Omega$ of the poro-elastic problem with dimensions in reference (undeformed) configuration $\Omega_0$. Additionally, we show the ground-truth permeability field $k_{\mathrm{gt}}(\bxgt,\bc)$ by a color map.}
\label{fig:poro_el_gt}
\end{figure}

Subsequently, we investigate a unit square, two-dimensional poro-elastic structure, depicted in Figure \ref{fig:poro_el_gt}. The structure has horizontal sliding \add{BCs} on its top and bottom, with an additional point fixture in the middle of the top boundary to make the system well-defined. We then inject a homogeneous inflow at the top boundary and assume that the fluid can only flow out over the bottom boundary and can otherwise not \emph{escape} the structure, \ie via the left or right boundary. Due to the fluid flow and its evolving pressure field, the structure deforms over time. We define a spatially variable ground-truth permeability field $k_{\mathrm{gt}}(\bxgt)$ (also depicted in Figure \ref{fig:poro_el_gt}).

Mathematically, the \add{BCs} are formulated as follows:
\begin{subequations}
\begin{align}
    \bdisp^{\text{s}}\cdot \bn_{\mathrm{top}} = 0, &\qquad \text{ on } \Gamma^{\text{s}}_{0,\text{top}} \times \left[t_0; t_\text{F}\right]
    \label{eqn: poro_elastic_bcs_a} \\
     \bdisp^{\text{s}} = \boldsymbol{0}, &\qquad \text{ at } P^{\text{s}}_{0} \times \left[t_0; t_\text{F}\right]
      \label{eqn: poro_elastic_bcs_point} \\
    \bdisp^{\text{s}} \cdot \bn_{\mathrm{bottom}} = 0, &\qquad \text{ on } \Gamma^{\text{s}}_{0,\text{bottom}} \times \left[t_0; t_\text{F}\right]
    \label{eqn: poro_elastic_bcs_b} \\
    \left( \bF \bS \right) \cdot \bN_{0,\mathrm{left,right}} = \bnil, &\qquad \text{ on } \Gamma^\text{s}_{0,\text{left}} \cup \Gamma^\text{s}_{0,\text{right}} \times \left[t_0; t_\text{F}\right]
    \label{eqn: poro_elastic_bcs_c} \\
    \left(\bu^\text{f} - \bu^\text{s} \right)\cdot \bn_{\mathrm{left, right}} = 0, & \qquad \text{ on } \Gamma^\text{f}_{t,\text{left}} \cup \Gamma^\text{f}_{t,\text{right}} \times \left[t_0; t_\text{F}\right]
    \label{eqn: poro_elastic_bcs_d} \\
    p^\text{f} = 0, &\qquad \text{ on } \Gamma^{\text{f}}_{t, \text{bottom}} \times \left[t_0; t_\text{F}\right]
     \label{eqn: poro_elastic_bcs_e}\\
     p^{\text{f}} = 0.02\cdot\pi\cdot\sin(0.4\cdot\pi\cdot t), &\qquad \text{ on } \Gamma^{\text{f}}_{t, \text{top}} \times \left[t_0; t_\text{F}\right]
     \label{eqn: poro_elastic_bcs_f}
\end{align}
\end{subequations}
where $\bdisp^\text{s}$ is the structural displacement field. $\bn$ and $\bN_0$ denote the outward boundary normal vectors in the spatial and reference configurations, respectively. \add{The pressure inflow and outflow conditions at the top and bottom, as well as the impervious boundary conditions on the left and right,  are chosen in a fashion that resembles a channel flow.}
Complementing the \add{BCs}, the initial conditions are homogeneous in the whole domain:
\begin{subequations}
\begin{align}
    \bdisp^{\text{s}} = \boldsymbol{0}, &\qquad \text{ in } \Omega^{\text{s}}_{0} \times \left[t_0\right]
    \label{eqn: poro_elastic_ics_d} \\
    \bu^{\text{s}} = \boldsymbol{0}, &\qquad \text{ in } \Omega^{\text{s}}_{0} \times \left[t_0\right]
    \label{eqn: poro_elastic_ics_us} \\
     \bu^{\text{f}} = \boldsymbol{0}, &\qquad \text{ in } \Omega^{\text{f}}_{0} \times \left[t_0\right]
      \label{eqn: poro_elastic_ics_uf} \\
    p^{\text{f}} = 0, &\qquad \text{ in } \Omega^{\text{f}}_{0} \times \left[t_0\right]
    \label{eqn: poro_elastic_ics_pf}
\end{align}
\end{subequations}

\FloatBarrier

\subsubsection{Bayesian inverse problem, BMFIA setup, ground-truth, and observables}
\label{sec: setup_poro_elastic}
We investigate the following \add{BIP}: 
Given are noisy observations $\Yobs$ (quantitative details below) of the previously described poro-elastic problem's vector-valued velocity field over time, we want to infer the permeability field, \ie obtain the fields posterior $p(\bx|\Yobs)$. 
We employ BMFIA to find an \add{MF} posterior approximation $\pmf(\bx|\Yobs)$.

\begin{figure}[htbp]
    \centering
    \includegraphics[scale=0.9]{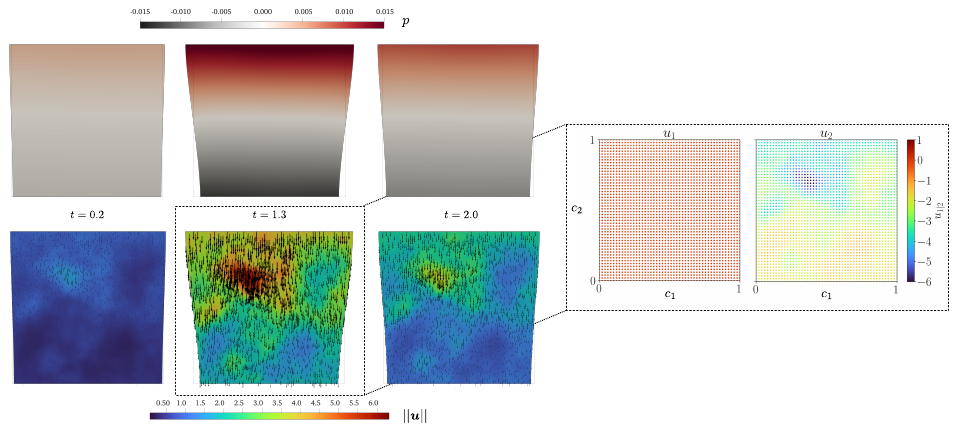}
    \caption{Ground-truth output fields (pressure, displacement, and velocity) at different physical time points. \textbf{Left top row}: Pressure field (color-coded) and displacement field (morphing of the domain) over time. \textbf{Left bottom row}: Velocity field and displacement field (morphing of the domain) over time. The black-dashed box around the velocity field at $t=1.3$ marks the time point at which we generate the observations (\textbf{right side}). The resulting pixel observations of the $u_1$ and $u_2$ velocity fields are shown in the dashed box on the right.}
    \label{fig: poro_elastic_observables}
\end{figure}

The ground-truth outputs in their deformed configurations at several time points and the observed, noise-corrupted velocity field at $t=1.3$ are depicted in Figure \ref{fig: poro_elastic_observables}.
We keep the Bayesian calibration procedure mainly analogous to the former example in Section \ref{sec: darcy_flow} and Table \ref{tab:bmfia_settings}, except for the deviating settings, summarized in Table \ref{tab:deviating_bmfia_settings}:
\begin{table}[htbp]
\centering
\caption{BMFIA settings (deviating from Table \ref{tab:bmfia_settings}) for poro-elastic problem}
\label{tab:deviating_bmfia_settings}
\begin{tabular}{lll} 
\toprule
\textbf{Settings} & \textbf{Sub-settings} & \textbf{Value} \\
\midrule
SVI settings & & \\
& Max number of solver calls & 10000 \\
\addlinespace
\add{MF} likelihood & & \\
 & Initial number of HF/LF simulations & 300\\
 & Number of refinement samples & 0 \\
\addlinespace
\add{MF} conditional & & \\
& Hyper prior for $\mathcal{D}$ & $\delta\sim\mathcal{U}\left[3\cdot10^{-3},6\cdot10^{-3}\right]$\\
& Learning rate & $1\cdot10^{-3}$\\
& Number of epochs & 10000\\
Prior & & \\
& Prior mean & 0.5 \\
\bottomrule
\end{tabular}
\end{table}
We only extended the SVI iterations as we expected the problem to be more challenging and used 300 instead of 100 training samples. Additionally, we adjusted the hyper-prior for the Gaussian Markov prior's precision parameter $\delta$ so that the resulting samples have \emph{reasonable} coverage for the problem. (This procedure is computationally very cheap and just a heuristic in which we check that we do not generate any too-extreme permeability field samples, but have a good enough spread of different candidates to explore the LF-HF mapping efficiently.) One further difference from the first demonstration is the now necessary grid-to-grid interpolation for the permeability fields on the LF and HF models. We can safely assume that the interpolation has a negligible effect.


\subsubsection{Choice of LF model and solver, HF solver and relative solver costs}
\label{sec:poro_elastic_solver_and_costs}
The HF model discretizes the poro-elastic governing equations \eqref{eqn: strong_form_poroelasticity_a} to \eqref{eqn: poro_elastic_ics_pf} with \add{FEM} according to the formulation by Vuong et al. \cite{vuong2015general}. We solve the resulting nonlinear system of equations with the Newton-Raphson method and use one-step-$\theta$ time-stepping with $\theta=0.5$. We assemble the system stiffness matrix monolithically, including all coupling terms, and use a direct solver for the linear system of equations in each Newton step. The poro-elastic framework is realized and freely available in our multi-physics C++ code 4C \cite{4C}.

We prescribe the same fluid BCs for the LF model and use the identical computational domain: An inflow BC on the top boundary, no penetration BCs on the left and right sides, and a zero-pressure BC on the bottom boundary. The inflow velocity on the top boundary in the y-direction is fixed at a higher value of $u_{2}^{f}=-0.5$ compared to the HF model's maximum inflow of $u_{2}^{f}=-0.063$, to showcase that BMFIA gives reliable results despite quantitative differences in BCs. We emphasize that the robustness of BMFIA \wrt the quantitative BC values is an important feature. Especially for coupled problems, choosing quantitatively similar LF BCs does not necessarily correspond to overall similar model behavior.

The computational specifics of the LF and HF model discretizations are summarized in Table \ref{tab: model_fem_poro_elastic}, along with their average solution times\footnote{The HF and LF simulations were run on different CPUs. The AMD EPYC 9354 offers approximately 10\% higher single-core performance than the Intel i7-8700K \cite{cpu_perf}. These scores reflect empirical single-thread throughput. Table \ref{tab: model_fem_poro_elastic} shows the raw performance without a CPU performance normalization.}.
\begin{table}[htbp]
\centering
\caption{Discretizations of the employed LF (porous medium) and HF (poro-elastic medium) model in the BMFIA approach}
\label{tab: model_fem_poro_elastic}
\begin{tabular}{lcc} 
\toprule
\textbf{Settings} & \textbf{HF} (Transient poro-elastic & \textbf{LF} (steady-state\\
 & media flow) & porous media flow)\\
\midrule
Element type & quadrilateral & quadrilateral\\
Polynomial degree & 2 for all three fields & $\bu$: 2, $p$: 1\\
Number of elements & 6586 (structure) + 6586 (poro-fluid) & 1024\\
Number of DoFs & 13172 (structure) + 19758 (poro-fluid) & 9531\\
$\dim(\bx)$ & interpolated from LF & 1089\\
Number of CPU cores used & 16 & 1\\
CPU type & AMD EPYC 9354 32-Core Proc. & Intel Core i7-8700K CPU\\
 & @ 2.80GHz (boost up to 3.5GHz) & @ 3.70GHz \\
Approx. cumul. CPU time [s] & 6700 & 0.2 \\
Approx. wall time [s] & 419 & 0.2\\
\midrule
\textit{HF/LF cost ratio cumul. CPU time} & \multicolumn{2}{c}{$\boldsymbol{\sim 33500}$}\\
\textit{HF/LF cost ratio wall time} & \multicolumn{2}{c}{$\boldsymbol{\sim 2095}$}\\
\bottomrule
\end{tabular}
\end{table}
Due to the flexibility in the LF model design, we can drastically reduce the computational cost by a \textbf{speed-up factor of approximately 33500}. This allows us to run the LF model evaluations in the inference phase of the algorithm on a desktop computer or even a laptop, even if thousands of simulation runs are necessary. The inference phase would be infeasible for such complex and coupled HF simulations. 
\FloatBarrier
\subsubsection{Practical notes on approximating the MF conditional for more complex LF-HF relationships}
\label{sec:complex_mf}
\begin{figure}[htbp]
    \centering
    \begin{tikzpicture}
    \node[anchor=south west,inner sep=0] (background) at (0,0) {\includegraphics[scale=0.27]{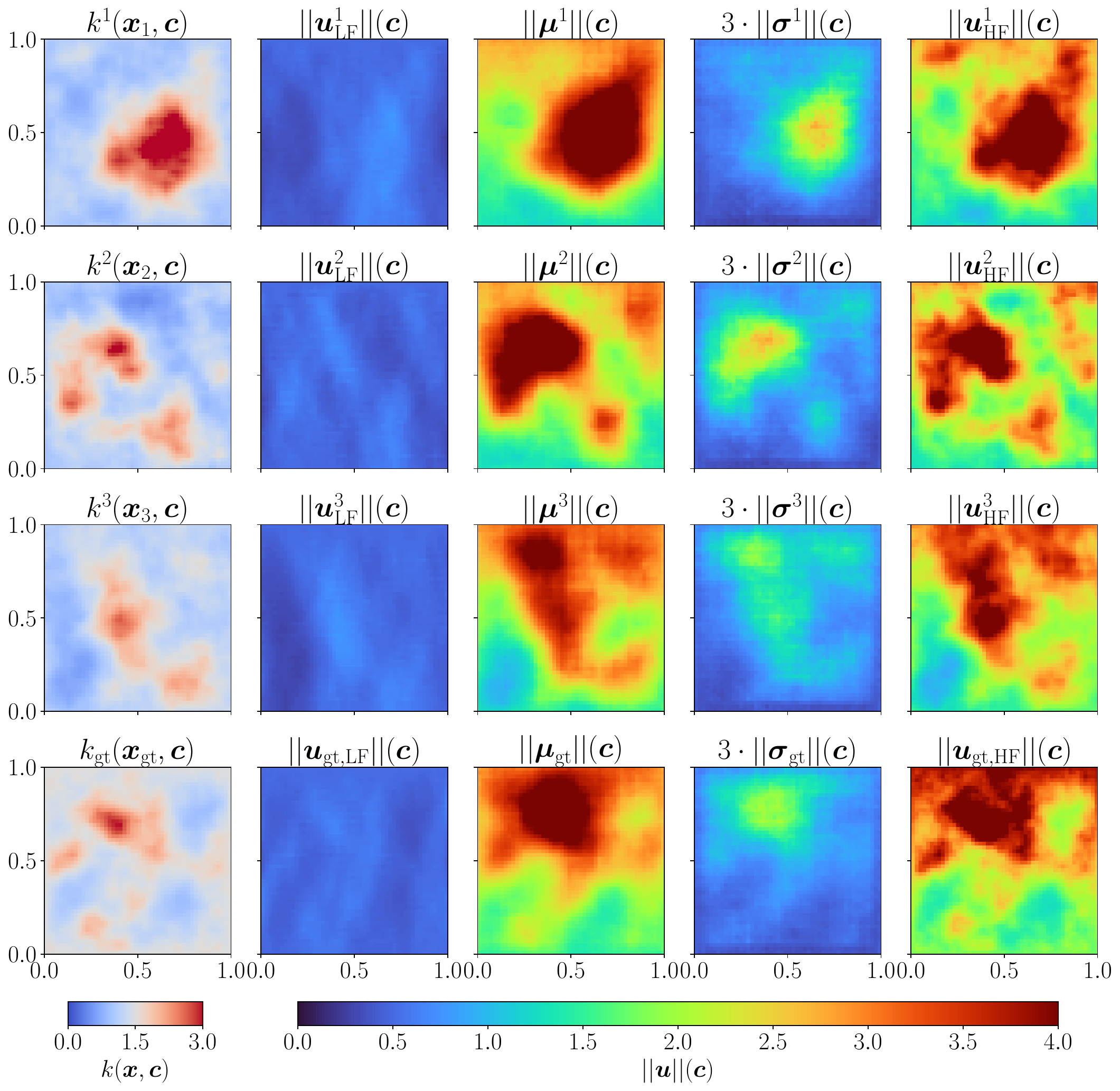}}; 
    \begin{scope}[x={(background.south east)},y={(background.north west)}]
        \node at (0.11, 1.02) {inputs};    
        \node at (0.31, 1.02) {LF-out};    
        \node at (0.6, 1.02) {Pred. MF conditional};    
        \node at (0.9, 1.02) {HF-out};    
        \node at (-0.1, 0.87) {$\shortstack{\text{input}\\ \text{sample 1}}$};    
        \node at (-0.1, 0.65) {$\shortstack{\text{input}\\ \text{sample 2}}$};    
        \node at (-0.1, 0.43) {$\shortstack{\text{input}\\ \text{sample 3}}$};    
        \node at (-0.1, 0.21) {$\shortstack{\text{ground-truth}\\ \text{sample}}$};  
    \end{scope}
    \end{tikzpicture}    
    \caption{Comparison of the velocity fields for the HF (poro-elastic medium) and LF (porous medium) model, as well as the prediction of the \add{MF} conditional model, given different permeability fields as inputs (row-wise). \textbf{First, left column (inputs)}: Permeability field samples (not part of training data) that serve as input for the different models. \textbf{Second column (LF-out)}: Resulting velocity fields of the porous media model (LF). \textbf{Third and fourth columns (Pred. MF conditional)}: Mean and standard deviation predictions for the HF velocity fields using the Gaussian convolutional autoencoder. \textbf{Last, right column (HF-out)}: Resulting velocity fields of the poro-elastic model (HF).}
    \label{fig: poro_elastic_LF_HF_comp}
\end{figure}

Figure \ref{fig: poro_elastic_LF_HF_comp} demonstrates the vast deviation in the LF and HF flow fields for four random testing samples and the actual ground-truth for the permeability field (not part of the training), along with the mean prediction and three times the standard deviation prediction of the Gaussian convolution autoencoder model. The settings \add{for training and the BMFIA procedure} are specified in Table \ref{tab:bmfia_settings} and \ref{tab:deviating_bmfia_settings}. \add{The training time for the convolutional autoencoder (here for a geometrical 2D domain) was in all investigated cases around four to six minutes on a single \emph{NVIDIA GeForce GTX 1080 Ti} GPU, and therefore neglebile compared to the HF model evaluation costs.} the fine-grained and localized differences in the flow fields in the HF model due to the additional nonlinear elastic coupling, which are not present in the LF model. 

\add{The} higher \add{conditional} variance within the \add{MF} conditional exacerbates instabilities and noise in the gradients of the \add{MF} likelihood \wrt $\bx$. While advanced machine learning architectures may initially improve point-wise predictive accuracy, they frequently lead to irregular or noisy gradients of the \add{MF} posterior likelihood \wrt $\bx$. Despite rigorous regularization, such gradient behaviors can severely complicate Bayesian inference procedures, particularly in the \emph{small-data regime}. \Eg a u-net \cite{ronneberger2015u} variant of the probabilistic autoencoder led to a more confident and more accurate mean point prediction, but unfortunately, destabilized the subsequent SVI procedure, leading to diverging BMFIA posterior predictions. We made similar findings for other initially promising architectures such as Fourier Neural Operators (FNO) \cite{li2020fourier} or deep operator networks (deep-o-nets) \cite{lu2021learning}, at least in their simpler variants. A more detailed investigation is outside the scope of this paper. Still, it should be a focus for future research on this topic, primarily as many different model versions, architectures, loss functions, and regularization measures exist that can handle the complex gradient landscapes more effectively. It would be exciting to investigate if this holds for second-order schemes, which are not yet common in the machine learning community, as recent studies \cite{kiyani2025optimizer} showed promising results under similar conditions. 

Finally, we address the \add{naive} attempt to directly learn a mapping from the HF model's input $\bx$ to the output $\Yhf$ in the sense of a classical surrogate. At first glance, this relationship might seem reasonable when looking at Figure \ref{fig: poro_elastic_LF_HF_comp}. However, in a \emph{small data regime}, the complexity of this mapping is too high and almost certainly leads to worse prediction results, even when treated probabilistically. In Appendix \ref{sec:counter_intuitive_mf}, we demonstrate the inferior prediction quality of a direct surrogate attempt for the interested reader. We can only make robust predictions in a small data regime because of the \add{MF} embedding. This is a powerful argument for the proposed BMFIA approach.

\FloatBarrier
\subsubsection{BMFIA posterior distribution for the poro-elastic problem}
\label{sec:poro_elastic_posterior}
\begin{figure}[htbp]
    \centering
    \begin{tikzpicture}
    \node[anchor=south west,inner sep=0] (background) at (0,0) {\includegraphics[scale=0.42]{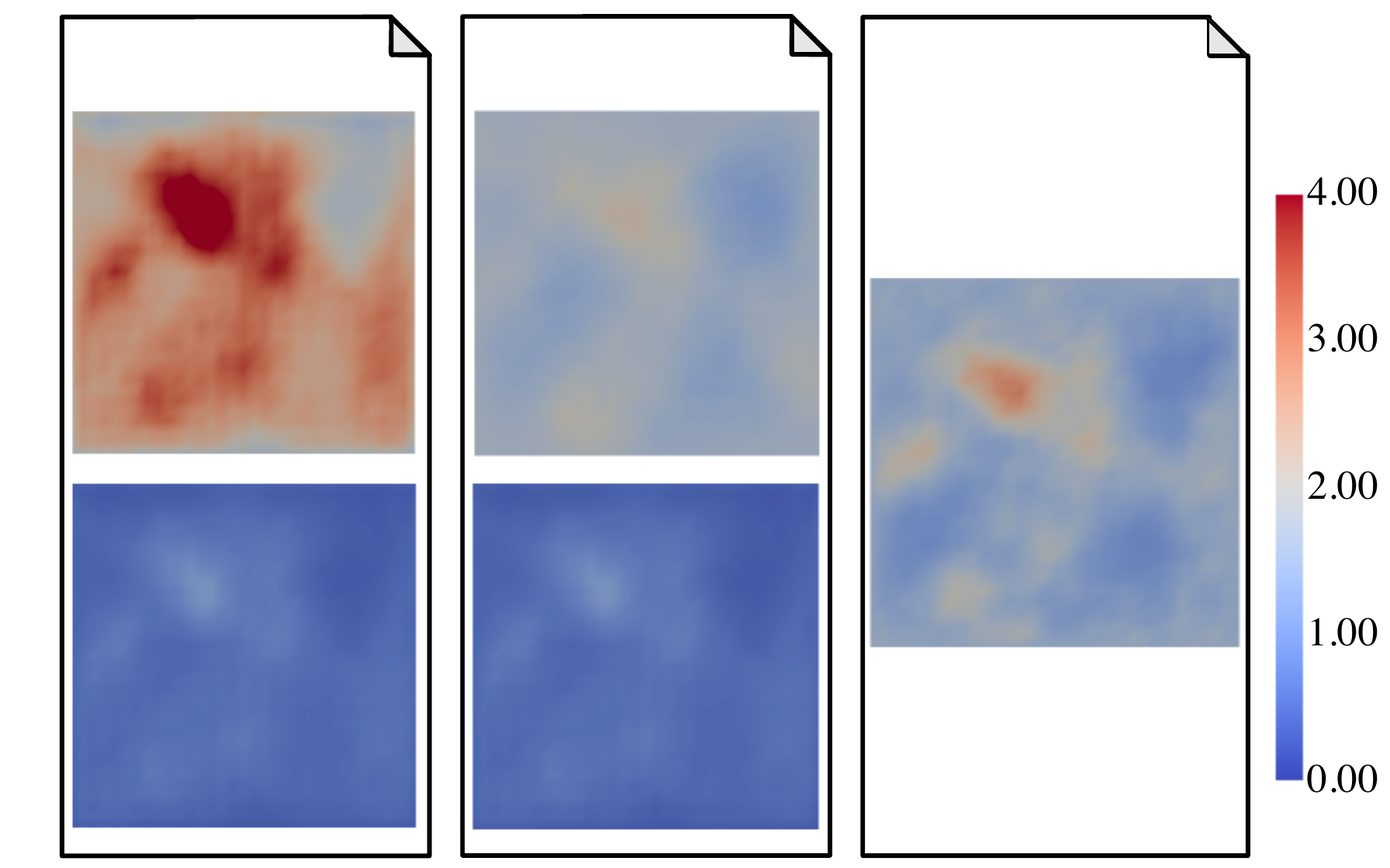}}; 
    \begin{scope}[x={(background.south east)},y={(background.north west)}]    
        \node at (0.17, 0.91) {LF};   
        \node at (0.46, 0.91) {BMFIA};   
        \node at (0.75, 0.91) {HF - gt.};   
        \node at (0.93, 0.83) {$k$};   
        \node[rotate=90] at (0.02, 0.69) {mean};   
        \node[rotate=90] at (0.02, 0.23) {$2\cdot\mathbb{STD}$};   
        \end{scope}
    \end{tikzpicture}
    \caption{Comparison of posterior distributions (SNR of 50) for the isotropic permeability field $k(\bx,\bc)$. The posteriors' mean functions (upper row) and $2\cdot \mathbb{STD}$ (bottom row) are depicted. \textbf{Left block:} Posterior distribution using only the porous medium LF model. \textbf{Middle block:} BMFIA posterior that approximates the HF posterior (here unavailable) using $\ntrain=300$. \textbf{Right block:} HF ground-truth field $\txgt(\bxgt,\bc)$ (normally unavailable). For a comparison of the computational costs, see Table \ref{tab: posterior_costs_ex2}.}
    \label{fig: poro_elastic_posterior}
\end{figure}

\begin{figure}[htbp]
    \centering
    \begin{tikzpicture}
    \node[anchor=south west,inner sep=0] (background) at (0,0) {\includegraphics[scale=0.23]{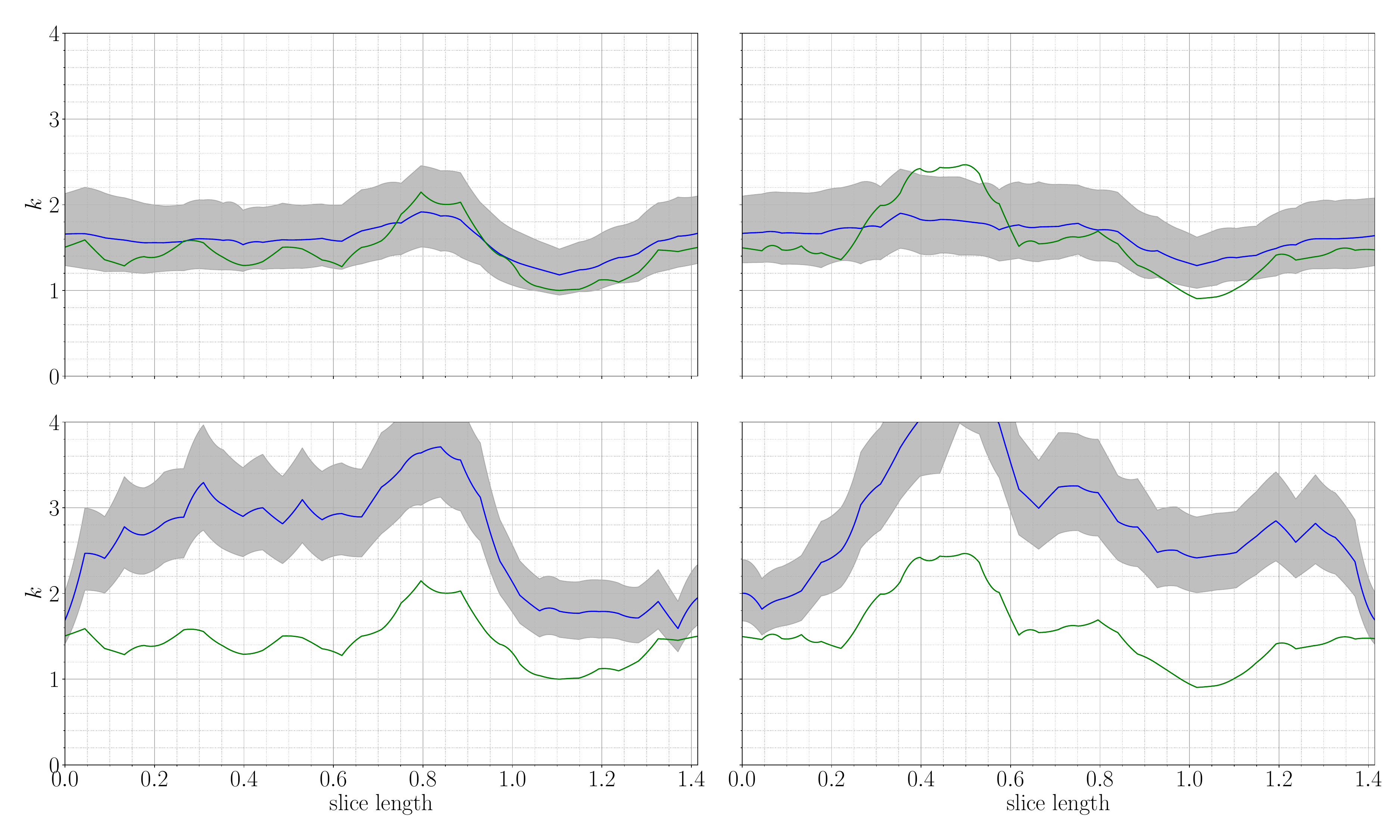}}; 
    \begin{scope}[x={(background.south east)},y={(background.north west)}]    
        \node at (0.28, 1.0) {(bottom-left) $\rightarrow$ (top-right)};    
        \node at (0.75, 1.0) {(top-left) $\rightarrow$ (bottom-right)};    
        \node[rotate=90] at (-0.01, 0.74) {BMFIA};    
        \node[rotate=90] at (-0.01, 0.29) {LF};    
    \end{scope}
    \end{tikzpicture}
    \caption{Diagonal slices through the posterior permeability field of Figure \ref{fig: poro_elastic_posterior} with the ground-truth permeability (\usebox{\darkgreenline}), posteriors' median or 50\% percentile (\usebox{\blueline}), and $90\%$ credible bounds (area between the 5\% and 95\% percentile) (\usebox{\lightgreybox}). \textbf{Top row:} BMFIA posterior. \textbf{Bottom row:} Posterior slices when directly using the LF model. \textbf{Left column:} Slice from bottom left to top right of $\Omega$. \textbf{Right column:} Slice from top left to bottom right of $\Omega$.}
    \label{fig: poro_elastic_posterior_slices}
\end{figure}

We now assess the BMFIA posterior distribution for the permeability field of the poro-elastic problem. To generate the BMFIA posterior, we used the settings listed in Table \ref{tab:bmfia_settings} and \ref{tab:deviating_bmfia_settings}. Note that the HF posterior itself is prohibitive for this demonstration due to the absence of HF model gradients in combination with the high stochastic dimension. Hence, we only inspect how well the BMFIA posterior $\pmf(\bx|\Yobs)$ encapsulates the ground-truth field $\txgt(\bxgt,\bc)$ \add{(Note that in contrast, in the former simpler the HF posterior could be calculated as its adjoint was available.)}. Figure \ref{fig: poro_elastic_posterior} compares the LF posterior, the BMFIA posterior, and the (usually unknown) HF ground-truth (from left to right). By LF posterior, we mean the posterior distribution we get when we naively use the porous media model (LF model) and the poro-elastic observations (no HF simulations). Obviously, the results are unsatisfactory and do not encapsulate the ground-truth. The model error of the LF model is too large to get any meaningful posterior distribution. On the contrary, when we use the LF model inside BMFIA, the \add{MF} posterior restores the detailed localized features of the HF ground-truth and almost perfectly encapsulates the latter. As expected, we get a bit more conditional variance in the posterior due to the noisier \add{Mf} relationship. This is desired as we prefer uncertainty over confident but biased posteriors.
For a better quantitative comparison, Figure \ref{fig: poro_elastic_posterior_slices} shows the LF and BMFIA posterior and the ground-truth in diagonal slices through the domain. In this visualization, the high quality of the BMFIA posterior is even more recognizable.
\FloatBarrier

Even for the very simplified LF model, which completely neglects the nonlinear elastic coupling, we can achieve high-quality posterior approximations using BMFIA at considerably lower computational costs. This last, successful demonstration for a large-scale nonlinear coupled poro-elastic model paves the way for further deployment of BMFIA in the context of challenging \add{BIPs} that involve coupled nonlinear physics and can not be solved \add{yet} with state-of-the-art methods.

Eventually, Table~\ref{tab: posterior_costs_ex2} compares the wall times for the LF, BMFIA, and hypothetical HF (assuming HF model gradients are available for the SVI routine, which is usually not the case) posteriors under hardware budgets of 32, 64, 128, and 256 available CPU cores. For BMFIA, we conducted a training phase of 300 LF and 300 HF simulations according to Section \ref{sec: setup_poro_elastic}. LF simulations ran on a single core and took 0.2 seconds each, while each HF simulation ran on 16 cores for a single simulation\footnote{For this comparison, we kept the parallelization of one HF evaluation at 16 cores for all CPU scenarios considered.} using 419 seconds of wall time. The training phase is embarrassingly parallel, and its duration depends on the hardware constraints.
Inference was performed using a batch-sequential SVI procedure, with six forward simulations and six adjoint runs per batch. This resulted in 1667 forward and 1667 adjoint batch-sequential solves in the SVI procedure (equivalent to 10000 forward simulation calls). Since inference is batch-sequential, parallelization can only be conducted within the batch and according to the available hardware constraints.

In the case of, \eg 64 available CPU cores, this leads to $2\cdot1667=3334$ sequential steps with 0.29 seconds run time each, resulting in 967 seconds wall time for the LF posterior. Note that for each batch-sequential step, only 6 cores were exhausted for the LF runs. For the HF model, we also have 3334 batch-sequential steps, but require 16 cores for one simulation run. Table~\ref{tab: posterior_costs_ex2} summarizes the wall time costs for the considered computational resources.
BMFIA achieves a substantial reduction in wall time compared to a hypothetical HF inference, especially if the available computational resources allow for exploitation of the embarrassingly parallel training phase. While the theoretical computational cost savings are already a significant advantage of BMFIA, the more important benefit pertains to sole reliance on LF adjoints or LF model gradients in the inference phase. This characteristic allows us to solve complex, nonlinear, coupled physics-based inference problems, which were inaccessible before.

\begin{table}[htbp]
\centering
\caption{Computational costs for obtaining the different posterior distributions, shown in Figure \ref{fig: poro_elastic_posterior} and \ref{fig: poro_elastic_posterior_slices}. The costs are given as the total wall time per posterior [s].}
\label{tab: posterior_costs_ex2}
\begin{tabular}{ccccc } 
\toprule
\textbf{Num. avail. CPUs} & \textbf{LF posterior} [s] & \textbf{BMFIA posterior} [s] & \textbf{HF posterior} [s] & \textbf{HF/BMFIA cost ratio}\\
\midrule
32& 967 & 63818.4& 4190838& $65.6\times$\\
64& 967 & 32393.4& 2095419 & $64.7\times$\\
128 & 967 & 16680.9& 1396946 & $83.7\times$\\
256 & 967 & 8824.7& 1396946 & $ 158.3\times$\\
\bottomrule
\end{tabular}
\end{table}

\FloatBarrier
\section{Conclusion and outlook}
\label{sec:conclusion_outlook}
We proposed a new inference method called \emph{Bayesian multi-fidelity inverse analysis (BMFIA)} to efficiently solve high-dimensional \add{BIPs} in combination with computationally expensive, non-differentiable nonlinear multi-physics models. Non-differentiable (legacy) codes are widespread in engineering, biomechanics, or applied physics. These codes often implement complex, large-scale, nonlinear coupled physical systems that have organically evolved and matured over extensive temporal spans. Solving \add{BIPs} with such code frameworks involves two significant challenges that we address with \emph{BMFIA}. First, an HF simulation run is computationally expensive, which limits the feasible number of likelihood evaluations. Second, the solution of a \emph{high-dimensional} (spatial) \add{BIP} relies on model gradients \wrt the input parameters $\bx$ to approximate the posterior in a reasonable time. Such gradients are generally not provided by legacy codes. Mostly, retrospective automated differentiation is not an option due to the complexity of existing codes and the infeasible amount of necessary code restructuring. Furthermore, adjoint-based model gradients \wrt input parameters are usually nonexistent or non-trivial when implementing coupled physics problems. Finite differences or surrogate-based approaches underlie the curse of dimensionality and are hence also prohibitive.

Instead, we learn a stochastic relationship, the so-called \emph{\add{MF} conditional density} between the HF model's (legacy code) output and simpler, computationally cheaper LF models in a \emph{small data regime}, \ie 100 to 300 HF and LF simulations. Afterwards, the \add{MF} conditional can be integrated into the likelihood formulation to yield a new, \add{MF} likelihood, which shifts the entire iterative and typically expensive solution process to the cheaper and simplified LF model, which we design and select, such that adjoint-based gradients can be derived. We demonstrated in Section \ref{sec:demonstration} that despite very inaccurate LF models, BMFIA reconstructs accurate approximations to the HF posterior using only very few HF simulations. We highlight the results for a spatial \add{BIP} of a coupled nonlinear poro-elastic model in Section \ref{sec: poro_elastic}: we showcased that BMFIA gives a reliable and accurate posterior approximation even when only a very crude single physics porous media model is used. The selected LF model solves 2095 times faster (wall time) than the coupled HF model and exhibits adjoints. 

BMFIA comprises the following key advantages: As BMFIA \textbf{reallocates most of the computational work to the cheaper LF model}, it is much \textbf{more efficient} than direct sampling on the expensive HF model (which is mainly prohibitive). Furthermore, the initial training phase of the stochastic LF-HF relationship (\add{MF} conditional) and its potential refinement give \textbf{control over the number of expensive HF simulations} as demonstrated in Section \ref{sec:demonstration}, BMFIA provides an \textbf{accurate approximation of the HF posterior} (usually unknown) despite the inaccurate and cheap LF models deployed (see Section \ref{sec:demonstration}). As we can now exploit gradients from simplified, decoupled models, BMFIA  \textbf{expands solvability frontiers for high-dimensional inverse problems in combination with costly legacy codes}. The \textbf{design of cheaper and simplified LF models can be conducted flexibly}, as the HF and the LF models only need to share a statistical dependence, which is only a weak requirement in practice. Additionally, coarser numerical discretization improves the computational speed-up further.

\add{Stabilizing} the training of the \add{MF} conditional model demands further investigation, as already pointed out in Section \ref{sec:small_data_approx}. Solutions include second-order optimization schemes to mitigate issues related to rugged optimization landscapes and poor gradient behavior, particularly in small data regimes. \add{It would be interesting to investigate an adaptive smoothing procedure for the input field within the inference procedure (a smoothing filter), to slowly increase the effective dimension of the field. This could accelerate variational posterior convergence and improve spurious local optima and gradient behavior in early iterations.}
Furthermore, other modern probabilistic machine learning architectures, especially in operator learning models, are a promising direction. We already conducted preliminary studies using Fourier neural operators (FNOs) and deep-operator networks (deep-o-nets). These were hampered by a worse gradient behavior \wrt differentiation in the \add{MF} likelihood. Therefore, we believe such developments must be conducted simultaneously with a more robust optimization and regularization.  
Moreover, incorporating physical domain knowledge explicitly into the conditional model could improve robustness beyond the current purely data-driven formulation. One promising idea is merging the \add{LF} model with the approximation of the \add{MF} conditional model into a single, physics-informed, probabilistic machine-learning model, eliminating the explicit construction of adjoint models. \add{Another very interesting further research direction is the extension of BMFIA to handle heterogeneous multi-physics data in $\Yobs$, which can significantly improve the information content and quality of the posterior: While we already demonstrated a multi-physics model in the second numerical demonstration, the observed data $\Yobs$ only encompassed one physical field (flow field) and the LF model reflected only the single-physics porous media flow, allowing us to derive an adjoint formulation for the likelihood gradients. However, in future investigations, one-way coupled LF models could be a natural extension to reflect a multi-physics character in an approximate manner, while still exhibiting adjoint-based gradients. Other ideas involve using several single-physics LFs, their model outputs stacked in $\Zlf$, which could be loosely coupled, \eg via BCs and hence still provide model gradients. Further comparisons with other state-of-the-art multi-fidelity methods, such as MLMC approaches, would also be promising, especially for challenging multi-physics problems.}
\add{A sensitivity study for the quality of the MF posterior \wrt BMFIA settings, parameterizations, and different LF model \emph{quality} could also be helpful to quantify the influence of the BMFIA setup.}
From an application perspective, we will extend and apply the proposed framework to realistic, clinically relevant, and three-dimensional inverse problems in biomechanics.

\section*{Acknowledgements}
The authors gratefully acknowledge financial support from the Deutsche Forschungsgemeinschaft (DFG, German Research Foundation) in the project WA1521/26-1, by BREATHE, a Horizon—ERC–2020–ADG project (grant agreement No. 101021526-BREATHE). Furthermore, we thank our colleagues Maximilian Bergbauer and Sebastian Pröll for their helpful support with the deal.II library and Gil Robalo Rei and Atul Agrawal for insightful discussions on probabilistic modeling and Bayesian inference.

\printbibliography

\appendix

\section{Physical analogy for BMFIA and the MF conditional}
\label{sec:physical_analogy}
\add{
Consider a controlled optical experiment designed to study an object’s spatially varying emission field $\bx$ (\eg a high-dimensional light intensity distribution, structured illumination pattern, or spatially varying material reflectivity).
The experiment can be performed using two optical setups:
\begin{enumerate}
\item \textbf{HF setup} - An expensive, high-precision optical system producing highly accurate images $\Yhf$ for a given object configuration or emission field $\bx$.
\item \textbf{LF setup} - A cheaper, faster optical system that produces nonlinearly blurred and distorted $\Ylf$ images. 
\end{enumerate}}

\add{
We can control $\bx$ (the object's emission field) in the laboratory, systematically varying the emission or geometry of the test object over 200-300 distinct configurations. For each configuration, we record both HF and LF images. From this paired dataset, we manufacture an optical \emph{correction plate} (\eg a crystal-based diffractive element) that physically implements the multi-fidelity conditional mapping in an analog manner: given an LF observation, it produces not a single corrected image but a diffraction spectrum of possible HF images that could plausibly correspond to it. This spectrum embodies the probabilistic mapping $p(\Yhf|\Ylf)$, including nonlinear shifts and deformations, rather than a simple Gaussian blurring.\\
In a real investigation outside the laboratory, we observe a noisy HF measurement $\Yobs=\Yhf+\mathrm{noise}$ for an unknown high-dimensional emission field $\bx$ which we seek to reconstruct. We then return to the laboratory and use the cheaper LF optical setup in combination with our manufactured correction plate to efficiently evaluate many candidate configurations 
$\bx$. In the BMFIA framework, these corrected LF outputs are incorporated into a Bayesian inversion loop, where stochastic sampling, potentially augmented with gradient information from the LF setup (assuming the LF simplicity allows for direct analog gradient measurements), iteratively searches for the most probable $\bx$ samples (in other words the posterior $p(\bx|\Yobs)$) given the noisy observation $\Yobs$.\\
The correction plate ensures that the LF images are adjusted in a probabilistic manner, yielding an HF-like distribution rather than a single overconfident estimate. 
While the resulting MF posterior is blurrier than a hypothetical HF posterior obtained using the HF system exclusively (not feasible due to time and cost constraints and absence of model gradients), it remains informative and crucially avoids overconfident errors. 
Given budgetary and practical constraints, this probabilistically corrected LF-based reconstruction represents the best achievable estimate of the posterior distribution for the emission field $\bx$.}

\section{Details on the probabilistic convolutional autoencoder}
\label{sec:details_autoencoder}
\add{
We use the following network architecture for the probabilistic convolutional autoencoder: 
Starting at the encoder, we first conduct a convolution with a window size of three and expand the feature space to 16 dimensions while keeping the spatial resolution of the input. While we demonstrate problems in 2D Euclidean space in this manuscript, we note that a natural extension is arbitrary 2D and 2D geometries. For simple hypercubic domains, the present convolutional autoencoder structure generalizes directly to 3D geometries. However, for arbitrary or more complex geometries, one would require more flexible architectures such as graph-based neural networks or operator-learning approaches, which we see as promising directions for extending BMFIA in future work. Afterwards, we push the extended feature space through an \emph{Exponential-Linear-Unit} (ELU) activation function and conduct a max-pooling operation with a window size of two, halving the size of the hyper-cube. We then repeat the convolution step and double the feature dimension to 32 with a subsequent ELU activation function. Another max-pooling layer halves the hyper-cube size again, finally leading to a last convolution stage with 64 feature dimensions and ELU activation. We then flatten the features to one vector and reduce the size to a bottleneck of dimension 200 with a fully connected neural network. The decoder then mirrors the encoder architecture inversely up to the last layer. This probabilistic layer maps to $2\cdot d$ output channels corresponding to the parameterization of a mean vector and standard deviation vector for a $d$-dimensional mean-field Gaussian distribution (per pixel). The ELU activation function promotes smoothness. Further regularization measures include standardizing all training data $\mathcal{D}$, a dropout rate of 30\% in the dense layers, and batch normalization before all activation functions. The max-pooling layers can be exchanged for average-pooling if more smoothness is required. In the probabilistic layer, we conduct an exponential reparameterization of the variance parameter and add a small nugget variance of $10^{-5}$ for stability. For vector-valued predictions, we only parameterize the diagonal of the covariance matrix per pixel, as we found that the full covariance leads to instability in the small data regime. The negative log-likelihood function of the output Gaussian serves as a natural loss function for training the probabilistic convolutional autoencoder. We achieved good results using either the simple stochastic gradient descent (SGD) approach, which often generalized better \cite{wilson2017marginal, keskar2017improving}, or the Adam optimizer \cite{kingma2014adam}.
}

\section{Variational Bayes Expectation Maximization for the MF Likelihood}
\label{sec:vb_em_tau}

We start by considering the MF log-likelihood initially dependent on the unknown precision parameter $\tau$. Subsequently, we use the following expression interchangeably: $\lmf(\bx)=\log \pmf(\Yobs|\bx)$, as well as $\lmf(\bx,\tau)=\log \pmf(\Yobs|\bx,\tau)$. To account for uncertainty in $\tau$, we introduce a probabilistic model treating $\tau$ as a latent variable with a specified prior $p(\tau)$:
\begin{subequations}
\begin{align}
\lmf(\bx)&=\log \pmf(\Yobs|\bx) = \log \int \pmf(\Yobs|\bx, \tau)\cdot p(\tau) \dd \tau \\
&= \log \Ex{p(\tau)}{\pmf(\Yobs|\bx,\tau)} \\
&= \Ex{p(\tau|\Yobs,\bx)}{\log \left(\frac{\pmf(\Yobs|\bx,\tau)\cdot p(\tau)}{p(\tau|\Yobs,\bx)}\right)} \label{eqn:latent_mf_loglik}
\end{align}
\end{subequations}
Equation \eqref{eqn:latent_mf_loglik} can be derived using Jensen's inequality:
\begin{equation}
    \begin{array}{ll}
     \log \pmf(\Yobs|\bx) &= \log \int \pmf(\Yobs|\bx, \tau)\cdot p(\tau) \dd\tau\\
    & = \log \int \frac{ \pmf(\Yobs|\bx, \tau)\cdot p(\tau) } {q(\tau)} q(\tau)\dd\tau  \\
                         & \ge \Ex{q(\tau)}{\log \left(\frac{ \pmf(\Yobs|\bx, \tau)\cdot p(\tau)} {q(\tau)}\right)} \qquad \textrm{(Jensen's~ inequality)}
    \end{array}
\end{equation}
The optimal $q(\tau)$ is the posterior $p(\tau|\Yobs,\bx)=\frac{\pmf(\Yobs|\bx, \tau)\cdot p(\tau)}{\pmf(\Yobs|\bx)}$, which, upon substitution, converts the inequality to an equality.

We furthermore adopt an uninformative Gamma prior for $\tau$:
\begin{equation}
p(\tau)=\Gamma(\tau|a_0,b_0), \quad a_0=b_0=10^{-9},
\end{equation}
preferring small precision values and thus larger variance in the likelihood.

Since the Gaussian \add{MF} log-likelihood has $\tau$ as an additive term in its covariance matrix (see Equation \eqref{eqn:margMFGauss}):
\begin{equation*}
\lmf(\bx, \tau) = \log\pmf(\Yobs|\bx,\tau)=\log\left(\mathcal{N}\left(\Yobs \Big| M\left(\Zlf(\bx)\right), \underbrace{\tau^{-1}\cdot I + \Kmat\left(\Zlf(\bx)\right)}_{\Kmf(\Zlf(\bx),\tau)}\right)\right),
\end{equation*}
a conjugate prior and therefore an analytical posterior solution for $p(\tau|\Yobs,\bx)$ is unavailable. Instead of fixing or heuristically estimating $\tau$, approaches known to cause instability, we utilize the Variational Bayes Expectation-Maximization (VB-EM) \cite{dempster1977maximum, beal2003variational} framework. VB-EM iteratively computes the expectation with respect to the posterior $p(\tau|\Yobs,\bx)$, effectively capturing the average impact of $\tau$ on the MF log-likelihood. This probabilistic and iterative averaging leads to more robust inference outcomes.

Specifically, in the \textbf{E-step} of Algorithm~\ref{alg:vb_em_tau}, we approximate the posterior $p(\tau|\Yobs,\bx)$ variationally using a log-normal distribution:
\begin{equation}
q(\tau | \bphi_{\tau}) \approx p(\tau | \Yobs, \bx), \quad \tau = \exp(\tilde{\tau}), \quad q(\tilde{\tau} | \bphi_{\tau}) = \mathcal{N}(\tilde{\tau} | \mu_{\tau}(\bphi_{\tau}), \sigma_{\tau}^2(\bphi_{\tau})).
\end{equation}
We iteratively optimize the variational parameters $\bphi_{\tau}$ by (see Algorithm \ref{alg:vb_em_tau}):
\begin{enumerate}
\item Sampling $\tau_j \sim q(\tau | \bphi_{\tau})$ via the reparameterization trick.
\item Computing gradients of the MF log-likelihood $\lmf(\bx, \tau)$ and prior $p(\tau)$ with respect to $\tau$.
\item Applying the chain rule to derive gradients with respect to $\bphi_{\tau}$.
\item Updating $\bphi_{\tau}$ using stochastic optimization (e.g., ADAM optimizer).
\end{enumerate}

After convergence, we use the optimized variational posterior samples $\{\tau_j\}|_{j=1}^{\ntau} \sim q(\tau|\bphi_{\tau})$ in the \textbf{M-step} of Algorithm~\ref{alg:vb_em_tau}. Here, we marginalize over $\tau$ per sample $\bx_i$ to estimate the gradients of the MF log-likelihood with respect to the \add{LF} outputs $\Ylf$. Note that at this point we already have LF samples $\Zlf_i=\Zlf(\bx_i)$, so that we can drop the dependence on $\bx$ in the following (using Fisher's identity\cite{fisher1925theory} for the gradient\footnote{$\Ex{p(\bz|\bw)}{\nabla_{\bw}\log p(\bz|\bw)}=0$, the gradient (or \emph{score function}) of a log-density, when averaged over its own distribution, equals zero.}):
\begin{equation}
\nabla_{\Ylf}\lmf(\Zlf_i) \approx \frac{1}{n_{\tau}}\sum_{j=1}^{n_{\tau}} \nabla_{\Ylf}\lmf(\Zlf_i,\tau_j), \quad \{\tau_j\}|_{j=1}^{\ntau} \sim q(\tau | \bphi_{\tau}).
\end{equation}
Thus, the VB-EM algorithm effectively captures the probabilistic uncertainty in $\tau$ by averaging its influence on the MF log-likelihood. This method ensures stable inference and accurate parameter updates within our VI framework. Further methodological details can be found in Section~\ref{sec: algo_summary}.
\begin{algorithm}[htbp]
\SetAlgoLined
Approximate posterior of lik. precision $\tau$ with SVI: $q(\tau|\bphi_{\tau})\approx p(\tau|\Yobs,\bx_i)$ \textbf{[E-step]}

\While{not converged}{
    Sample $\ntau$ samples from variational distribution $q(\tau|\bphi_\tau)$ using the reparameterization trick:
    \[
    \{\tau_j\}_{j=1}^{\ntau} =\mu_\tau(\bphi_\tau) + \sigma_\tau(\bphi)\cdot \{r_{\tau_j}\}|_{i=j}^{\ntau} \gets \{r_{\tau,j}\}|_{i=j}^{\ntau}\sim \mathcal{N}\left(r|0,1\right)
    \]

    Plug $\tau$-samples into precision prior and MF log-likelihood:
    \[
        \{\nabla_\tau\lmf\left(\Zlf_i,\tau_j\right), \nabla_\tau p(\tau_j)\}|_{j=1}^{n_\tau}\gets \Zlf_i, \{\tau_j\}|_{j=1}^{\ntau}
    \]
    
    Apply chain rule form reparameterization to yield gradients \wrt $\bphi_\tau$:
    \[
    \{\nabla_{\bphi_\tau}\lmf\left(\Zlf_i,\tau_j\right), \nabla_{\bphi_\tau} p(\tau_j)\}|_{j=1}^{n_\tau}\gets \{\nabla_\tau\lmf\left(\Zlf_i,\tau_j\right) + \nabla_\tau p(\tau_j)\}|_{j=1}^{\ntau}
    \]

    Formalize and evaluate $\tau$-ELBO gradients:
    \[\nabla_{\bphi_\tau}\mathrm{ELBO}^{q(\tau|\bphi_{\tau})}\gets \{\nabla_\tau\lmf\left(\Yobs|\Zlf_i,\tau_j\right) + \nabla_\tau p(\tau_j)\}|_{j=1}^{n_\tau}\]

    Update variational parameters $\bphi_\tau$ (\eg SGD, ADAM):
    
    \[\bphi_{\tau}\gets \text{Stochastic optimizer step }\left(\bphi_\tau,\nabla_{\bphi_\tau}\mathrm{ELBO}^{q(\tau|\bphi_{\tau})}\right)\]
}
Sample from (converged) variational posterior approximation for precision $q(\tau|\bphi_\tau)$:
\[\{\tau_j\}_{j=1}^{n_\tau}\sim q(\tau|\bphi_{\tau})\]

Evaluate partial gradient of MF log-likelihood \wrt $\Ylf$, then marginalize $\tau$ \textbf{[M-step]}:
\[
\nabla_{\Ylf}\lmf(\Zlf_i)\gets \{\nabla_{\Ylf} \lmf(\Zlf_i,\tau_j)\}|_{j=1}^{n_\tau}\gets \Zlf_i, \{\tau_j\}|_{j=1}^{n_\tau}
\]
\Return $\nabla_{\Ylf}\lmf(\Zlf_i)$
\caption{VB-EM for MF log-likelihood precision hyper-parameter $\tau$, per sample $\bx_i$}
\label{alg:vb_em_tau}
\end{algorithm}

\FloatBarrier
\section{SVI convergence plots of BMFIA numerical demonstrations}
\label{sec:convergence_plots}
\begin{figure}[htbp]
    \centering
    \includegraphics[scale=0.35]{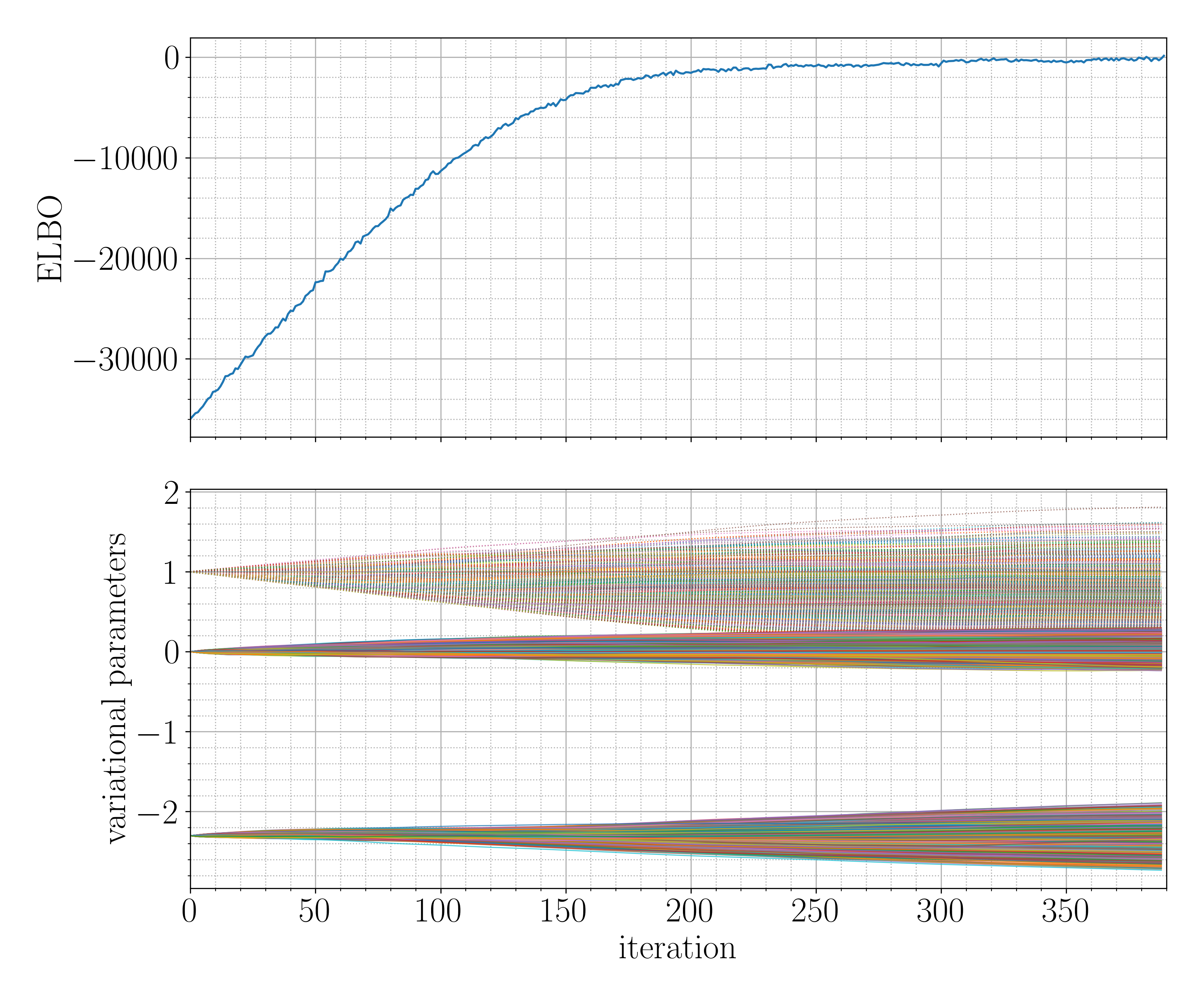}
    \caption{Convergence plots for the SVI procedure with VB-EM for BMFIA with the \lfm model in Section \ref{sec: darcy_flow} (case: $\ntrain=100$ with two times $\nrefine=10$). \textbf{Top:} Convergence of the evidence lower bound (ELBO) over SVI iterations. \textbf{Bottom:} Convergence of the individual variational parameters $\bphi$ over iterations. The bottom parameter cluster encodes the diagonal (variance) entries of the Cholesky factor, the middle cluster encodes the off-diagonal Cholesky entries, and the upper parameter cluster (dotted) encodes the mean values of the variational distribution.}
    \label{fig: svi_convergence_ex1}
\end{figure}

\begin{figure}[htbp]
    \centering
    \includegraphics[scale=0.35]{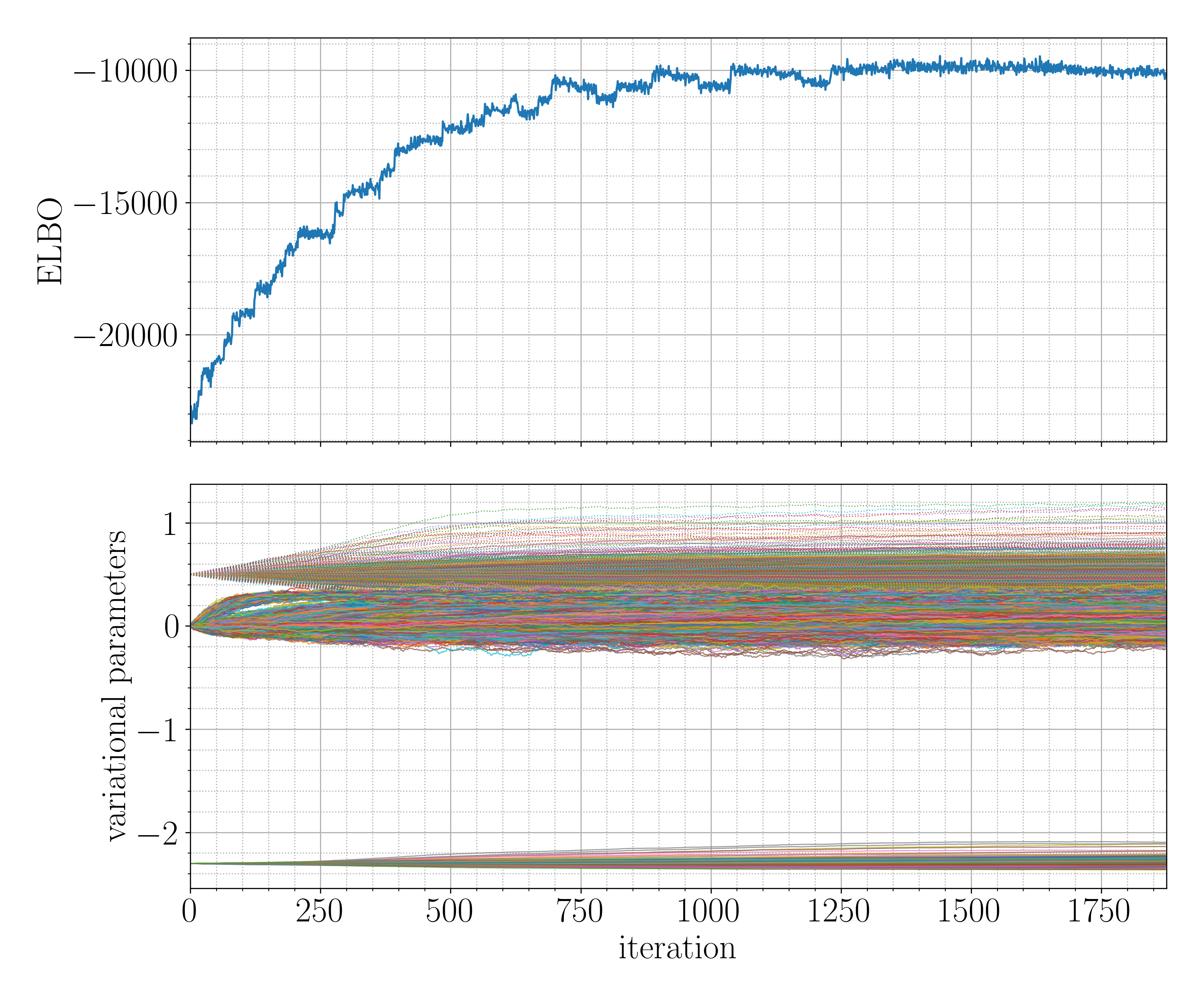}
    \caption{Convergence plots for the sparse SVI procedure with VB-EM for the poro-elastic reconstruction problem in Section \ref{sec: poro_elastic} using BMFIA up to the first 1900 iterations. \textbf{Top:} Convergence of the ELBO over SVI iterations. \textbf{Bottom:} Convergence of the individual variational parameters $\bphi$ over iterations. The bottom parameter cluster encodes the diagonal (variance) entries of the Cholesky factor, the middle cluster encodes the off-diagonal Cholesky entries, and the upper parameter cluster (dotted) encodes the mean values of the variational distribution.}
    \label{fig: svi_convergence_ex2}
\end{figure}

In Figure \ref{fig: svi_convergence_ex1}, we show the convergence of the employed VB-EM inference in BMFIA with two refinements for the first numerical demonstration (Section \ref{sec: darcy_flow}).
Furthermore, Figure \ref{fig: svi_convergence_ex2} shows the convergence of the \add{SVI} for the poro-elastic demonstration in Section \ref{sec: poro_elastic}, where we used BMFIA to solve a spatial reconstruction problem for a coupled poro-elastic medium model.
\FloatBarrier

\section{Convergence study over amount of training data for porous media example}
\label{sec:convergence_study}
\begin{figure}[htbp]
    \centering
    \includegraphics[scale=0.25]{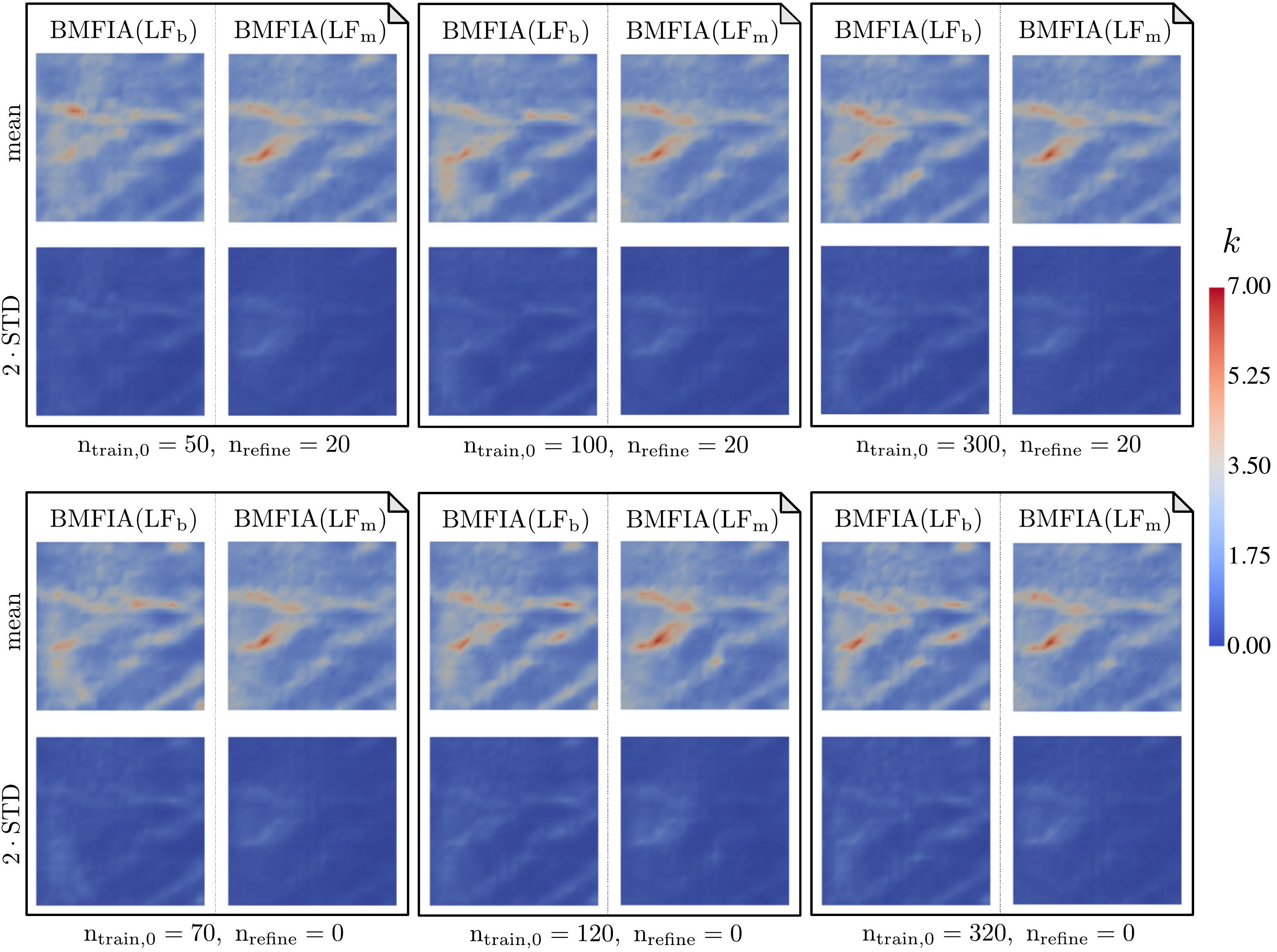}
    \caption{Convergence of the BMFIA posterior using a different number of training data $\ntrain$ and refinement data $\nrefine$. \textbf{Top block:} BMFIA posterior for an initial training amount of $\ntrain\in\{50,100,300\}$ with subsequent two times ten refinement samples ($\nrefine=20$) after 100 and 300 SVI epochs. \textbf{Bottom block:} Equivalent total training data without subsequent refinement.}
    \label{fig: bmfia_posterior_training_points}
\end{figure}

We additionally conducted a convergence study over the training data and refinement amount $\nrefine$ for the first numerical demonstration from Section \ref{sec: darcy_flow}, which uses the single-physics porous medium model. In Figure \ref{fig: bmfia_posterior_training_points}, we show the resulting BMFIA posteriors for $\ntrain\in\{50,100,300\}$ with two times refinement of 10 samples each ($\nrefine=20$) after 100 and 300 SVI epochs in the top block. Without further refinement, the bottom block shows the equivalent total training amount of $\ntrain\in\{70,120,320\}$. All BMFIA posterior distributions result in very accurate posteriors with only slight differences. 
This is especially remarkable for the very low number of HF data used in the cases of $\ntrain=50,\ \nrefine=20$, respectively $\ntrain=70$. Upon closer examination, we summarize the following three points:

First and most apparent, increasing the total amount of training data generally leads to better BMFIA posteriors. However, saturation effects are observed rather quickly. In this study, the case of $\ntrain=100,\ \nrefine=20$ offers the best accuracy-to-cost ratio compared to the HF posterior in Figure \ref{fig: bmfia_posterior_fields}. 
Second, increasing the training data has a more pronounced effect using the bad \lfb model. Initial artifacts for $\ntrain=70$, respectively, $\ntrain=120$ are mostly resolved for $\ntrain=320$. Hence, the robust performance of the \lfb, even for a minimal amount of 70 HF simulations, plus the convergence behavior for more training data, makes a strong argument for the proposed BMFIA method.
Third, we compare the two training strategies, refinement versus no refinement. We observe slightly better results for the refinement strategy. However, for the very small initial training size of $\ntrain=50$ and refinement size of $\nrefine=20$, the initial training amount cannot fully explore the \add{MF} conditional adequately, so the refinement does not bring any benefit. For all other cases, refinement brings a minor improvement, but the BMFIA method provides stable results even without it.
Last, our current choice for a probabilistic machine learning model, namely, a probabilistic version of a convolutional autoencoder (see Figure \ref{fig:cae} or Equation \eqref{eqn:margMFGauss}), reflects the epistemic uncertainty only to a limited extent. Hence, we do not see an increase in posterior uncertainty for a smaller number of training data but rather a disorientation of the posterior itself. This is not yet optimal, but can be mitigated by a more elaborate hyper-prior model and locally adaptable precision scaling in the Gaussian Markov prior model. We leave this finding for future research.

\FloatBarrier

\section{Examples of inferior BMFIA posterior quality}
\label{sec:counter_intuitive_mf}

As discussed in Section~\ref{sec: svi} and illustrated in the poro-elastic case study (Section~\ref{sec:poro_elastic_posterior}), the quality of BMFIA strongly depends on appropriate model choices.
In this section, we present two examples leading to inferior posterior quality: First, reducing regularization increases variance and destabilizes the posterior. Second, neglecting the LF model output and learning a direct mapping from $\bx$ to $\Yhf$ removes important guidance. Although this setup maintains the probabilistic BMFIA formulation, the posterior becomes diffuse \add{and underestimates the posterior uncertainty}.

\begin{figure}[htbp]
    \centering
    \begin{tikzpicture}
    \node[anchor=south west,inner sep=0] (background) at (0,0) {\includegraphics[scale=0.42]{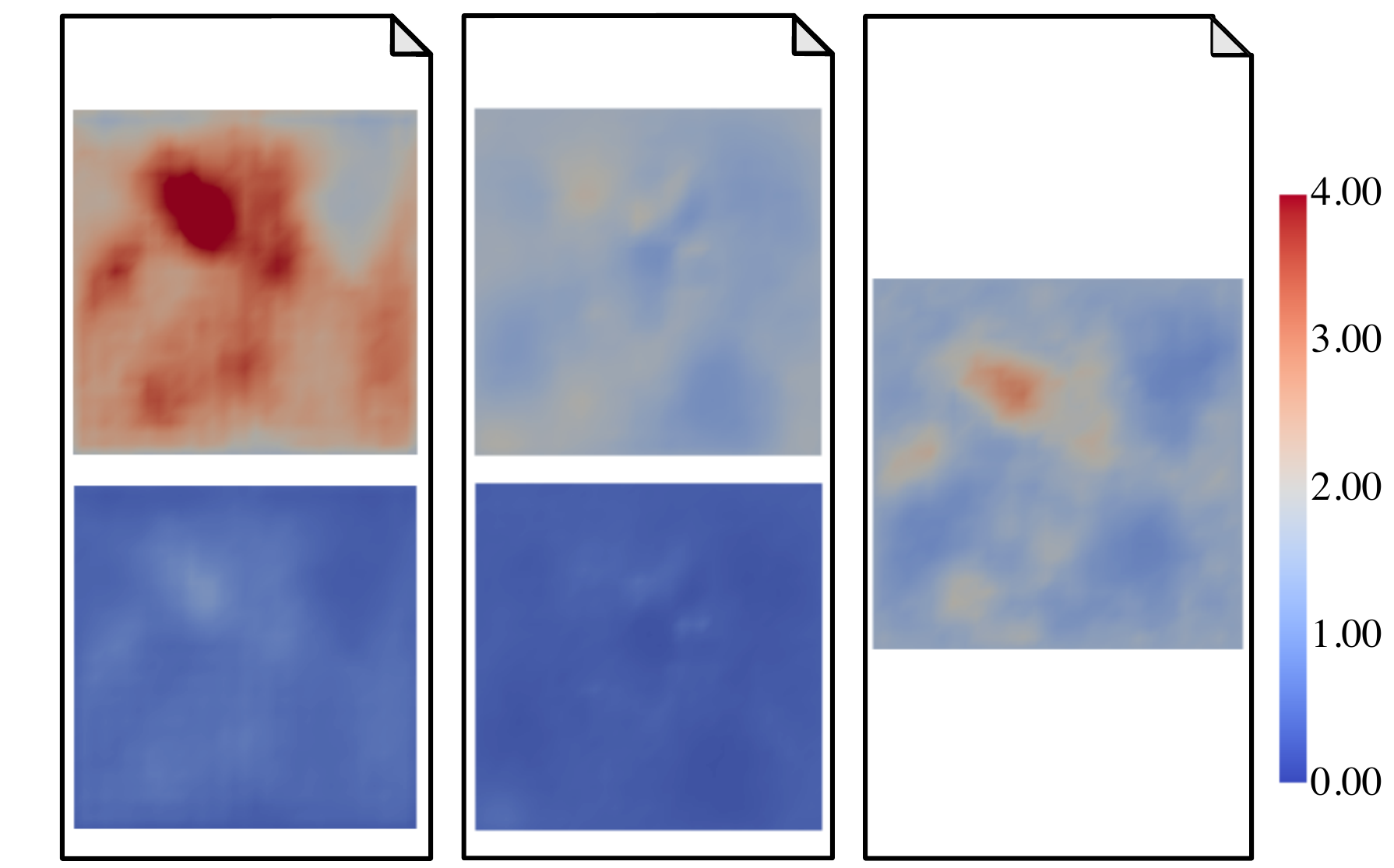}}; 
    \begin{scope}[x={(background.south east)},y={(background.north west)}]    
        \node at (0.17, 0.91) {LF};   
        \node at (0.46, 0.91) {BMFIA};   
        \node at (0.75, 0.91) {HF - gt.};   
        \node at (0.93, 0.83) {$k$};   
        \node[rotate=90] at (0.02, 0.69) {mean};   
        \node[rotate=90] at (0.02, 0.23) {$2\cdot\mathbb{STD}$};   
        \end{scope}
    \end{tikzpicture}
    \caption{\textbf{Counter example to Figure \ref{fig: poro_elastic_posterior}} using only 10\% dropout in the \add{MF} conditional approximation in BMFIA: Comparison of posterior distributions (SNR of 50) for the isotropic permeability field $k(\bx,\bc)$. The posteriors' mean functions (upper row) and $2\cdot \mathbb{STD}$ (bottom row) are depicted. \textbf{Left block:} Posterior distribution using only the porous medium LF model. \textbf{Middle block:} BMFIA posterior (calculated with only 10\% dropout) that approximates the HF posterior (here unavailable) of the poro-elastic model using $\ntrain=300$. \textbf{Right block:} HF ground-truth field $\txgt(\bxgt,\bc)$ (normally unavailable).}
    \label{fig: dropout_10}
\end{figure}

We first lower the dropout rate from 30\% to 10\% in the dense layers of the probabilistic convolutional autoencoder model. This can lead to better point-wise predictions of the \add{MF} conditional model, but leads to noisier model gradients \wrt variations in $\bx$. All other settings are analogous to Section \ref{sec: poro_elastic}. We show the effect on the BMFIA posterior results in Figure \ref{fig: dropout_10} as the sample-wise predictions are almost unaltered.
Figure \ref{fig: dropout_10} clearly shows the inferior posterior quality compared to Figure \ref{fig: poro_elastic_posterior} due to the noisier gradients.

\begin{figure}[htbp]
    \centering
    \begin{tikzpicture}
    \node[anchor=south west,inner sep=0] (background) at (0,0) {\includegraphics[scale=0.42]{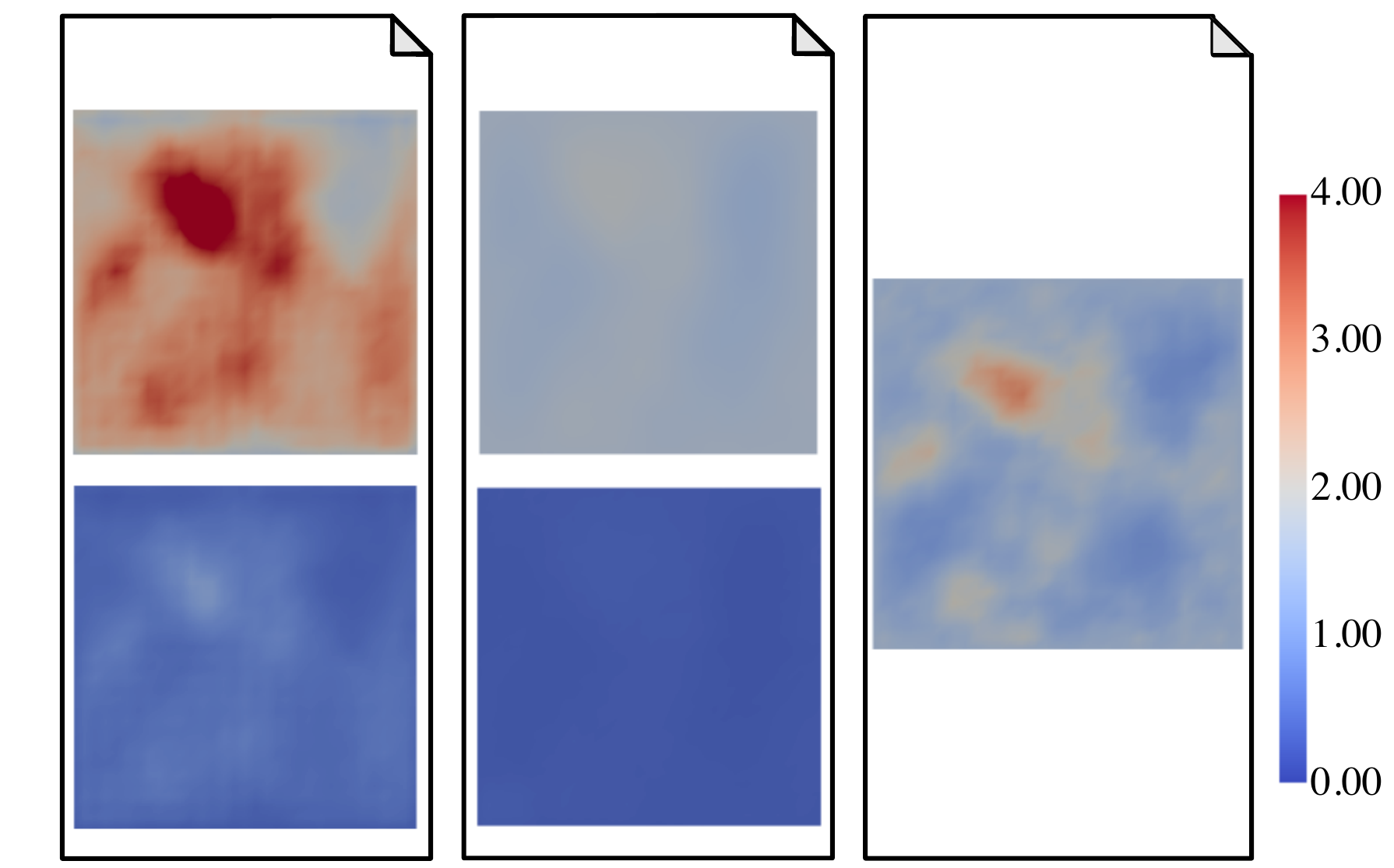}}; 
    \begin{scope}[x={(background.south east)},y={(background.north west)}]    
        \node at (0.17, 0.91) {LF};   
        \node at (0.46, 0.95) {BMFIA};   
        \node at (0.46, 0.90) {without LF output $\Ylf$};   
        \node at (0.75, 0.91) {HF - gt.};   
        \node at (0.93, 0.83) {$k$};   
        \node[rotate=90] at (0.02, 0.69) {mean};   
        \node[rotate=90] at (0.02, 0.23) {$2\cdot\mathbb{STD}$};   
        \end{scope}
    \end{tikzpicture}
    \caption{\textbf{Counter example to Figure \ref{fig: poro_elastic_posterior}} neglecting the LF model output and only using the (interpolated) input field $X$. This approach resembles building a direct but probabilistic surrogate and using it within the BMFIA formulation, which is different from sampling on a direct (deterministic) surrogate. The resulting posterior loses detail due to the large conditional uncertainty in the (probabilistic) mapping $X\mapsto \Yhf$ \add{and underestimates the uncertainty.}.
    Comparison of posterior distributions (SNR of 50) for the isotropic permeability field $k(\bx,\bc)$. The posteriors' mean functions (upper row) and $2\cdot \mathbb{STD}$ (bottom row) are depicted. \textbf{Left block:} Posterior distribution using only the porous medium LF model. \textbf{Middle block:} BMFIA posterior neglecting the LF model, using $\ntrain=300$. \textbf{Right block:} HF ground-truth field $\txgt(\bxgt,\bc)$ (normally unavailable).}
    \label{fig: direct_surrogate}
\end{figure}

To further clarify the role of the LF model output $\Ylf$ in our BMFIA formulation, we now consider a reduced setting where we construct a probabilistic surrogate directly from the inputs $\bx$ to the HF outputs, ignoring the LF outputs entirely by setting $\Zlf = X$. While this may seem attractive due to its simplicity, it fundamentally lacks the \emph{guidance} provided by the statistically correlated LF model, and as a result, suffers from the \emph{curse of dimensionality}.
Importantly, even though we learn the mapping directly from $\bx$ to $\Yhf$, this approach should not be confused with a classical direct surrogate. Classical surrogates typically learn a deterministic mapping and subsequently generate samples by evaluating this mapping, thereby neglecting model uncertainty and often leading to biased or overconfident posteriors. In contrast, \emph{we continue to use the BMFIA framework}, which maintains a probabilistic treatment of the surrogate.
\add{However, the resulting} posterior distribution becomes diffuse, lacks fine detail \add{and underestimates the posterior uncertainty}, as shown in Figure~\ref{fig: direct_surrogate}.

\end{document}